\documentclass[12pt,english,preprint]{revtex4-1}
\usepackage[T1]{fontenc}
\usepackage[latin9]{inputenc}
\usepackage{float}
\usepackage{textcomp}
\usepackage{mathrsfs}
\usepackage{amsmath}
\usepackage{amssymb}
\usepackage{graphicx}
\usepackage{multirow}
\usepackage{array}

\makeatletter
\newcommand{\lsim}{\lesssim}

\pdfpageheight\paperheight
\pdfpagewidth\paperwidth

\providecommand{\tabularnewline}{\\}

\makeatother

\usepackage{babel}

\begin{document}
\allowdisplaybreaks[4]

\preprint{APS/123-QED}
\vspace*{1cm}
\title{Two and Three Pseudoscalar Production in $e^{+}e^{-}$ annihilation
and their contributions to $(g-2)_{\mu}$ \vspace*{1.5cm}}

\author{Wen Qin}
\email{wqin@hnu.edu.cn}

\affiliation{School of Physics and Electronics, Hunan University, Changsha 410082,
China}
\affiliation{Hunan Provincial Key Laboratory of High-Energy Scale Physics and
Applications, Hunan University, Changsha 410082, China}
\author{Ling-Yun Dai}
\email{dailingyun@hnu.edu.cn}

\affiliation{School of Physics and Electronics, Hunan University, Changsha 410082,
China}
\affiliation{Hunan Provincial Key Laboratory of High-Energy Scale Physics and
Applications, Hunan University, Changsha 410082, China}
\author{Jorge Portol\'es}
\email{Jorge.Portoles@ific.uv.es}

\affiliation{IFIC, CSIC - Universitat de Val\`encia, Apt. Correus 22085, E-46071 Val\`encia, Spain \vspace*{1cm}}
\date{\today}\vspace*{1cm}

\begin{abstract}
A coherent study of $e^{+}e^{-}$ annihilation into two ($\pi^{+}\pi^{-}$,$K^{+}K^{-}$)
and three ($\pi^{+}\pi^{-}\pi^{0},\pi^{+}\pi^{-}\eta$) pseudoscalar
meson production is carried out within the framework of resonance
chiral theory in energy region $E\lesssim 2 \, \mathrm{GeV}$. The work
of {[}L. Y. Dai, J. Portol\'es, and O. Shekhovtsova,
Phys. Rev. D 88,056001 (2013){]} is revisited with the latest experimental
data and a joint analysis of two pseudoscalar meson production. Hence, we evaluate the
lowest order hadronic vacuum polarization contributions of those two and three pseudoscalar processes
to the anomalous magnetic moment of the muon. We also estimate some higher-order additions led by
the same hadronic vacuum polarization. Combined with
the other contributions from the standard model, the theoretical prediction
differs still by $(21.6 \pm 7.4)\times 10^{-10}$ (2.9$\sigma$) from the experimental value.
\end{abstract}

\maketitle

\section{\label{sec:level1}Introduction}

It is well known that Quantum Chromodynamics (QCD) is successful in
describing strong interactions. In the high energy region, the correlation
functions could be well determined by perturbative QCD. However, the
situation becomes more complicated in the low energy region, as the strong coupling
constant increases when the energy decreases. Fortunately, at the
very low energy region $E\ll M_{\rho}$ {[}$M_{\rho}$ being the mass
of the $\rho(770)${]}, the spontaneous chiral symmetry breaking of
QCD generates the pseudoscalar octet of Goldstone bosons, which are treated
as degrees of freedom in the effective field theory (EFT) of QCD:
chiral perturbation theory ($\chi$PT) \cite{Weinberg:1978kz,Gasser:1983yg}.
However, $\chi$PT is not the EFT in the intermediate energy region, $M_{\rho} \lsim E \lsim 2 \, \mbox{GeV}$,
where it is populated by dense spectra of resonances.
Resonance chiral theory (R$\chi$T) is a reasonable approach to
extent the working regime of $\chi$PT by including the resonances
as new degrees of freedom \cite{Ecker:1988te,Ecker:1989yg,Cirigliano:2006hb,Portoles:2010yt}.
The construction of the lagrangian is guided by Lorentz invariance and by
chiral  and discrete symmetries, i.e. C-, P-parity conservation.
The lack of a coupling that may guide a perturbative expansion in the calculations of the amplitudes,
is compensated by a model of the large-$N_C$ setting (being $N_C$ the number of colours)
\cite{tHooft:1973alw,tHooft:1974pnl,Witten:1979kh}.
As in $\chi$PT, this approach produces the relevant operators in the lagrangian, in terms
of Goldstone bosons, resonances and external fields, but
leaves undetermined their coupling constants.
\par
One may use experimental data to obtain information of the couplings. Meanwhile, there is one
theoretical tool that has proven efficient in this task: one can extract information of the coupling constants
by matching the perturbative Green functions of QCD currents, using the operator product expansion (OPE) at leading order,
with those constructed in the R$\chi$T framework
\cite{Knecht:2001xc,RuizFemenia:2003hm,Cirigliano:2004ue,Cirigliano:2005xn,Husek:2015wta,Dai:2019lmj,Kadavy:2020hox}.
Actually, R$\chi$T can also match, by construction, with $\chi$PT by integrating out the resonances in the Lagrangian
\cite{Ecker:1989yg,Guo:2007ff}, allowing to relate their coupling constants, too.
Indeed R$\chi$T is successful in dealing with the lightest resonances and their
interaction with the lightest pseudoscalars. It has been well applied
in the study of hadron tau decays
\cite{Jamin:2008qg,Dumm:2009kj,Dumm:2009va,Guo:2010dv,Escribano:2013bca,Nugent:2013hxa,Miranda:2020wdg}, two-photon transition form factors \cite{Chen:2012vw,Xiao:2015uva,Dai:2017tew},
and $e^{+}e^{-}$ annihilation in the nonperturbative regime of
QCD \citep{Dubinsky:2004xv,Dai:2013joa}.
\par
Low-energy processes with many hadrons in the final state involve final-state interactions (FSI) that are notoriously difficult to deal with in a model independent way.
 The use of dispersive approaches to deal with them is possible in some instances, namely when
good phenomenological data are available (see for instance
Refs.~\cite{Niecknig:2012sj,Schneider:2012ez,Danilkin:2014cra,Albaladejo:2017hhj,Isken:2017dkw,Colangelo:2018jxw,Yao:2020bxx} for some recent work).
In the framework of R$\chi$T, this is also achievable as we did in Ref.~\cite{Dai:2013joa},
where both vector-meson dominance and the anomalous terms were considered
in a coherent analysis of the $e^{+}e^{-}\to\pi^{+}\pi^{-}\pi^{0},\pi^{+}\pi^{-}\eta$
channels, in the energy region populated by many hadron resonances up to $E\lesssim 2.3 \, \mbox{GeV}$. Here we will revisit that work
and extend it to two pseudoscalar production in the light of the new data.
\par
Recent interest on $e^{+}e^{-}$ annihilation into two and three
pesudoscalars is driven by their contribution to the anomalous magnetic moment of the muon
$a_{\mu}=(g_{\mu}-2)/2$, with $g_{\mu}$ the muon Land\'e factor. The theoretical prediction of $a_{\mu}$ has become
a major tour de force in the last years because,
on the experimental side, it has been measured with high
precision, $a_{\mu}^{\mathrm{exp}}=11659208.9(6.3)\times10^{-10}$
{[}\citealp{Bennett:2006fi},\citealp{Zyla:2020zbs}{]}, and there seems to be a $3.3 \, \sigma$ \cite{Zyla:2020zbs} or $3.7 \, \sigma$ \cite{Aoyama:2020ynm}  discrepancy from the standard model (SM)
prediction. This fact paves the possibility of bringing out new physics contributions.
Within the standard model \cite{Jegerlehner:2017gek,Aoyama:2020ynm}, the most important contribution, the electromagnetic one,
is accurately calculated up to tenth-order $\alpha_{e}^{5}$,
$a_{\mu}^{\mathrm{QED}}=11658471.8931(104)\times10^{-10}$, with very
small uncertainty \citep{Aoyama:2012wk,Aoyama:2019ryr}. The electroweak contribution
at the two-loop level is also well determined as $a_{\mu}^{\mathrm{EW}}=15.36(0.1)\times10^{-10}$
\cite{Jackiw:1972jz,Knecht:2002hr,Czarnecki:2002nt,Gnendiger:2013pva}.
The hadronic contribution is considered as the major source of uncertainty and has two components:
hadronic light-by-light scattering (HLBL) and hadronic vacuum polarization (HVP).
The HLBL cannot be directly estimated from experimental
input, and a combination of different theoretical models has estimated
it as $a_{\mu}^{\mathrm{HLBL}}=9.2(1.8)\times10^{-10}$ \cite{Aoyama:2020ynm,Prades:2009tw,Colangelo:2019uex,Danilkin:2019opj}.
The lattice calculations on HLBL and HVP could be found in, e.g. Refs.~\cite{Borsanyi:2020mff,Blum:2016lnc,Blum:2017cer,Asmussen:2019act}.
A comprehensive amplitude analysis on $\gamma\gamma\to\pi\pi,K\bar{K}$ is
done in Refs.\cite{Dai:2014lza,Dai:2014zta,Dai:2016ytz,Dai:2017cvz}.
They are indeed the constraints on HLBL where the photons are real.
HVP is the largest hadronic contribution and it is related
with the cross section of $e^{+}e^{-}\to \, \mbox{anything}$ throughout causality
and unitarity \footnote{ We note that in the early works \cite{Terazawa:1968jh,Terazawa:1969ih}, the upper limit of HVP contribution has been given.    }.
The present value for the leading order HVP contribution is $a_{\mu}^{\mathrm{HVP},\mathrm{LO}}=694.0(4.0)\times10^{-10}$
\citep{Davier:2019can}. And the next-to-leading order and next-to-next-to-leading
order HVP corrections are derived by considering also higher order
hadronic loops, $a_{\mu}^{\mathrm{HVP},\mathrm{NLO}}=-9.83(0.07)\times10^{-10}$\citep{Keshavarzi:2019abf},
$a_{\mu}^{\mathrm{HVP},\mathrm{NNLO}}=1.24(0.01)\times10^{-10}$ \citep{Kurz:2014wya}.
The computation of the HVP contribution relies heavily on the available experimental data and, consequently,
its improvement will come from the accurate measurement of the electron-positron cross-section.
\par
Comparing the theoretical predictions with the SM and the experimental measurement,
there is still a discrepancy, as commented above.
There are lots of experimental data available. However, there are discordances among different collaborations,
even those with the highest statistics datasets.
The study of three pseudoscalar production was carried out in Ref.~\citep{Dai:2013joa}, but
recently new experimental measurements of $e^{+}e^{-}\to\pi^{+}\pi^{-}\pi^{0},\pi^{+}\pi^{-}\eta$
have become available. SND \citep{Aulchenko:2015mwt} has given a new
measurement of $e^{+}e^{-}\to\pi^{+}\pi^{-}\pi^{0}$ in the energy
range $1.05-2.00$ GeV. BESIII \citep{Ablikim:2019sjw} provided a
measurement for $e^{+}e^{-}\to\pi^{+}\pi^{-}\pi^{0}$ in a wide energy
range between 0.7 and 3.0 GeV using the Initial State Radiation (ISR)
method. SND also measured $e^{+}e^{-}\to\eta\pi\pi$ channel with
$\eta$ in $\eta\rightarrow\gamma\gamma$ mode {[}\citealp{Aulchenko:2014vkn}{]}
and $\eta\rightarrow3\pi^{0}$ mode \citep{Achasov:2017kqm}, and
a combined results of these two modes were provided in \citep{Achasov:2017kqm}.
CMD3 \citep{Gribanov:2019qgw} also measured $e^{+}e^{-}\to\eta\pi\pi$
in $\eta\rightarrow\gamma\gamma$ mode, and the cross section values
combined with its previous measurements were provided.
Very recently, BESIII measured $e^{+}e^{-}\to\eta'\pi\pi$ above 2 GeV~\cite{Ablikim:2020wyk}.
Besides, there
are also new experimental measurements for the two pseudoscalar cases.
BaBar \citep{Lees:2012cj} measured $e^{+}e^{-}\to\pi^{+}\pi^{-}$
from threshold up to 3 GeV. KLOE has done three precise measurements
of $e^{+}e^{-}\to\pi^{+}\pi^{-}$ \cite{Ambrosino:2008aa,Ambrosino:2010bv,Babusci:2012rp},
using ISR below 1 GeV, and a combined results with all these three
measurements were provided in Ref.~\citep{Anastasi:2017eio}. There
are also precise measurements below 1 GeV, such as, SND \citep{Achasov:2020iys},
BESIII \citep{Ablikim:2015orh} and CLEO \citep{Xiao:2017dqv}. Before
2008, there are also lots of experiment datasets, CMD2 \cite{Aulchenko:2006na,Akhmetshin:2006wh,Akhmetshin:2006bx},
DM2 \citep{Bisello:1988hq} and CMD \& OLYA \citep{Barkov:1985ac}. In
contrast, the $e^{+}e^{-}\to K^{+}K^{-}$ process has a considerably
shorter history starting from SND \citep{Achasov:2000am} in 2001.
Later, SND updated the measurements in 2007 \citep{Achasov:2007kg},
and the most recent one in 2016 \citep{Achasov:2016lbc}. In 2019,
a high precision measurement has been given by BESIII~\cite{Ablikim:2018iyx}. There
are also some other measurements from BarBar \citep{Lees:2013gzt},
CMD2 \citep{Akhmetshin:2008gz} and CMD3 \citep{Kozyrev:2017agm}.\\
\indent In this paper, we give a coherent analysis of $e^{+}e^{-}$
annihilation into two pseudoscalars $\pi^{+}\pi^{-}$,$K^{+}K^{-}$
and three pseudoscalars $\pi^{+}\pi^{-}\pi^{0},\pi^{+}\pi^{-}\eta$
based on the former work \citep{Dai:2013joa}, combined with all the recent
experimental measurements. In Sec.~II we will briefly update the theoretical
framework and give the amplitudes calculated by R$\chi$T. In Sec.~III, we fit the amplitudes to the experimental data
up to 2.3 GeV. In Sec.~IV, the leading order HVP contribution to $g-2$ is estimated. Higher-order hadronic contributions
are considered in Sec.~V.
Finally, we collect our conclusions in Sec.~VI. An appendix collects detailed expressions for the involved form factors
and decay widths.

\section{\label{sec:level2} Theoretical Framework Updates and notations}
\subsection{R$\chi$T and further improvements on the form factors\label{subsec:Resonance-Chiral-Theory}}
Massless QCD exhibits a chiral symmetry that rules its effective field theory at low energy.
$\chi$PT, valid at $E \ll M_{\rho}$, provides the interaction between
the lightest octet of pseudoscalar mesons, and of these with external currents.
At higher energies we need to take into account the hadronic resonance states, and a
successful phenomenological approach is provided by R$\chi$T. Only the aspects of interest for our
case are collected here. We follow the language and notation of Ref.~\cite{Cirigliano:2006hb}.
\par
The structure of the lagrangian has, essentially, three pieces:
\begin{equation} \label{eq:rchtl1}
{\cal L}_{\mathrm{R}\chi\mathrm{T}} \, = \, {\cal L}_{\mathrm{GB}} \, + \, {\cal L}^{\mathrm{V}}_{\mathrm{kin}} \, + \, {\cal L}_{\mathrm{V-GB}} \,  .
\end{equation}
The first piece involves interaction terms with Goldstone bosons that cannot be generated by integrating out
the vector resonance states. They are characterized by a perturbative expansion in terms of momenta (and masses), as in $\chi$PT.
${\cal L}^{\mathrm{V}}_{\mathrm{kin}}$ involves the kinetic term of the  vector resonance states and ${\cal L}_{\mathrm{V-GB}}$ the interaction between Goldstone bosons and vector resonance fields. For the processes that we study in this work only the vector resonance fields will be needed. All of these lagrangians include also external fields coupled to scalar, pseudoscalar, vector, axial-vector or tensor currents. The lowest even-intrinsic-parity ${\cal O}(p^2)$ of the ${\cal L}_{\mathrm{GB}}$ Lagrangian is given by
\begin{equation}
\mathcal{L}_{(2)}^{\mathrm{GB}}\equiv\mathcal{L}_{(2)}^{\chi\mathrm{PT}}=\frac{F^{2}}{4}\left\langle u_{\mu}u^{\mu}+\chi_{+}\right\rangle \ ,\label{eq:CPT}
\end{equation}
being $F$ the decay constant of the pion and $\langle ... \rangle$ indicates the trace in the $SU(3)$ space.
The leading Wess-Zumino-Witten term describing the anomaly with
odd-intrinsic-parity is of ${\cal O}(p^4)$ \cite{Witten:1983tw,Wess:1971yu}. The explicit expression of interest for our work
is given by
\begin{equation}
\mathcal{L}_{(4)}^{\mathrm{GB}} \, = \, i\frac{N_{C}\sqrt{2}}{12\pi^{2}F^{3}} \, \varepsilon_{\mu\nu\rho\sigma} \, \left\langle \partial^{\mu}\Phi\partial^{\nu}\Phi\partial^{\rho}\Phi v^{\sigma}\right\rangle +\cdots\ \ ,\label{eq:WZW}
\end{equation}
where $v^{\sigma}$ is the external vector current and $\Phi$ the multiplet of Goldstone bosons.
Higher orders of the ${\cal L}_{\mathrm{GB}}$ lagrangian will not be considered, as we assume that their couplings are dominated by resonance contributions \footnote{Up to ${\cal O}(p^4)$ at least, this setting depends on the realization of the spin-1 resonance fields.
In Ref.~\cite{Ecker:1989yg}, it was proven that this assumption is correct if one uses the antisymmetric formulation for those fields, as we do.}.
\par
The kinetic term of the vector resonance field is given by
\begin{equation} \label{eq:vkin}
\mathcal{L}_{\mathrm{kin}}^{\mathrm{V}}=-\frac{1}{2}\left\langle \nabla^{\lambda}V_{\lambda\mu}\nabla_{\nu}V^{\nu\mu}\right\rangle +\frac{1}{4}M_{V}^{2}\left\langle V_{\mu\nu}V^{\mu\nu}\right\rangle \ ,
\end{equation}
Here the resonances are collected as $SU(3)_{V}$ octets and have the corresponding properties under chiral
transformations. The Lagrangian that involves the interaction between Goldstone bosons and vector resonances, ${\cal L}_{\mathrm{V-GB}}$, couples the later octets with a chiral tensor constituted by the pseudoscalar nonet and external fields. Hence these chiral tensors obey a chiral counting ${\cal O}(p^n)$. This allows us to assign a label $n$ to the different pieces as
${\cal L}_{(n)}^{V...}$, where the numerator indicates the resonance fields in the interaction terms.
We will consider
\begin{equation} \label{eq:vchit}
\mathcal{L}_{\mathrm{V-GB}}=\mathcal{L}_{(2)}^{\mathrm{V}}+\mathcal{L}_{(4)}^{\mathrm{V}}+\mathcal{L}_{(2)}^{\mathrm{VV}}\ .
\end{equation}
For instance, in the antisymmetric formulation for the spin-one vector resonances that we use,
\begin{eqnarray} \label{eq:lvp2}
{\cal L}_{(2)}^{\mathrm{V}} &=&  \langle \, V_{\mu \nu} \, \chi_{(2)}^{\mu \nu} \, \rangle \, , \nonumber \\
\chi_{(2)}^{\mu \nu} & = & \frac{F_V}{2 \sqrt{2}} \, f_{+}^{\mu \nu} \, + \, i \, \frac{G_V}{\sqrt{2}} \, u^{\mu} \, u^{\nu} \,
\end{eqnarray}
where $F_V$ and $G_V$ are coupling constants not determined by the symmetry. The rest of terms in Eq.~(\ref{eq:vchit}) are collected
in Ref.~\cite{Cirigliano:2006hb} for the even-intrinsic-parity terms and Refs.~\cite{RuizFemenia:2003hm,Dumm:2009kj,Dai:2013joa} for those of odd-intrinsic parity.
The coupling constants of the interaction terms of $\mathcal{L}_{\mathrm{V-GB}}$ could be extracted from the phenomenology involving those states. As commented in the introduction the matching between the leading order in the OPE expansion of specific Green functions of QCD and their expressions within R$\chi$T is also a useful tool that has been employed in the bibliography
\cite{Knecht:2001xc,RuizFemenia:2003hm,Cirigliano:2004ue,Cirigliano:2005xn,Husek:2015wta,Dai:2019lmj,Kadavy:2020hox}. We will implement this procedure as far as it helps in our task. In particular we will use the relations between couplings specified in Ref.~\cite{Dai:2013joa}.
\par
However, the large energy region of study cannot be described fully with only one multiplet of vector resonances $V_{\mu \nu}$. The lightest one is situated around $M_{\rho}$, i.e. under $1 \, \mbox{GeV}$. Two other vector multiplets populate the interval $1 \, \mbox{GeV} \lsim E \lsim 2 \, \mbox{GeV}$, that we will call $V'_{\mu \nu}$ and $V''_{\mu \nu}$. Their couplings to the pseudoscalar mesons will be defined with respect to the ones of the lightest multiplet as $\beta_{\pi\pi}'$,$\beta_{\pi\pi}''$,$\beta_{KK}'$,$\beta_{KK}''$, through their poles, as
\begin{equation} \label{eq:betasdf}
\frac{1}{M_{V}^{2}-x}\rightarrow\frac{1}{M_{V}^{2}-x}+\frac{\beta_{\pi\pi,KK}^{\prime}}{M_{V^{\prime}}^{2}-x}+\frac{\beta_{\pi\pi,KK}^{\prime\prime}}{M_{V^{\prime\prime}}^{2}-x}\ .
\end{equation}
\par
The $\rho-\omega$ mixing, required by the $e^{+}e^{-}\to\pi^{+}\pi^{-}$ process, is reconsidered. While a constant mixing angle
$\delta_0$ is enough to describe mixing in the three pseudoscalar case as discussed
in Ref.~\citep{Dai:2013joa}:
\begin{equation}
\left(\begin{array}{c}
|\bar{\rho}^{0} \rangle\\
|\bar{\omega} \rangle
\end{array}\right)=\left(\begin{array}{cc}
\cos\delta_{0} & -\sin\delta_{0}\\
\sin\delta_{0} & \cos\delta_{0}
\end{array}\right)\left(\begin{array}{c}
|\rho^{0} \rangle\\
|\omega \rangle
\end{array}\right)\ ,\label{eq:constant-delta}
\end{equation}
an energy dependent mixing angle is discussed in Ref.~\citep{Gasser:1982ap}, although in the non-relativistic limit and we need
to generalize it to the relativistic case. The energy dependent mixing angle could be parameterized as

\begin{align}
\left(\begin{array}{c}
|\bar{\rho}^{0}\rangle\\
|\bar{\omega}\rangle
\end{array}\right) & =\left(\begin{array}{cc}
{\cos\delta} & {\frac{M_{V}\Gamma_{\rho}\sin\delta}{-(M_{V}^{2}-s)+iM_{V}(\Gamma_{\rho}-\Gamma_{\omega})}}\\
{\frac{M_{V}\Gamma_{\rho}\sin\delta}{-(M_{V}^{2}-s)-iM_{V}(\Gamma_{\rho}-\Gamma_{\omega})}} & {\cos\delta}
\end{array}\right)\left(\begin{array}{c}
\left|\rho^{0}\right\rangle \\
\left|\omega\right\rangle
\end{array}\right)\nonumber \\
 & \equiv\left(\begin{array}{cc}
{\cos\delta} & {-\sin\delta^{\omega}(s)}\\
{\sin\delta^{\rho}(s)} & {\cos\delta}
\end{array}\right)\left(\begin{array}{c}
\left|\rho^{0}\right\rangle \\
\left|\omega\right\rangle
\end{array}\right)\ ,\label{eq:energy-dependent-delta}
\end{align}

where $\left|\rho^{0}\right\rangle ,|\omega \rangle$ denote
the physical states. Hence the energy dependence of the mixing angle is driven by the resonance propagators. Here  $M_V$ is the
mass of the nonet of vector resonances in the $SU(3)$ limit. We will take $M_V = M_{\rho}$.
%
%
%
For the two body final state processes $e^{+}e^{-}\to\pi^{+}\pi^{-},K^{+}K^{-}$,
we always take energy dependent mixing mechanism according to Eq.~(\ref{eq:energy-dependent-delta}).
For the three body cases, we adopt two ways. One is to take the same energy dependent $\rho-\omega$
mixing mechanism as that of the two body case. This will be Fit I. The
other is to use the constant mixing angle $\delta_0$. This will be our Fit II. Comparison of both fits will unveil the influence of
$\rho-\omega$ mixing in the analysis of data.

\subsection{Cross sections for two and three pseudoscalar final states}

The amplitude for three-meson production in $e^+ e^-$ collisions is driven by the hadronization of the electromagnetic
current, in terms of one vector form factor only:
\begin{equation} \label{eq:fvv3}
\langle \pi^+(p_1) \pi^-(p_2) P (p_3) | \left( {\cal V}_{\mu}^3 + {\cal V}_{\mu}^8 /\sqrt{3} \right) \, e^{i
{\cal L}_{\mbox{\tiny{QCD}}}} | 0 \rangle \, = \, i \, F_V^P(Q^2,s,t) \, \varepsilon_{\mu \nu \alpha \beta} \,p_1^{\nu}
p_2^{\alpha} p_3^{\beta} \, ,
\end{equation}
being ${\cal V}_{\mu}^i = \overline{q} \gamma_{\mu} (\lambda^i/2) q$ and $P= \pi, \eta$. The Mandelstam variables are defined
as $s= (Q-p_3)^2$, $t=(Q-p_1)^2$, with $Q=p_1+p_2+p_3$.
The cross section and amplitudes for the three pseudoscalar cases that we are considering,
namely $e^+ e^- \rightarrow \pi^+ \pi^- \pi^0$ and $e^+ e^- \rightarrow \pi^+ \pi^- \eta$, are
quite the same as specified in Ref.~\citep{Dai:2013joa}, except for a
small change in the treatment of $\rho-\omega$ mixing, as illustrated
in Sect.~\ref{subsec:Resonance-Chiral-Theory}. The corresponding expressions for the cross-section and the modified form
factors for the three pseudoscalar cases are collected in Appendix~\ref{app:A}.
\par
These form factors depend on several couplings of the R$\chi$T lagrangian that are not determined by the symmetry. However,
some of them or, at least, relations between them can be established by matching Green functions calculated in this framework with
their expressions at leading order OPE expansion of QCD, as it has been commented before. By implementing these short-distance relations our form factors satisfy both the chiral constraints in the low-energy region and the asymptotic constraints at the high energy limit ($Q^2 \rightarrow \infty$). Hence the only unknown couplings in these form factors will be $F_V$, $2 g_4+g_5$, $d_2$, $c_3$ and
$\alpha_V$ \cite{Dai:2013joa}, to be added to the $\beta_{\pi \pi, KK}'$ and $\beta_{\pi \pi, KK}''$ from Eq.~(\ref{eq:betasdf}) and
the mixing angles between the octet and singlet pseudoscalar ($\theta_P$) and vector ($\theta_V$) components, defined also in
\cite{Dai:2013joa}.
\par
Two-pseudoscalar final states in $e^+ e^-$ annihilation are given by the corresponding vector form factor
\begin{equation} \label{eq:fvv2}
\langle P^+(p_1) P^-(p_2) | \left( {\cal V}_{\mu}^3 + {\cal V}_{\mu}^8 /\sqrt{3} \right) \, e^{i {\cal L}_{\mbox{\tiny{QCD}}}} | 0 \rangle \, = \, (p_1 - p_2)_{\mu} \, F_V^P(Q^2) \, ,
\end{equation}
with $Q=p_1+p_2$ and $P=\pi,K$. The energy in the center of mass frame is given by $E_{cm}\equiv\sqrt{Q^{2}}$.
The cross sections  $\sigma_{\pi \pi} \equiv \sigma(e^{+}e^{-}\to\pi^{+}\pi^{-})$ and $\sigma_{KK} \equiv
\sigma(e^{+}e^{-}\to K^{+}K^{-})$ are given by
\begin{equation} \label{eq:cxross}
\sigma_{PP}= \alpha_{e}^{2} \, \frac{\pi}{3 \, Q^2} \, \left(1-4 \frac{m_{P}^{2}}{Q^2} \right)^{3/2} \, |F_V^P(Q^{2})|^{2} \, .
\end{equation}
The form factors $F_V^{\pi}(Q^{2})$ and $F_V^K(Q^{2})$ were thoroughly
studied in Ref.~\citep{Arganda:2008jj} (see also \cite{Guerrero:1997ku,Pich:2001pj,Miranda:2018cpf} for alternative parameterizations) in the case of tau decays. Hence we need to include now the new $\rho-\omega$ mixing mechanism, present in $e^+ e^-$ into hadrons. We also extend the described energy region by adding heavier vector multiplets, as commented above. Their expressions are:
\begin{align}
F_V^{\pi} & =\left(1+\frac{F_{V}G_{V}}{F^{2}}Q^{2}\left(BW(M_{\rho},\Gamma_{\rho,},Q^{2})\right.\right.\nonumber \\
 & \left.\ \ +\beta_{\pi\pi}^{'}BW(M_{\rho^{'}},\Gamma_{\rho^{'},},Q^{2})+\beta_{\pi\pi}^{''}BW(M_{\rho^{''}},\Gamma_{\rho^{''},},Q^{2})\right)\nonumber \\
 & \ \ (\frac{1}{\text{\ensuremath{\sqrt{3}}}}\sin\theta_{V}\sin\delta^{\rho}+\cos\delta)\cos\delta\nonumber \\
 & -\frac{F_{V}G_{V}}{F^{2}}Q^{2}\left(BW(M_{\omega},\Gamma_{\omega,},Q^{2})+\beta_{\pi\pi}^{'}BW(M_{\omega^{'}},\Gamma_{\omega^{'},},Q^{2})\right.\nonumber \\
 & \left.\left.+\beta_{\pi\pi}^{'"}BW(M_{\omega^{''}},\Gamma_{\omega^{''},},Q^{2})\right)(\frac{1}{\text{\ensuremath{\sqrt{3}}}}\sin\theta_{V}\cos\delta-\sin\delta^{\omega})\sin\delta^{\omega}\right)\nonumber \\
 & \ \ \exp\left[\frac{-s}{96\pi^{2}F^{2}}\left({\rm Re}\left[A[m_{\pi},M_{\rho},Q^{2}]+\frac{1}{2}A[m_{K},M_{\rho},Q^{2}]\right]\right)\right]\ ,\label{eq:sigam_pipi}
\end{align}
\begin{align}
F_V^{K} & =\Bigm(\frac{\cos\theta_{V}{}^{2}}{2}\frac{F_{V}G_{V}}{F^{2}}(1+8\sqrt{2}\alpha_{V}\frac{2m_{K}^{2}-m_{\pi}^{2}}{M_{V}^{2}})M_{\phi}^{2}\bigm(BW(M_{\text{\ensuremath{\phi}}},\Gamma_{\phi},Q^{2})+\beta_{KK}^{'}BW(M_{\phi^{'}},\Gamma_{\phi^{'},},Q^{2})\nonumber \\
 & \ \ +\beta_{KK}^{''}BW(M_{\phi^{''}},\Gamma_{\phi^{''},},Q^{2}))+\frac{X_{1}}{24}\frac{F_{V}G_{V}}{F^{2}}(1+8\sqrt{2}\alpha_{V}\frac{m_{\pi}^{2}}{M_V^{2}})M_{\omega}^{2}\bigm(BW(M_{\omega},\Gamma_{\omega},Q^{2})\nonumber \\
 & \ \ +\beta_{KK}^{'}BW(M_{\omega^{'}},\Gamma_{\omega^{'},},Q^{2})+\beta_{KK}^{''}BW(M_{\omega^{''}},\Gamma_{\omega^{''},},Q^{2})\Bigm)\exp\left[\frac{-q^{2}}{96\pi^{2}F^{2}}\left(\frac{3}{2} {\rm Re} (A[m_{K},M_{\rho},Q^{2}])\right)\right]\nonumber \\
 & \ \ +\frac{X_{2}}{24}\frac{F_{V}G_{V}}{F^{2}}M_{\rho}^{2}(1+8\sqrt{2}\alpha_{V}\frac{m_{\pi}^{2}}{M_V^{2}})\Bigm(BW(M_{\rho},\Gamma_{\rho},Q^{2})+\beta_{KK}^{'}BW(M_{\rho^{'}},\Gamma_{\rho^{'},},Q^{2})\nonumber \\
 & \ \ +\beta_{KK}^{''}BW(M_{\rho^{''}},\Gamma_{\rho^{''},},Q^{2})\Big)\exp\left[\frac{-q^{2}}{96\pi^{2}F^{2}}\left({\rm Re}\left[A[m_{\pi},M_{\rho},Q^{2}]+\frac{1}{2}A[m_{K},M_{\rho},Q^{2}]\right]\right)\right]\text{\ .}
\label{eq:sigma_KK}
\end{align}
The functions in Eqs. (\ref{eq:sigam_pipi},\ref{eq:sigma_KK}) are given by:
\begin{eqnarray} \label{eq:func2}
\left[BW(M_{V},\Gamma_{V},Q^{2})\right]^{-1} & = & M_{V}^{2}-iM_{V}\Gamma_{V}(Q^{2})-Q^{2} \, , \nonumber \\
A\left(m_{P},\mu,Q^{2}\right)& = & \ln\left(m_{P}^{2}/\mu^{2}\right)+\frac{8m_{P}^{2}}{Q^{2}}-\frac{5}{3}+\sigma_{P}^{3}\ln\left(\frac{\sigma_{P}+1}{\sigma_{P}-1}\right) \, , \nonumber \\
\sigma_{P} & \equiv & \sqrt{1-4m_{P}^{2}/Q^{2}} \, ,
\end{eqnarray}
and
\begin{eqnarray} \label{eq:func22}
X_{1} & = & - \, 16\sqrt{3}\cos\delta\sin\theta_{V}\sin\delta^{\omega}(Q^{2})-6\cos^{2}\delta\cos2\theta_{V}+12\sin^{2}\delta_{\omega}(Q^{2})+3\cos2\delta+3\ , \nonumber \\
X_{2} & = & - \, 6\cos2\theta_{V}\sin^{2}\delta^{\rho}(Q^{2})+16\sqrt{3}\cos\delta\sin\theta_{V}\sin\delta^{\rho}(Q^{2}) \nonumber \\
& & + \, 6\sin^{2}\delta^{\rho}(Q^{2})+6\cos2\delta+6\ ,
\end{eqnarray}
Notice that $X_{1}=12\sin^{2}\theta_{V}$ and $X_{2}=12$ in the isospin limit. The angles $\sin \delta^{\rho,\omega}$ related with the $\rho-\omega$ mixing are
defined in Eq.~(\ref{eq:energy-dependent-delta}).
The $Q^2$ dependence of resonance widths are a debated issue. A thorough proposal within the chiral framework was proposed
in Ref.~\cite{GomezDumm:2000fz} for wide resonances. We will use this result for $\Gamma_{\rho}(Q^2)$, while a parameterization
in terms of the on-shell widths, driven by the two-body phase-space decay will be employed for $\Gamma_{\rho', \rho''} (Q^2)$.
The precise expressions are collected in Ref.~\cite{Dai:2013joa}. Meanwhile the rest of resonances, that are quite narrow, will
be taken constant.
Notice that the two-body vector form factors do not include more unknown couplings to those of three-body form factors.

\section{\label{sec:level3} Combined fit to experimental data}

As we have seen R$\chi$T provides a controlled setting to extract information from
experimental data. Part of, but not all, of the couplings have been
constrained by demanding that Green functions, in this framework, match the asymptotic behaviour of QCD, within the leading
term of the OPE expansion, in the high energy limit.
The remaining coupling constants, the mixing angles and resonance masses and on-shell widths are left to be determined
from the experimental data of cross sections and widths involving vector resonances.
\par
The unknown couplings include $F_{V},2g_{4}+g_{5},d_{2},c_{3},\alpha_{V}$,
the phenomenological parameters, $\beta_{X}^{'}$ and $\beta_{X}^{''}$
with $X=\pi,\eta,\pi\pi,KK$, counting for the corresponding
strength of the couplings of the $V'$ and $V''$ \footnote{Notice that $X=\pi,\eta$
appear in the three pseudoscalar final state $\pi^{+}\pi^{-}\pi^{0}$
and $\pi^{+}\pi^{-}\eta$, respectively and $X=\pi\pi,KK$ denote
the two pseudoscalar final state $\pi^{+}\pi^{-}$and $K^{+}K^{-}$,
respectively.}. The mixing angles of the pseudoscalar singlet and octet
$\theta_{P}$, that of vector singlet and octet $\theta_{V}$, and
the $\rho-\omega$ mixing angle, the energy dependent $\delta$ and/or
constant $\delta_{0}$ are also left free. The masses and widths of
resonances belonging to heavier second and third multiplets are also fitted around the central values listed
in PDG \citep{Zyla:2020zbs}.
\par
The last thirty years of experimental work have been very fruitful getting results for the cross-sections we are interested
in, as collected in Sect.~\ref{sec:level1}. In order to get results for our parameters we decide to fit the experimental
data of cross-sections obtained by dedicated experiments in the last twenty years, i.e. we exclude data older than 2000, with one exception: BESIII \citep{Ablikim:2019sjw} measured the cross section of $e^{+}e^{-}\to\pi^{+}\pi^{-}\pi^{0}$ with high statistics above $1.05 \, \mbox{GeV}$, while it has a relatively large uncertainty below that energy. Thus we do not fit the data points below $1.05 \, \mbox{GeV}$ from this dataset.
In addition we also fit the PDG figures \cite{Zyla:2020zbs} for the decay widths of vector resonances whose expressions are collected in Appendix~\ref{app:A}.
\par
Two fits are performed: Fit I uses a
uniform energy dependent $\rho-\omega$ mixing according to Eq.~(\ref{eq:energy-dependent-delta}).
In Fit II, the two body final state cases take into account the energy
dependent $\rho-\omega$ mixing, while the other processes are carried
out with a constant $\rho-\omega$ mixing angle, see Eq.~(\ref{eq:constant-delta}). The comparison between cross-section data
and the fit is shown in Figure~\ref{fig:Three-pseudo-case} for the three-pseudoscalar case and Figure~\ref{fig:Two-pseudo-case} for
the two-pseudocalar case. The captions in the figures collect all data used in the plots and in the fits.

The global fit includes decay widths of related resonances and their results are shown in Table~\ref{tab:decay-widths}.
The reported errors are obtained, in quadrature, from two components: one arises from the
Bootstrap method by varying the central value of experimental data within its error bar,
and the other comes from the statistics with dozens of solutions which
could also fit to the experimental data sets well. The latter one is the dominant
source of error estimation.  The cyan bands of all the
solutions of Fit II can be found in figure~\ref{fig:Three-pseudo-case} and
Figure~\ref{fig:Two-pseudo-case}.
In general, both Fit I and Fit II provide overall reasonable approximations to the
experimental figures quoted in the PDG \citep{Zyla:2020zbs}.

\subsection{Analysis of the results}

\begin{figure}[H]
\includegraphics[width=0.5\textwidth,height=0.23\textheight]{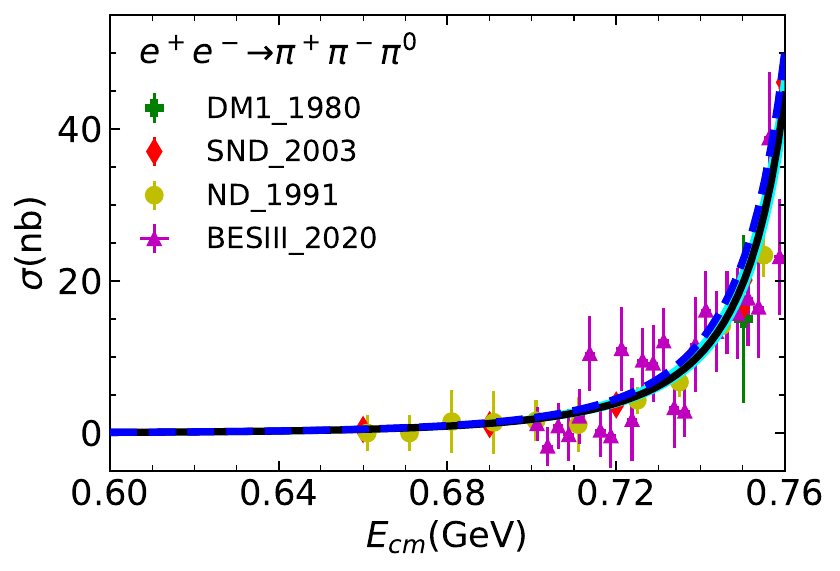}\includegraphics[width=0.5\textwidth,height=0.23\textheight]{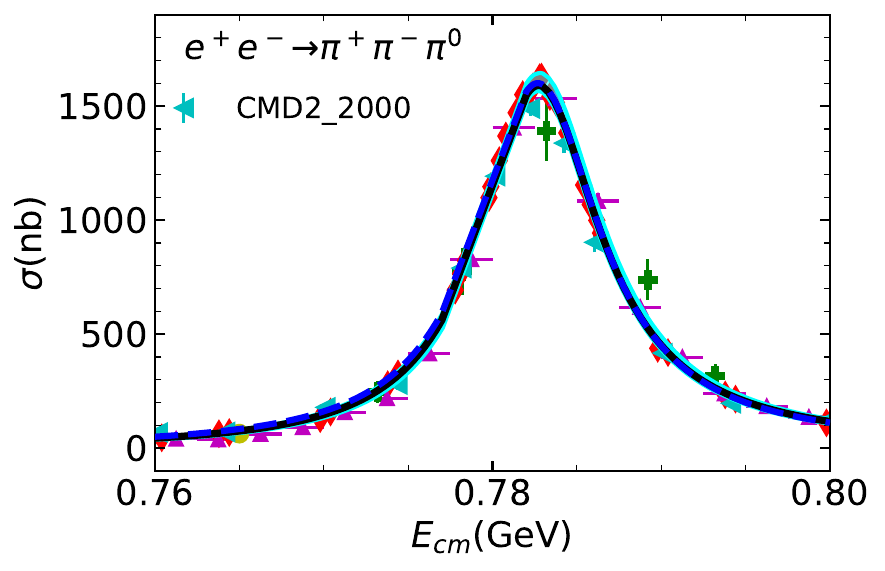}
\includegraphics[width=0.5\textwidth,height=0.23\textheight]{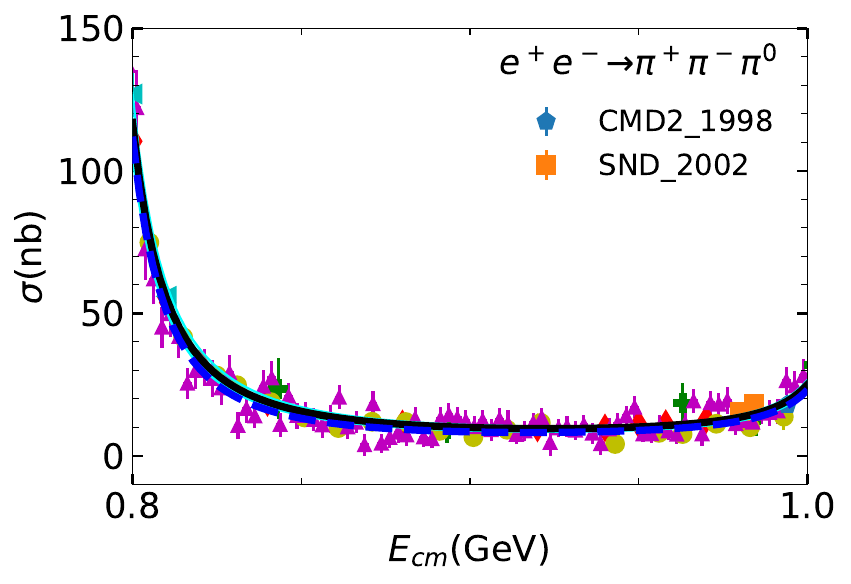}\includegraphics[width=0.5\textwidth,height=0.23\textheight]{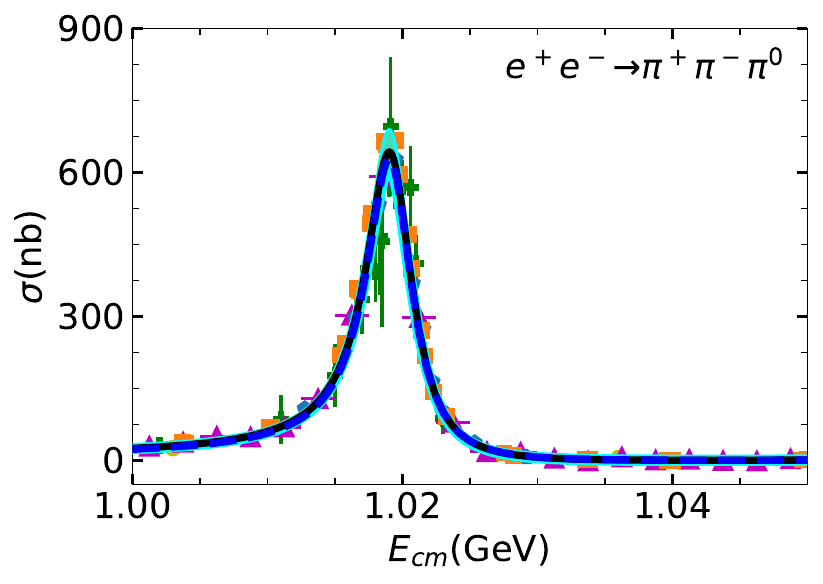}
\includegraphics[width=0.5\textwidth,height=0.24\textheight]{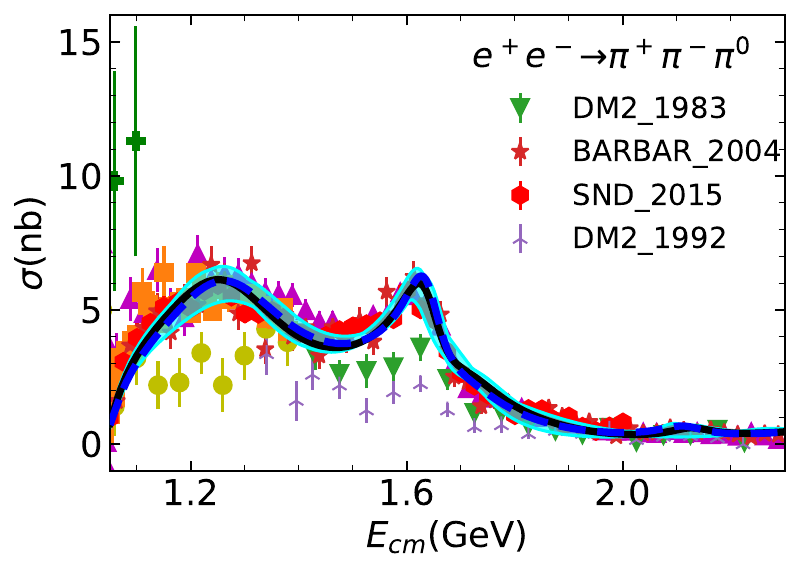}\includegraphics[width=0.5\textwidth,height=0.24\textheight]{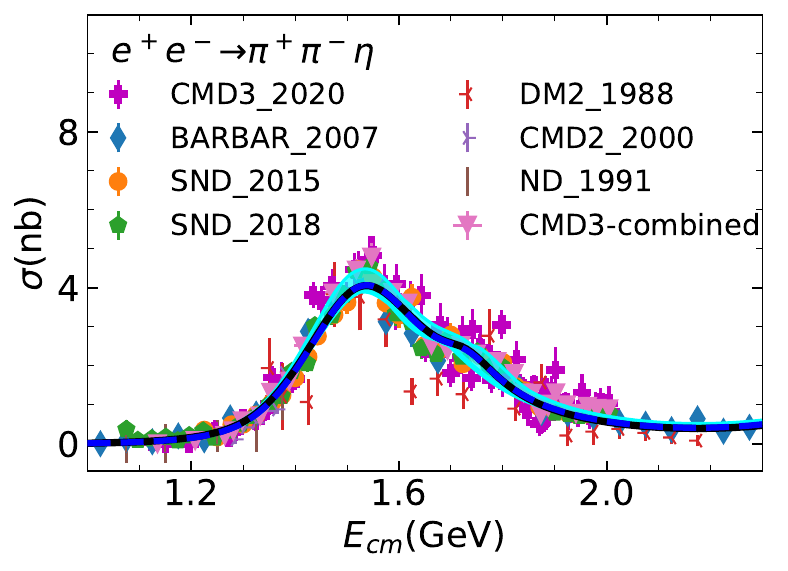}
\caption{Fit to the cross sections of $e^{+}e^{-}\to\pi\pi\pi,\pi\pi\eta$
of Fit I (dashed blue line) and Fit II (solid black line).  The cyan bands correspond to the uncertainty of Fit II.  The last graph is about $\eta\pi\pi$ channel and the others for $\pi\pi\pi$.
The experimental data displayed for $e^{+}e^{-}\to\pi\pi\pi$ are
from DM1 \citep{Cordier:1979qg}, ND \citep{Dolinsky:1991vq}, DM2 \citep{Antonelli:1992jx},
CMD2 \cite{Akhmetshin:2003zn,Akhmetshin:1998se,Akhmetshin:2000ca},
SND \citep{Achasov:2003ir,Achasov:2002ud,Aulchenko:2015mwt},
Babar \citep{Aubert:2004kj}, and BESIII \citep{Ablikim:2019sjw}.
The experimental data displayed for $e^{+}e^{-}\to\pi\pi\eta$ are
from DM2 \citep{Antonelli:1988fw}, ND \citep{Dolinsky:1991vq}, CMD2 \citep{Akhmetshin:2000wv},
Babar \citep{Aubert:2007ef}, SND \cite{Aulchenko:2014vkn,ACHASOV:2014nra,Achasov:2017kqm},
and CMD3 \citep{Gribanov:2019qgw}.
\label{fig:Three-pseudo-case}}
\end{figure}
\begin{figure}[H]
\includegraphics[width=0.49\textwidth,height=0.25\textheight]{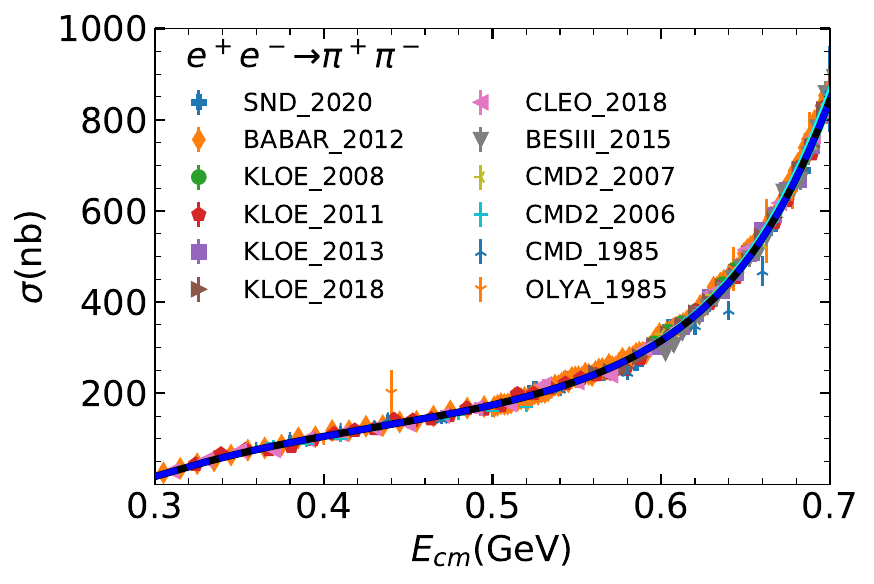}\includegraphics[width=0.52\textwidth,height=0.245\textheight]{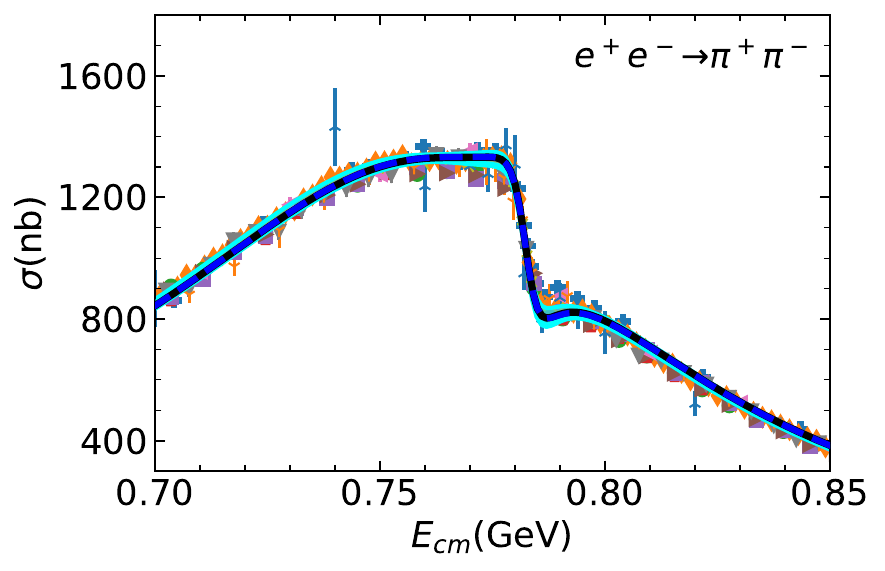}
\includegraphics[width=0.5\textwidth,height=0.25\textheight]{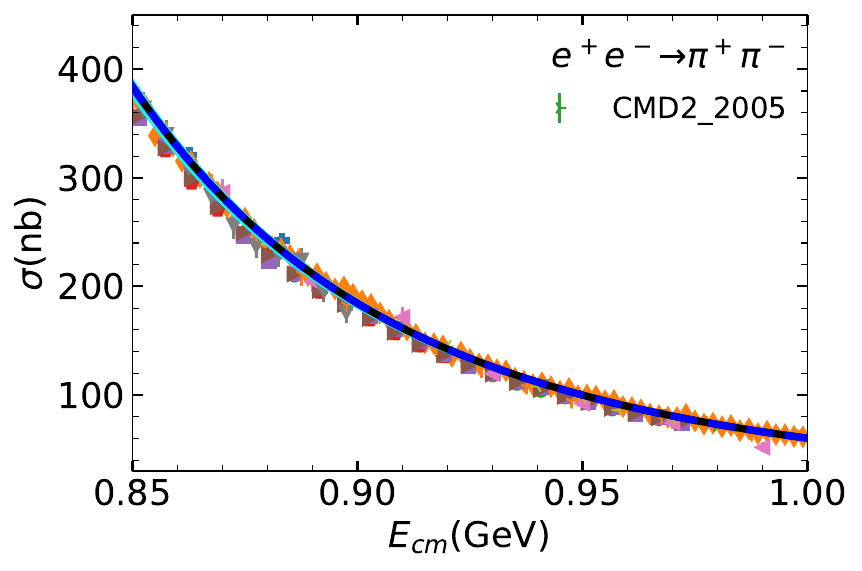}\includegraphics[width=0.5\textwidth,height=0.25\textheight]{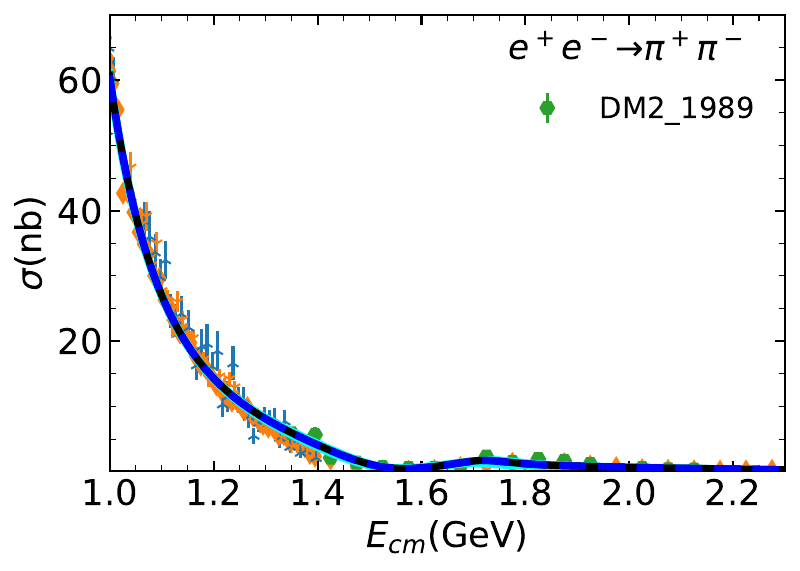}
\includegraphics[width=0.5\textwidth,height=0.25\textheight]{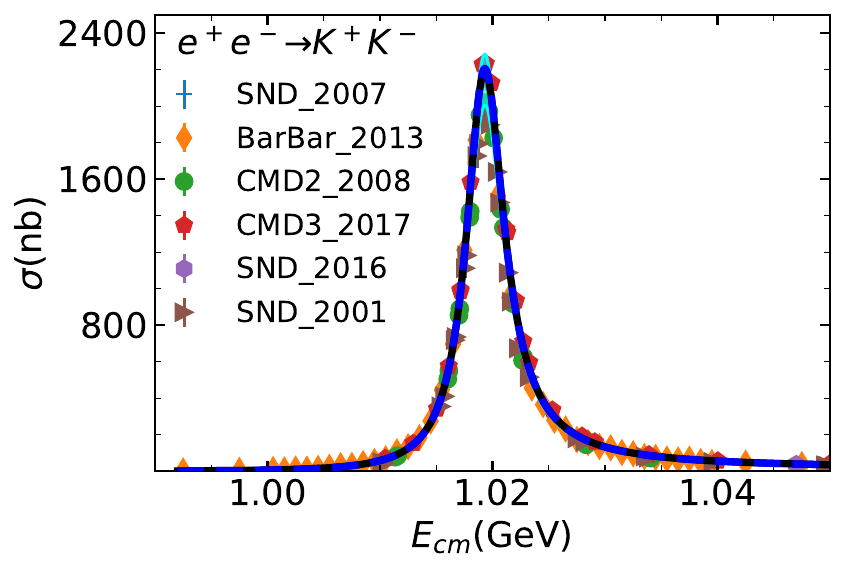}\includegraphics[width=0.5\textwidth,height=0.25\textheight]{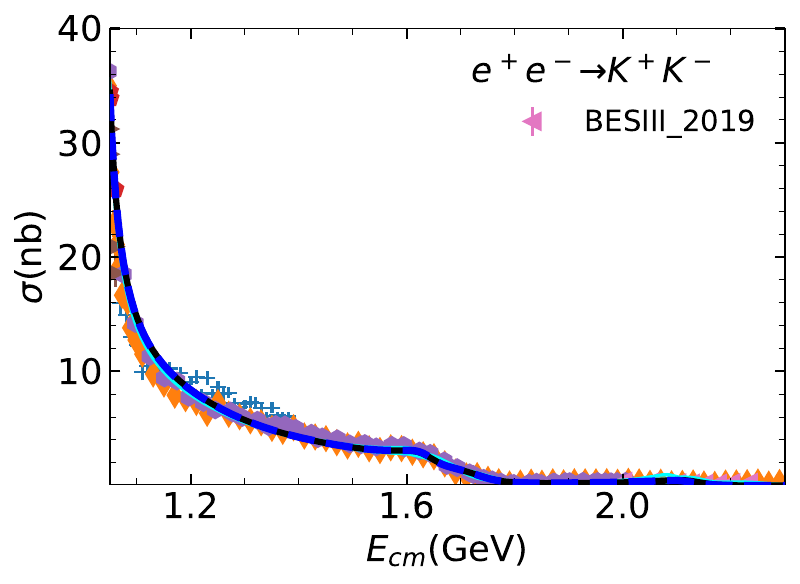}
\caption{Fit to the cross sections of $e^{+}e^{-}\to\pi^{+}\pi^{-},K^{+}K^{-}$
of Fit I (dashed blue line) and Fit II (solid black line).
The cyan bands corresponds to the uncertainty of Fit II.
The top four graphs are for $e^{+}e^{-}\to\pi^{+}\pi^{-}$ ,the bottom two
graphs are for $e^{+}e^{-}\to K^{+}K^{-}$.
The experimental data displayed for $e^{+}e^{-}\to\pi^{+}\pi^{-}$
are from BaBar \citep{Lees:2012cj}, KLOE \cite{Ambrosino:2008aa,Ambrosino:2010bv,Babusci:2012rp,Anastasi:2017eio},
SND \citep{Achasov:2020iys}, BESIII
\citep{Ablikim:2015orh}, CLEO
\citep{Xiao:2017dqv}, CMD2 \cite{Aulchenko:2006na,Akhmetshin:2006wh,Akhmetshin:2006bx},
DM2 \citep{Bisello:1988hq} and CMD \& OLYA \citep{Barkov:1985ac}. The
experimental data displayed for $e^{+}e^{-}\to K^{+}K^{-}$ are from
SND \citep{Achasov:2000am,Achasov:2007kg,Achasov:2016lbc},
BaBar \citep{Lees:2013gzt}, CMD2 \citep{Akhmetshin:2008gz},
CMD3 \citep{Kozyrev:2017agm}  and BESIII~\cite{Ablikim:2018iyx}.
\label{fig:Two-pseudo-case}}
\end{figure}

A comparison between our fitted parameters and those of Fit 4 in Ref.~\citep{Dai:2013joa}
is shown in Table~\ref{tab:fitted-parameters}. We also compare
the masses and widths of the resonances with those listed in PDG \citep{Zyla:2020zbs}.
The fitting procedure is carried out with MINUIT \citep{James:1975dr}.

The quoted errors in the fitted
parameters are provided by the Bootstrap method.
 In general, the parameters
in Fit I and Fit II are consistent with those of Fit 4 in Ref.~\citep{Dai:2013joa},
within a deviation of about $10\%$. $F_{V}$, $2g_{4}+g_{5}$,
$\theta_{V}$, $\delta_0$ and/or $\delta$ are mainly determined by
the experimental data under $1.05 \, \mbox{GeV}$, where it has higher statistics and precision.
However, the joint fit including the $e^{+}e^{-}\to K^{+}K^{-}$ process constrain $\theta_{V}$
strongly. This can be understood from the form factor in Eq.~(\ref{eq:sigma_KK}), where the cross
section around the $\phi$ peak increases with the descent of $\theta_{V}$.
In contrast, the cross section of $e^{+}e^{-}\to\pi^{+}\pi^{-}\pi^{0}$
around the $\phi$ peak decreases when $\theta_{V}$ goes down,
which could be deduced from the expressions in Appendix~\ref{app:A}.
As a consequence, $\theta_{V}$ is about $1^{\circ}$ larger than that of Ref.~\citep{Dai:2013joa}.
The inclusion of  $e^{+}e^{-}\to K^{+}K^{-}$ process also constrains $\alpha_{V}$,
the higher order correction to the $F_{V}$ coupling
arising from $SU(3)$ symmetry breaking. The cross section
of $e^{+}e^{-}\to K^{+}K^{-}$ increases with rising $\alpha_{V}$.
To confront the theoretical predictions to the experimental data of
the cross section of $e^{+}e^{-}\to K^{+}K^{-}$, $\alpha_{V}$ is fixed to be negative.
Notice that $\alpha_{V}$ is small as it is higher order correction.

The energy dependent $\rho-\omega$ mixing angle $\delta$ is determined
by the $e^{+}e^{-}\to\pi^{+}\pi^{-}$ process. From Eq.~(\ref{eq:sigam_pipi}),
the cross section of $e^{+}e^{-}\to\pi^{+}\pi^{-}$ is mainly determined
by $\delta$, since $F_{V}G_{V}/F^{2}=1$ is constrained by the high
energy behaviour and $\theta_{V}$ could be determined as above.
The two mechanisms of $\rho-\omega$ mixing adopted in Fit I and II
have almost no effects on the three body final state case. There is
only a very little difference reflected around the $\rho$ peak in the
$e^{+}e^{-}\to\pi^{+}\pi^{-}\pi^{0}$ process. In the energy region around their masses, $\rho$ and $\omega$ mix with
a relative phase that results in a larger mode of $|F_{V}^{\pi}|^{2}$.
Hence the magnitude of $F_{V}$ and $2g_{4}+g_{5}$ are smaller in
Fit I in comparison to Fit II and the results in Ref.~\citep{Dai:2013joa}.

The parameters related with the resonance multiplets are almost the
same in Fit I and Fit II, but some of them are different from those
of Ref.~\citep{Dai:2013joa}. They are mainly determined by the energy
region above $1.0 \, \mbox{GeV}$. Both $e^{+}e^{-}\to\pi^{+}\pi^{-}$ and $e^{+}e^{-}\to\eta\pi^{+}\pi^{-}$
processes are sensitive to the masses and widths of $\rho^{'}$ and $\rho^{''}$ in this energy region.
The $e^{+}e^{-}\to\pi^{+}\pi^{-}$ data gives
relative smaller masses and larger widths of $\rho^{'}$,
compared with those provided by the $e^{+}e^{-}\to\eta\pi^{+}\pi^{-}$ process.
Hence the combined fitted $\rho^{'}$ mass is about $30 \, \mbox{MeV}$ smaller
and the $\rho^{'}$ width is about $100 \, \mbox{MeV}$ larger than those in Ref.~\citep{Dai:2013joa}.
The mass and width of $\rho^{''}$ also changes slightly. Consequently, the relative
weights of the $e^{+}e^{-}\to\eta\pi^{+}\pi^{-}$ process $\beta'_{\eta}$
and $\beta''_{\eta}$ have sizable changes compared with those
in Ref.~\citep{Dai:2013joa}. Meanwhile,
the strengths of the $e^{+}e^{-}\to\pi^{0}\pi^{+}\pi^{-}$
process $\beta'_{\pi}$ and $\beta''_{\pi}$ are
similar. Notice that in the two-body processes $e^{+}e^{-}\to\pi^{+}\pi^{-}$
and $e^{+}e^{-}\to K^{+}K^{-}$, the parameters $\beta_{\pi\pi}^{'('')}$  and $\beta_{KK}^{'('')}$
turn out to be very small with magnitudes $\lesssim \, 0.2$, as
expected by lowest meson dominance \cite{Moussallam:1994xp,Moussallam:1997xx,Knecht:1999gb,Knecht:2001xc,Bijnens:2003rc}.

{\footnotesize{}}
\begin{table}[H]
\centering{}{\footnotesize{}}%
\begin{tabular}{c|cccc}
\hline
{\footnotesize{}Width} & {\footnotesize{}$\ \ \ $ Fit 1$\ \ \ $} & {\footnotesize{}$\ \ \ $ Fit II$\ \ \ $} & {\footnotesize{}$\ \ \ $ Ref.~ \citep{Dai:2013joa}$\ \ \ $} & {\footnotesize{}$\ \ \ $ PDG \citep{Zyla:2020zbs}$\ \ \ $}\tabularnewline
\hline
{\footnotesize{}$\Gamma_{\rho^{0}\to\pi\pi\pi}\;\;(10^{-5}\,\mbox{GeV})$} &  {\footnotesize{}0.86$\pm$0.31} & {\footnotesize{}0.64$\pm$0.49} & {\footnotesize{}0.93} & {\footnotesize{}1.49$_{-0.73}^{+0.94}$}\tabularnewline
{\footnotesize{}$\Gamma_{\omega\to\pi\pi\pi}\;\;(10^{-3}\,\mbox{GeV})$} & {\footnotesize{}7.43$\pm$0.78} & {\footnotesize{}7.96$\pm$0.74} & {\footnotesize{}7.66} & {\footnotesize{}7.58$\pm$0.05}\tabularnewline
{\footnotesize{}$\mathrm{\mathsf{\mathbf{\mathsf{\Gamma_{\phi\to\pi\pi\pi}}}}}\;\;(10^{-4}\,\mbox{GeV})$} & {\footnotesize{}9.08$\pm$1.57} & {\footnotesize{}9.00$\pm$1.14} & {\footnotesize{}6.25} & {\footnotesize{}6.53$\pm$0.14}\tabularnewline
{\footnotesize{}$\Gamma_{\rho\to ee}\;\;(10^{-6}\,\mbox{GeV})$} & {\footnotesize{}5.56$\pm$0.66} & {\footnotesize{}5.81$\pm$0.52} & {\footnotesize{}6.54} & {\footnotesize{}6.98$\pm$0.07}\tabularnewline
{\footnotesize{}$\Gamma_{\omega\to ee}\;\;\;(10^{-7}\,\mbox{GeV})$} & {\footnotesize{}7.28$\pm$0.85} & {\footnotesize{}7.60$\pm$0.65} & {\footnotesize{}6.69} & {\footnotesize{}6.25$\pm$0.13}\tabularnewline
{\footnotesize{}$\mathscr{\mathit{\mathtt{\Gamma_{\phi\to ee}\;}}}\;\;(10^{-6}\,\mbox{GeV})$} & {\footnotesize{}0.82$\pm$0.09} & {\footnotesize{}0.86$\pm$0.08} & {\footnotesize{}1.20} & {\footnotesize{}1.26$\pm$0.01}\tabularnewline
{\footnotesize{}$\Gamma_{\rho\to\pi\pi}\;\;(10^{-1}\,\mbox{GeV})$} & {\footnotesize{}1.30$\pm$0.17} & {\footnotesize{}1.24$\pm$0.11} & {\footnotesize{}1.14} & {\footnotesize{}1.48$\pm$0.01}\tabularnewline
{\footnotesize{}$\Gamma_{\omega\to\pi\pi}\;\;(10^{-4}\,\mbox{GeV})$} & {\footnotesize{}1.33$\pm$0.47} & {\footnotesize{}1.23$\pm$0.11} & {\footnotesize{}1.61} & {\footnotesize{}1.30$\pm$0.05}\tabularnewline
{\footnotesize{}$\Gamma_{\phi\to\pi\pi}\;\;(10^{-7}\,\mbox{GeV})$} & {\footnotesize{}1.82$\pm$0.20} & {\footnotesize{}1.91$\pm$0.18} & {\footnotesize{}2.66} & {\footnotesize{}3.10$\pm$0.55}\tabularnewline
{\footnotesize{}$\Gamma_{\rho^{0}\to\pi^{0}\gamma}\;\;(10^{-5}\,\mbox{GeV})$} & {\footnotesize{}4.60$\pm$0.64} & {\footnotesize{}5.38$\pm$0.64} & {\footnotesize{}5.96} & {\footnotesize{}6.95$\pm$0.89}\tabularnewline
{\footnotesize{}$\Gamma_{\rho^{+}\to\pi^{+}\gamma}\;\;(10^{-5}\,\mbox{GeV})$} & {\footnotesize{}4.46$\pm$0.62} & {\footnotesize{}4.53$\pm$0.37} & {\footnotesize{}4.81} & {\footnotesize{}6.65$\pm$0.74}\tabularnewline
{\footnotesize{}$\Gamma_{\omega\to\pi^{0}\gamma}\;\;(10^{-4}\,\mbox{GeV})$} & {\footnotesize{}3.97$\pm$0.47} & {\footnotesize{}4.07$\pm$0.35} & {\footnotesize{}4.43} & {\footnotesize{}7.13$\pm$0.19}\tabularnewline
{\footnotesize{}$\Gamma_{\phi\to\pi^{0}\gamma}\;(10^{-6}\,\mbox{GeV})$} & {\footnotesize{}9.01$\pm$2.26} & {\footnotesize{}9.17$\pm$1.30} & {\footnotesize{}7.34} & {\footnotesize{}5.52$\pm$0.21}\tabularnewline
{\footnotesize{}$\Gamma_{\rho\to\eta\gamma}\;\;(10^{-5}\,\mbox{GeV})$} & {\footnotesize{}3.95$\pm$0.69} & {\footnotesize{}4.32$\pm$0.38} & {\footnotesize{}4.85} & {\footnotesize{}4.43$\pm$0.31}\tabularnewline
{\footnotesize{}$\Gamma_{\omega\to\eta\gamma}\;\;(10^{-6}\,\mbox{GeV})$} & {\footnotesize{}4.42$\pm$0.77} & {\footnotesize{}3.77$\pm$0.48} & {\footnotesize{}4.13} & {\footnotesize{}3.82$\pm$0.34}\tabularnewline
{\footnotesize{}$\Gamma_{\phi\to\eta\gamma}\;\;(10^{-5}\,\mbox{GeV})$} & {\footnotesize{}5.92$\pm$0.78} & {\footnotesize{}6.10$\pm$0.48} & {\footnotesize{}6.57} & {\footnotesize{}5.54$\pm$0.11}\tabularnewline
{\footnotesize{}$\Gamma_{\eta'\to\rho\gamma}\;\;(10^{-5}\,\mbox{GeV})$} & {\footnotesize{}4.51$\pm$1.34} & {\footnotesize{}5.10$\pm$1.10} & {\footnotesize{}5.37} & {\footnotesize{}5.66$\pm$0.10}\tabularnewline
{\footnotesize{}$\Gamma_{\eta'\to\omega\gamma}\;(10^{-6}\,\mbox{GeV})$} & {\footnotesize{}6.24$\pm$1.77} & {\footnotesize{}5.52$\pm$0.94} & {\footnotesize{}5.12} & {\footnotesize{}4.74$\pm$0.13}\tabularnewline
{\footnotesize{}$\Gamma_{\phi\to\eta'\gamma}\;\;(10^{-7}\,\mbox{GeV})$} & {\footnotesize{}3.07$\pm$0.71} & {\footnotesize{}3.36$\pm$0.44} & {\footnotesize{}3.93} & {\footnotesize{}2.64$\pm$0.09}\tabularnewline
\hline
\end{tabular}{\footnotesize{}\caption{\label{tab:decay-widths} Decay widths involving vector resonances
compared with the Fit 4 of Ref.~\citep{Dai:2013joa} and PDG~\citep{Zyla:2020zbs}.}
}
\end{table}
Since $d_{2}$, $c_{3}$ and $\theta_{P}$
are mainly correlated with the $e^{+}e^{-}\to\eta\pi^{+}\pi^{-}$
process, they also have sizable changes, while masses and widths of other resonance multiplets are quite the
same. In summary, and as shown in Table~\ref{tab:fitted-parameters}, the fitted masses and
widths of heavier multiplets are closer to the experimental
average values in PDG~\citep{Zyla:2020zbs}, due to a combination of updated
experimental measurements and the constraints from $\pi^{+}\pi^{-}$
and $K^{+}K^{-}$ processes.

Notice however, that the masses and widths of $\rho'$ and $\rho''$ obtained here correspond to the
specific definition of the energy dependent width propagator shown
in Eq.~(40) of Ref.~\citep{Dai:2013joa}, which may not be used by the
experimentalists. Hence a precise comparison with the experimental
determinations is not straightforward.
\par
Finally $\theta_{V}$ and $\alpha_{V}$ change sizeably with respect to the results of Ref.~\citep{Dai:2013joa}
due to the inclusion of the process $e^{+}e^{-}\to K^{+}K^{-}$,
so that the partial widths sensitive to $\theta_{V}$ and $\alpha_{V}$
become worse. Nevertheless, these partial widths turn out to be bearable
with the experimental data from PDG~\citep{Zyla:2020zbs}, considering
the incertitude associated with the theoretical framework of large-$N_C$
expansion implemented in the framework of R$\chi$T.
In addition, the difference of partial widths of $\Gamma_{\rho^{0}\to\pi\pi\pi}$
in Fit I and Fit II are caused by the different parametrization of the
$\rho-\omega$ mixing. The $\rho^0$ decays in Fits. I and II have different mixing angles and also the former one is energy dependent, see eq.(2.9).
\par
The comparison of our solutions for the three pseudoscalar case with
experimental data is shown in Figure~\ref{fig:Three-pseudo-case}, and
that of the two pseudoscalar case is shown in Figur~\ref{fig:Two-pseudo-case}.
The results of Fit I are shown in blue dashed lines and those of Fit
II are shown in solid black lines. In general, Fit I and Fit II are
almost indistinguishable. There is slight difference shown around
the $\rho$ peak in $e^{+}e^{-}\to\pi^{+}\pi^{-}\pi^{0}$ process
at $0.6 \, < \, E \, < \, 1$ (GeV), due to the different parametrization of $\rho-\omega$
mixing adopted. Noted that Fit II is a little better in this region,
since there is one more parameter $\delta_{0}$ and the energy dependent mixing mechanism designed for the
$\pi\pi$ scattering may not be suitable for the three pion case, where the three body re-scattering needs to be considered.. Fit II seems also a little better
at the $\phi$ peak in the $e^{+}e^{-}\to\pi^{+}\pi^{-}\pi^{0}$ process. This is because that, $F_{V}$ and $2g_{4}+g_{5}$
in Fit II are allowed to have larger values than in Fit I, which can
slightly compensate the $\phi$ peak in $e^{+}e^{-}\to\pi^{+}\pi^{-}\pi^{0}$.
As illustrated above, the $\theta_{V}$ and $\alpha_{V}$ constrained
by the $e^{+}e^{-}\to K^{+}K^{-}$ will suppress the $\phi$ peak
in $e^{+}e^{-}\to\pi^{+}\pi^{-}\pi^{0}$. The high energy behaviour
of $e^{+}e^{-}\to\pi^{+}\pi^{-}$, as shown in Figure~\ref{fig:Two-pseudo-case},
is balanced with the $e^{+}e^{-}\to\eta\pi^{+}\pi^{-}$ process through
the mass and width of $\rho^{'}$.

{\footnotesize{}}
\begin{table}[H]
\vspace{-1cm}
\centering{}{\footnotesize{}}%
\begin{tabular}{c|cccc}
\hline
 & {\footnotesize{}$\ \ \ \ \ \ \ \ \ \ \ $Fit I$\ \ \ \ \ \ \ \ \ \ \ $} & {\footnotesize{}$\ \ \ \ \ \ \ \ \ \ \ $Fit II$\ \ \ \ \ \ \ \ \ \ \ $} & {\footnotesize{}$\ \ \ \ \ \ \ $Ref.~ \citep{Dai:2013joa}$\ \ \ \ \ \ \ $} & {\footnotesize{}$\ \ \ \ \ $PDG \citep{Zyla:2020zbs}$\ \ \ \ \ $}\tabularnewline
\hline
{\footnotesize{}$F_{V}\,\mbox{(GeV)}$} & {\footnotesize{}0.139$\pm$0.001} & {\footnotesize{}0.142$\pm$0.001} & {\footnotesize{}0.148$\pm$0.001} & {\footnotesize{}-}\tabularnewline
{\footnotesize{}$2g_{4}+g_{5}$} & {\footnotesize{}-0.442$\pm$0.001} & {\footnotesize{}-0.492$\pm$0.002} & {\footnotesize{}-0.493$\pm$0.003} & {\footnotesize{}-}\tabularnewline
{\footnotesize{}$d_{2}$} & {\footnotesize{}0.0273$\pm$0.0005} & {\footnotesize{}0.0276$\pm$0.0006} & {\footnotesize{}0.0359$\pm$0.0007} & {\footnotesize{}-}\tabularnewline
{\footnotesize{}$c_{3}$} & {\footnotesize{}0.00432$\pm$0.00012} & {\footnotesize{}0.00435$\pm$0.00013} & {\footnotesize{}0.00689$\pm$0.00017} & {\footnotesize{}-}\tabularnewline
{\footnotesize{}$\alpha_{V}$} & {\footnotesize{}-0.00120$\pm$0.00012} & {\footnotesize{}-0.00113$\pm$0.00014} & {\footnotesize{}0.0126$\pm$0.0007} & {\footnotesize{}-}\tabularnewline
{\footnotesize{}$\theta_{V}(^{\circ})$} & {\footnotesize{}39.61$\pm$0.01} & {\footnotesize{}39.56$\pm$0.01} & {\footnotesize{}38.94$\pm$0.02} & {\footnotesize{}-}\tabularnewline
{\footnotesize{}$\theta_{P}(^{\circ})$} & {\footnotesize{}-19.39$\pm$0.09} & {\footnotesize{}-19.61$\pm$0.10} & {\footnotesize{}-21.37$\pm$0.26} & {\footnotesize{}-}\tabularnewline
{\footnotesize{}$\delta_{0}(^{\circ})$} & {\footnotesize{}-} & {\footnotesize{}1.70$\pm$0.05} & {\footnotesize{}2.12$\pm$0.06} & {\footnotesize{}-}\tabularnewline
{\footnotesize{}$\delta(^{\circ})$} & {\footnotesize{}-1.83$\pm$0.04} & {\footnotesize{}-1.80$\pm$0.01} & {\footnotesize{}-} & {\footnotesize{}-}\tabularnewline
{\footnotesize{}$\beta_{\pi}'$} & {\footnotesize{}-0.434$\pm$0.005} & {\footnotesize{}-0.454$\pm$0.003} & {\footnotesize{}-0.469$\pm$0.008} & {\footnotesize{}-}\tabularnewline
{\footnotesize{}$\beta_{\pi}''$} & {\footnotesize{}0.239$\pm$0.002} & {\footnotesize{}0.224$\pm$0.005} & {\footnotesize{}0.225$\pm$0.007} & {\footnotesize{}-}\tabularnewline
{\footnotesize{}$\beta'_{\eta}$} & {\footnotesize{}-0.452$\pm$0.008} & {\footnotesize{}-0.438$\pm$0.006} & {\footnotesize{}-0.174$\pm$0.017} & {\footnotesize{}-}\tabularnewline
{\footnotesize{}$\beta''_{\eta}$} & {\footnotesize{}-0.0213$\pm$0.0031} & {\footnotesize{}-0.0233$\pm$0.0023} & {\footnotesize{}-0.0968$\pm$0.0139} & {\footnotesize{}-}\tabularnewline
{\footnotesize{}$\beta_{\pi\pi}'$} & {\footnotesize{}-0.0625$\pm$0.0007} & {\footnotesize{}-0.0625$\pm$0.0009} & {\footnotesize{}-} & {\footnotesize{}-}\tabularnewline
{\footnotesize{}$\beta_{\pi\pi}''$} & {\footnotesize{}0.0115$\pm$0.0006} & {\footnotesize{}0.0118$\pm$0.0007} & {\footnotesize{}-} & {\footnotesize{}-}\tabularnewline
{\footnotesize{}$\beta_{KK}'$} & {\footnotesize{}-0.0652$\pm$0.0023} & {\footnotesize{}-0.0712$\pm$0.0040} & {\footnotesize{}-} & {\footnotesize{}-}\tabularnewline
{\footnotesize{}$\beta_{KK}''$} & {\footnotesize{}-0.202$\pm$0.003} & {\footnotesize{}-0.197$\pm$0.005} & {\footnotesize{}-} & {\footnotesize{}-}\tabularnewline
{\footnotesize{}$M_{\rho'}\,\mbox{(GeV)}$} & {\footnotesize{}1.517$\pm$0.001} & {\footnotesize{}1.519$\pm$0.002} & {\footnotesize{}1.550$\pm$0.012} & {\footnotesize{}1.465(25)}\tabularnewline
{\footnotesize{}$\Gamma_{\rho'}\,\mbox{(GeV)}$} & {\footnotesize{}0.340$\pm$0.006} & {\footnotesize{}0.340$\pm$0.001} & {\footnotesize{}0.238$\pm$0.018} & {\footnotesize{}0.400(60)}\tabularnewline
{\footnotesize{}$M_{\omega'}\,\mbox{(GeV)}$} & {\footnotesize{}1.256$\pm$0.006} & {\footnotesize{}1.253$\pm$0.003} & {\footnotesize{}1.249$\pm$0.003} & {\footnotesize{}1.410(60)}\tabularnewline
{\footnotesize{}$\Gamma_{\omega'}\,\mbox{(GeV)}$} & {\footnotesize{}0.310$\pm$0.005} & {\footnotesize{}0.310$\pm$0.003} & {\footnotesize{}0.307$\pm$0.007} & {\footnotesize{}0.290(190)}\tabularnewline
{\footnotesize{}$M_{\phi'}\,\mbox{(GeV)}$} & {\footnotesize{}1.640$\pm$0.003} & {\footnotesize{}1.640$\pm$0.003} & {\footnotesize{}1.641$\pm$0.005} & {\footnotesize{}1.680(20)}\tabularnewline
{\footnotesize{}$\Gamma_{\phi'}\,\mbox{(GeV)}$} & {\footnotesize{}0.083$\pm$0.001} & {\footnotesize{}0.090$\pm$0.002} & {\footnotesize{}0.086$\pm$0.007} & {\footnotesize{}0.15(5)}\tabularnewline
{\footnotesize{}$M_{\rho''}\,\mbox{(GeV)}$} & {\footnotesize{}1.720$\pm$0.004} & {\footnotesize{}1.720$\pm$0.001} & {\footnotesize{}1.794$\pm$0.012} & {\footnotesize{}1.720(20)}\tabularnewline
{\footnotesize{}$\Gamma_{\rho''}\,\mbox{(GeV)}$} & {\footnotesize{}0.150$\pm$0.001} & {\footnotesize{}0.150$\pm$0.005} & {\footnotesize{}0.297$\pm$0.033} & {\footnotesize{}0.25(10)}\tabularnewline
{\footnotesize{}$M_{\omega''}\,\mbox{(GeV)}$} & {\footnotesize{}1.683$\pm$0.005} & {\footnotesize{}1.725$\pm$0.010} & {\footnotesize{}1.700$\pm$0.011} & {\footnotesize{}1.670(30)}\tabularnewline
{\footnotesize{}$\Gamma_{\omega''}\,\mbox{(GeV)}$} & {\footnotesize{}0.400$\pm$0.002} & {\footnotesize{}0.400$\pm$0.003} & {\footnotesize{}0.400$\pm$0.013} & {\footnotesize{}0.315(35)}\tabularnewline
{\footnotesize{}$M_{\phi''}$ (GeV)} & {\footnotesize{}2.114$\pm$0.010} & {\footnotesize{}2.126$\pm$0.025} & {\footnotesize{}2.086$\pm$0.022} & {\footnotesize{}2.160(80)}\tabularnewline
{\footnotesize{}$\Gamma_{\phi''}$ (GeV)} & {\footnotesize{}0.108$\pm$0.014} & {\footnotesize{}0.100$\pm$0.014} & {\footnotesize{}0.108$\pm$0.017} & {\footnotesize{}0.125(65)}\tabularnewline
\hline
\end{tabular}{\footnotesize{}\caption{\label{tab:fitted-parameters} Fitted parameters of Fits I and II
compared with the Fit 4 of Ref. \citep{Dai:2013joa} and PDG~\citep{Zyla:2020zbs}. The uncertainty of the parameters are coming from the Bootstrap method. }
}
\end{table}
{\footnotesize\par}

\section{\label{sec:g-2-th} Leading-order hadronic vacuum polarization contributions to $a_{\mu}$}
The Hadronic Vacuum Polarization (HVP) corrections to $a_{\mu}= (g_{\mu} - 2)/2$ are related to the $e^{+}e^{-}\to\text{ hadrons }$ cross
sections through the optical theorem and analyticity \cite{Gourdin:1969dm,Jegerlehner:2017gek}. The leading
order HVP correction can be expressed as
\begin{equation}
a_{\mu}^{\mbox{\tiny had}}=\left(\frac{\alpha_{e}(0)m_{\mu}}{3\pi}\right)^{2}\int_{s_{\mbox{\tiny thr}}}^{\infty}\mathrm{d}s\frac{\hat{K}(s)}{s^{2}}R_{\mathrm{h}}(s)\text{\ ,}\label{eq:amu}
\end{equation}
where
\begin{eqnarray} \label{eq:amusup}
\alpha_{e}=\frac{e^{2}}{4\pi}, & \qquad & \qquad  R_{\mathrm{h}}(s)=\frac{3s}{4\pi\alpha_{e}^{2}(s)} \, \sigma\left(e^{+}e^{-}\rightarrow\text{ hadrons }\right)\ ,
\end{eqnarray}
and the kernel function is defined as,
\begin{equation} \label{eq:amuker}
\hat{K}(s)=\frac{3s}{m_{\mu}^{2}}\left.\left[\frac{\left(1+x^{2}\right)(1+x)^{2}}{x^{2}}\left(\ln(1+x)-x+\frac{x^{2}}{2}\right)\right.\left.+\frac{x^{2}}{2}\left(2-x^{2}\right)+\frac{1+x}{1-x}x^{2}\ln x\right]\text{\ ,}\right.
\end{equation}
with
\begin{eqnarray} \label{eq:amusup2}
x=\frac{1-\beta_{\mu}(s)}{1+\beta_{\mu}(s)}, & \qquad & \qquad \beta_{\mu}(s)=\sqrt{1-\frac{4m_{\mu}^{2}}{s}}\ .
\end{eqnarray}
Notice that the lower limit in the integration in Eq.~(\ref{eq:amu}) depends on the starting contribution and its ${\cal O}(\alpha_e)$ order. Hence $s_{\mbox{\tiny thr}} = m_{\pi^0}^2$ when including the $\pi^0 \gamma$ contribution and
$s_{\mbox{\tiny thr}} = 4 m_{\pi}^2$ when starting in the $\pi \pi$ contribution.
\par
It is interesting to notice the $1/s^{2}$ enhancement factor (leading order) of  contributions of low energies in $a_{\mu}^{\mbox{\tiny had}}$  (\ref{tab:amu}). Thus the kernel gives higher weight, in particular, to the lowest lying resonance $\rho (770)$ that couples
strongly to $\pi^+ \pi^-$. This is the reason why the pion pair production $e^+ e^- \rightarrow \pi^+ \pi^-$ gives, by far,
the largest contribution to $a_{\mu}^{\mbox{\tiny had}}$.
However, we are in the position to determine the contributions to the muon anomalous magnetic moment relevant
to the three and two pseudoscalar final states that we discussed above. They
are shown as $a_{\mu}^{C}$ with different energy regions in Table~\ref{tab:amu}.

Here $a_{\mu}^{C}$ ($C=\pi\pi,KK,\pi\pi\pi,\eta\pi\pi$)
denotes for the lowest order hadronic vacuum polarization contribution
of $e^{+}e^{-}\to\pi\pi,KK,\pi\pi\pi,\eta\pi\pi$, respectively. The
error bars for $a_{\mu}^{C}$ are given by the combination of the
uncertainty coming from the Bootstrap method and the statistics
from dozens of solutions that also fit to the experimental data sets well.
{\footnotesize{}}
\begin{table}[H]
\centering{}{\footnotesize{}}%
\begin{tabular}{ccccccc}
\hline
{\footnotesize{}$a_{\mu}^{C}\times10^{-10}$} & {\footnotesize{}Ref. \citep{Colangelo:2018mtw}} & {\footnotesize{}Ref. \citep{Dai:2013joa}} & {\footnotesize{}Ref. \citep{Hoferichter:2019mqg}} & {\footnotesize{}Ref. \citep{Davier:2019can}} & {\footnotesize{}Fit I} & {\footnotesize{}Fit II}\tabularnewline
\hline
{\footnotesize{}$a_{\mu}^{\pi\pi}|_{\leq0.63\mathrm{GeV}}$} & {\footnotesize{}132.8(0.4)(1.0)} & {\footnotesize{}-} & {\footnotesize{}-} & {\footnotesize{}-} & {\footnotesize{}132.11$\text{\ensuremath{\pm}}$0.63} & {\footnotesize{}132.11$\text{\ensuremath{\pm}}$0.67}\tabularnewline
{\footnotesize{}$a_{\mu}^{\pi\pi}|_{\leq1\mathrm{GeV}}$} & {\footnotesize{}495.0(1.5)(2.1)} & {\footnotesize{}-} & {\footnotesize{}-} & {\footnotesize{}-} & {\footnotesize{}498.48$\text{\ensuremath{\pm}}$2.34} & {\footnotesize{}498.47$\text{\ensuremath{\pm}}$2.33}\tabularnewline
{\footnotesize{}$a_{\mu}^{\pi\pi}|_{\leq1\mathrm{.8GeV}}$} & {\footnotesize{}-} & {\footnotesize{}-} & {\footnotesize{}-} & {\footnotesize{}507.85$\pm$0.83$\pm$3.23$\pm$0.55} & {\footnotesize{}508.89$\text{\ensuremath{\pm}}$2.45} & {\footnotesize{}508.89$\text{\ensuremath{\pm}}$2.45}\tabularnewline
{\footnotesize{}$a_{\mu}^{\pi\pi}|_{\leq2.3\mathrm{GeV}}$} & {\footnotesize{}-} & {\footnotesize{}-} & {\footnotesize{}-} & {\footnotesize{}-} & {\footnotesize{}509.13$\text{\ensuremath{\pm}}$2.48} & {\footnotesize{}509.13$\text{\ensuremath{\pm}}$2.48}\tabularnewline
\hline
{\footnotesize{}$a_{\mu}^{KK}|_{\leq\mathrm{.1.1GeV}}$} & {\footnotesize{}-} & {\footnotesize{}-} & {\footnotesize{}-} & {\footnotesize{}-} & {\footnotesize{}20.73$\text{\ensuremath{\pm}}$0.94} & {\footnotesize{}20.74$\text{\ensuremath{\pm}}$0.88}\tabularnewline
{\footnotesize{}$a_{\mu}^{KK}|_{\leq\mathrm{.1.8GeV}}$} & {\footnotesize{}-} & {\footnotesize{}-} & {\footnotesize{}-} & {\footnotesize{}23.08$\pm$0.20$\pm$0.33$\pm$0.21} & {\footnotesize{}24.35$\text{\ensuremath{\pm}}$1.02} & {\footnotesize{}24.36$\text{\ensuremath{\pm}}$0.97}\tabularnewline
{\footnotesize{}$a_{\mu}^{KK}|_{\leq\mathrm{2.3GeV}}$} & {\footnotesize{}-} & {\footnotesize{}-} & {\footnotesize{}-} & {\footnotesize{}-} & {\footnotesize{}24.43$\text{\ensuremath{\pm}}$1.03} & {\footnotesize{}24.44$\text{\ensuremath{\pm}}$1.01}\tabularnewline
\hline
{\footnotesize{}$a_{\mu}^{\pi\pi\pi}|_{\leq1.8\mathrm{GeV}}$} & {\footnotesize{}-} & {\footnotesize{}48.55} & {\footnotesize{}46.2(8)} & {\footnotesize{}46.21$\pm$0.40$\pm$1.10$\pm$0.86} & {\footnotesize{}48.55$\text{\ensuremath{\pm}}$1.42} & {\footnotesize{}48.54$\text{\ensuremath{\pm}}$1.39}\tabularnewline
{\footnotesize{}$a_{\mu}^{\pi\pi\pi}|_{\leq2.3GeV}\mathrm{}$} & {\footnotesize{}-} & {\footnotesize{}-} & {\footnotesize{}-} & {\footnotesize{}-} & {\footnotesize{}48.76$\text{\ensuremath{\pm}}$1.45} & {\footnotesize{}48.75$\text{\ensuremath{\pm}}$1.43}\tabularnewline
\hline
{\footnotesize{}$a_{\mu}^{\eta\pi\pi}|_{\leq1.8\mathrm{GeV}}$} &  & {\footnotesize{}1.135} & {\footnotesize{}-} & {\footnotesize{}1.19$\pm$0.02$\pm$0.04$\pm$0.02} & {\footnotesize{}1.28$\ensuremath{\pm}$0.10} & {\footnotesize{}1.29$\text{\ensuremath{\pm}}$0.09}\tabularnewline
{\footnotesize{}$a_{\mu}^{\eta\pi\pi}|_{\leq2.3GeV}$} & {\footnotesize{}-} & {\footnotesize{}-} & {\footnotesize{}-} & {\footnotesize{}-} & {\footnotesize{}1.52$\text{\ensuremath{\pm}}$0.12} & {\footnotesize{}1.53$\text{\ensuremath{\pm}}$0.12}\tabularnewline
\hline
{\footnotesize{}$a_{\mu}^{\mathrm{HVP}\mathrm{.LO}}$} & {\footnotesize{}-} & {\footnotesize{}-} & {\footnotesize{}-} & {\footnotesize{}694.0$\pm$4.0} & {\footnotesize{}699.46$\pm$3.41} & {\footnotesize{}699.47$\pm$3.39}\tabularnewline
{\footnotesize{}$a_{\mu}^{\mathrm{SM}}$} &  &  &  & {\footnotesize{}11659183.1$\pm$4.8} & {\footnotesize{}11659187.3$\pm$3.8} & {\footnotesize{}11659187.3$\pm$3.9}\tabularnewline
{\footnotesize{}$\Delta a_{\mu}$} &  &  &  & {\footnotesize{}$26.0\pm7.9(3.3\sigma)$} & {\footnotesize{}$21.6\pm7.4(2.9\sigma)$} & {\footnotesize{}$21.6\pm7.4(2.9\sigma)$}\tabularnewline
\hline
\end{tabular}{\footnotesize{}\caption{Our predictions of muon anomalous magnetic
moment, where other contributions are from Refs. \citep{Davier:2019can,Aoyama:2020ynm}
and references therein. We compared the $a_{\mu}^{C}$ , $a_{\mu}^{\mathrm{HVP}\mathrm{,LO}}$,
$a_{\mu}^{\mathrm{SM}}$ and $\Delta a_{\mu}$ with Refs.~\cite{Dai:2013joa,Hoferichter:2019mqg,Davier:2019can,Colangelo:2018mtw}.
The experimental value is measured as $a_{\mu}^{\mathrm{exp}}=11659208.9\pm 6.3$
\cite{Bennett:2006fi}.\label{tab:amu}}
}
\end{table}
{\footnotesize\par}
It is noted that, although different parameterizations of the $\rho-\omega$
mixing are adopted in Fit I and Fit II, the individual contributions
of each channel are almost the same.
A look back to the Figure~\ref{fig:Three-pseudo-case}
shows that
the results of Fit I are slightly different from the ones of Fit II around the
$\rho$ peak in the $e^{+}e^{-}\to\pi^{+}\pi^{-}\pi^{0}$ process (see
the first three graphs). However,
the total contributions to $a_{\mu}^{\pi\pi\pi}|_{\leq 1.8 \, \mathrm{GeV}}$ are
almost the same, as the contribution of Fit I is slightly larger than
that of Fit II on the left hand side of he $\rho$ peak, but it is
in the opposite situation on the right hand side of $\rho$ peak.
They tend to cancel between each other. Since there is little difference between the
two fits, we will discuss below with Fit II. The $a_{\mu}^{C}$ evaluated
here are consistent with those in Refs.~\citep{Colangelo:2018mtw,Hoferichter:2019mqg,Davier:2019can}, within their uncertainty.
In addition, $a_{\mu}^{\pi\pi\pi}|_{\leq 1.8 \, \mathrm{GeV}}$ is also
consistent with that evaluated based on the cross section fitted
in Ref.~\citep{Dai:2013joa}. On the other hand, a slightly larger $a_{\mu}^{\eta\pi\pi}|_{\leq 1.8 \,\mathrm{GeV}}$
is obtained compared with that of Refs.~\citep{Dai:2013joa,Davier:2019can}.
One has to note that
the $e^{+}e^{-}\to\eta\pi^{+}\pi^{-}$ process has a threshold at about
$0.73 \, \mbox{GeV}$, and therefore has a larger dependence on the resonance multiplets.
\par
As explained above
the largest contribution of the hadronic vacuum polarization comes
from $e^{+}e^{-}\to\pi^{+}\pi^{-}$. In our theoretical framework,
the cross section of $e^{+}e^{-}\to\pi^{+}\pi^{-}$ below $1 \,  \mbox{GeV}$ is
almost fixed with a small dependence on $\delta$, while other parameters
contribute little. Hence the $e^{+}e^{-}\to\pi^{+}\pi^{-}$ cross-section
shares little uncertainty from the parameters. For the $e^{+}e^{-}\to K^{+}K^{-}$
process, only $\theta_{V}$ and $\alpha_{V}$ are sensitive, but $\theta_{V}$
and $\alpha_{V}$ are in tension with the $\phi$ peak in $e^{+}e^{-}\to\pi^{+}\pi^{-}\pi^{0}$.
Hence, there is a dedicated balance between these two data sets, which causes considerable uncertainty.
Since we have fitted up to $E=2.3$ GeV, we also listed the corresponding
$a_{\mu}^{C}|_{\leq 2.3 \, \mbox{GeV}}$ in Table \ref{tab:amu}.
\par
The total contribution is
\begin{equation} \label{eq:amufinal}
a_{\mu}^{\mathrm{\tiny HVP}\mathrm{\tiny ,LO}} \, = \,( 699.47 \, \pm \, 3.38) \times 10^{-10}
\end{equation}
from Fit II, in combination with the left
channels fitted in Ref.~\citep{Davier:2019can}.
Note that the four contributions we consider here provide the largest
uncertainty among all the channels. Combined with the other contributions
(QED \citep{Aoyama:2012wk}, EW \cite{Jackiw:1972jz,Knecht:2002hr,Czarnecki:2002nt,Gnendiger:2013pva},
NLHVP \citep{Kurz:2014wya}, NNLHVP \citep{Kurz:2014wya}, HLBL \cite{Zyla:2020zbs,Prades:2009tw,Colangelo:2019uex,Danilkin:2019opj})
within the SM, we also give an estimation of the anomalous magnetic moment
of muon in SM. It is about $4.2\times10^{-10}$ larger in total than
that in Ref.~\citep{Davier:2019can}. Hence our estimation of the discrepancy
$\Delta a_{\mu}$ between the theoretical prediction in SM and that
measured by experiment is $0.4\sigma$ smaller than that in Ref.~\citep{Davier:2019can}.
Our estimation of $\Delta a_{\mu}=(21.6 \pm 7.4)\times 10^{-10}$ is $2.9\sigma$ smaller
than that of the experimental value.

\section{\label{sec:nlo-g-2-th} Higher-order hadronic vacuum polarization contributions to $a_{\mu}$}

We can also consider the contribution of the hadronic vacuum polarization to higher-order
corrections to the leading result of the previous Sec.~\ref{sec:g-2-th}. These have already been computed in the past at
next-to-leading (NLO) order \cite{Krause:1996rf} and nex-to-next-to-leading (NNLO) order \cite{Kurz:2014wya}. In our case, however,
we will only consider the contribution of two and three pseudoscalars to HVP, as we have obtained in Sec.~\ref{sec:level3}.
\par
NLO contributions correspond to  ${\cal O}(\alpha_e^3)$  with one and two HVP insertions. They are given by
\begin{eqnarray}
a_{\mu}^{(2a,2b)}&=&\frac{1}{3}\left(\frac{\alpha_{e}(0)}{\pi}\right)^{3}\int_{4m_{\pi}^{2}}^{\infty} \frac{ds}{s} \, R_{\mathrm{h}}(s) \ K^{(2a,2b)}(s)\,,
\nonumber\\[2mm]
a_{\mu}^{(2c)}&=&\frac{1}{9}\left(\frac{\alpha_{e}(0)}{\pi}\right)^{3} \iint_{4m_{\pi}^{2}}^{\infty}
\frac{ds}{s} \frac{ds^{\prime}}{s^{\prime}} R_{\mathrm{h}}(s) \, R_{\mathrm{h}}\left(s^{\prime}\right) \, K^{(2c)}\left(s,s^{\prime}\right)\,,
\end{eqnarray}
respectively, where $R_{\mathrm{h}}(s)$ has been defined in Eq.~(\ref{eq:amusup}). The label notation and the kernels $K^{(2a,2b,2c)}$ can be read from Ref.~\cite{Krause:1996rf}. Notice that the lower limit in the integral is taken to be $4 m_{\pi}^2$ as we are only including the contributions of  cross-sections of two and three pseudoscalars.
\par
 ${\cal O}(\alpha_e^4)$  with up to three HVP insertions corresponds to the NNLO case. Their contributions
can be computed as
\begin{eqnarray}
a_{\mu}^{(3a,3b,3bLBL)}&=&\frac{1}{3}\left(\frac{\alpha_{e}(0)}{\pi}\right)^{4}\int_{4m_{\pi}^{2}}^{\infty} \frac{ds}{s}
\, R_{\mathrm{h}}(s) \, K^{(3a,3b,3bLBL)}(s)\,,
\nonumber\\[2mm]
a_{\mu}^{(3c)}&=&\frac{1}{9}\left(\frac{\alpha_{e}(0)}{\pi}\right)^{4} \iint_{4m_{\pi}^{2}}^{\infty}
\frac{ds}{s} \frac{ds^{\prime}}{s^{\prime}} R_{\mathrm{h}}(s) \, R_{\mathrm{h}}\left(s^{\prime}\right) \, K^{(3c)}\left(s,s^{\prime}\right)\,,
\\[2mm]
a_{\mu}^{(3d)}&=&\frac{1}{27}\left(\frac{\alpha_{e}(0)}{\pi}\right)^{4} \iiint_{4m_{\pi}^{2}}^{\infty}
\frac{ds}{s} \frac{ds^{\prime}}{s^{\prime}} \frac{ds^{\prime\prime}}{s^{\prime \prime}} R_{\mathrm{h}}(s) \,
R_{\mathrm{h}}\left(s^{\prime}\right) \, R_{\mathrm{h}}\left(s^{\prime\prime}
\right) \, K^{(3d)}\left(s,s^{\prime},s^{\prime\prime}\right)\,. \nonumber \label{eq:nlo;gm2}
\end{eqnarray}
Here the label notation and the different kernels $K^{(3a,3b,3bLBL,3c,3d)}$ follow from Ref.~\cite{Kurz:2014wya}.
\noindent {\footnotesize{}}
\begin{table}[t]
\centering{}{\small{}}%
\begin{tabular}{c|cccc|cc}
\hline
$\times10^{-12}$ & {\small{}$\pi\pi$} & {\small{}$KK$} & {\small{}$\pi\pi\pi$} & {\small{}$\pi\pi\eta$} & \multicolumn{1}{c}{{\small{}Our total}} & \multicolumn{1}{c}{{\small{}Total \citep{Kurz:2014wya}}}\tabularnewline
\hline
{\small{}2a } & {\small{}-1369$\pm$8} & {\small{}-79.8$\pm$2.8} & {\small{}-145$\pm$3} & {\small{}-5.93$\pm$0.46} & {\small{}-1600$\pm$9} & {\small{}-2090}\tabularnewline
{\small{}2b } & {\small{}776$\pm$5} & {\small{}37.6$\pm$1.3} & {\small{}74.7$\pm$1.8} & {\small{}2.37$\pm$0.18} & {\small{}891$\pm$5} & {\small{}1068}\tabularnewline
{\small{}2c } & \multicolumn{4}{c|}{{\small{}22.4$\pm$0.2}} & {\small{}22.4$\pm$0.2} & {\small{}35}\tabularnewline
{\small{}$a_{\mu}^{\mathrm{NLO}}$} &  &  &  &  & {\small{}-687$\pm$10} & {\small{}\textendash 987$\pm$9}\tabularnewline
\hline
{\small{}3a } & {\small{}45.4$\pm$0.3} & {\small{}3.11$\pm$0.11} & {\small{}5.20$\pm$0.12} & {\small{}0.267$\pm$0.021} & {\small{}54.0$\pm$0.3} & {\small{}80}\tabularnewline
{\small{}3b } & {\small{}-24.8$\pm$0.2} & {\small{}-1.62$\pm$0.06} & {\small{}-2.78$\pm$0.06} & {\small{}-0.131$\pm$0.010} & {\small{}-29.3$\pm$0.2} & {\small{}-41}\tabularnewline
{\small{}3bLBL } & {\small{}58.0$\pm$0.3} & {\small{}3.47$\pm$0.12} & {\small{}6.19$\pm$0.14} & {\small{}0.268$\pm$0.021} & {\small{}67.9$\pm$0.4} & {\small{}91}\tabularnewline
{\small{}3c} & \multicolumn{4}{c|}{{\small{}-2.34$\pm$0.02}} & {\small{}-2.34$\pm$0.02} & {\small{}-6}\tabularnewline
{\small{}3d} & \multicolumn{4}{c|}{{\footnotesize{}0.0249$\pm$0.0004}} & {\footnotesize{}0.0249$\pm$0.0004} & {\small{}0.05}\tabularnewline
{\small{}$a_{\mu}^{\mathrm{NNLO}}$} &  &  &  &  & {\small{}90.3$\pm$0.5} & {\small{}124$\pm$1}\tabularnewline
\hline
\end{tabular}\caption{Our estimation of the higher-order HVP contributions
 to $a_{\mu}^{\mbox{\tiny had}}$ using Fit II results and with and upper limit of integration of $\sqrt{s} = 2.3 \, \mbox{GeV}$.
The sum of the four processes considered here is given in the penultimate
column, while the contributions of all channels, estimated in Ref.~\citep{Kurz:2014wya},
are listed in the last column. \label{tab:amu-NLO}}
\end{table}
{\footnotesize\par}

\par
Our results are shown in Table~\ref{tab:amu-NLO}. Since Fit I and Fit II are almost indistinguishable, we would just derive the higher order HVP corrections with Fit II. We also quote the results of Ref.~\cite{Kurz:2014wya}, although we remind that
the later include all the cross-sections but not only the two- and three-pseudoscalar contributions (with $\sqrt{s} \leq 2.3 \, \mbox{GeV}$) to HVP that we have computed.  Hence, the difference between both results can be considered as an estimate of the HVP contributions, that we have not included, and of the higher-energy contribution of the two- and three-pseudoscalar channels. The errors have been estimated in the same way as the leading order contributions to $a_{\mu}^C$.
It is found that these four processes (with the quoted energy upper limit) account for roughly 70 percent of the higher-order HVP corrections to $a_{\mu}^{\mbox{\tiny had}}$.

\section{Conclusions}

Combined with the latest experimental data available for $e^{+}e^{-}$ annihilation
into three pseudoscalar cases $e^{+}e^{-}\to\pi\pi\pi,\pi\pi\eta$,
we carried out joint fits including the two pseudoscalar cases $e^{+}e^{-}\to\pi^{+}\pi^{-},K^{+}K^{-},$ within
the framework of R$\chi$T in the energy region up to $E \lesssim 2 \, \mbox{GeV}$.
Taking into account the possible different mixing mechanisms
of $\rho-\omega$ in the three and two pseudoscalar cases, two fits
have been performed. In Fit I, we apply a uniform energy dependent $\rho-\omega$
mixing parametrization. In Fit II, the energy dependent $\rho-\omega$
mixing parametrization is only used in the two pseudoscalar channel,
while a constant mixing angle is used in the three body case. Overall
very reasonable fits for both cases are found. There
is no relevant difference between Fit I and Fit II except for a small difference
around the $\rho$ peak in the $\pi^{+}\pi^{-}\pi^{0}$ case. This indicates that
the $\rho-\omega$ mixing mechanism that plays an important role in
the two pion case may not be exactly the one to be applied in the
three body case. However, it will not affect much the descriptions
in the three body case, as well as their contribution to the HVP.
Our results have been obtained within a QCD-based phenomenological theory framework with a
joint fit of four different channels that restrict mutually each other.
\par
The main hadronic contributions to the muon anomalous magnetic moment come
from the lower energy region $E \, < \, 1.05 \, \mbox{GeV}$ of the hadronic vacuum polarization input, where few parameters are
dominant. Hence, reliable predictions can be made within our theoretical
framework from our previous analyses of the two- and three-pseudoscalar contributions to the $e^+ e^-$ cross-section.
Accordingly we have computed the leading-order HVP contribution to the anomalous magnetic moment of the muon by
including the four main channels, studied previously, in our estimate.
The central value of these four channels to HVP is about $5\times10^{-10}$
larger than that of Ref.~\citep{Davier:2019can}. In consequence, the
discrepancy between SM prediction and the experimental measurement
decreases to $(21.6 \pm 7.4)\times 10^{-10}$. As an aside, we have also computed the NLO and NNLO HVP contributions to the anomalous magnetic moment of the muon as given by the two- and three-pseudoscalar contributions to the cross-section.

\section*{Acknowledgments}

We thank for the useful discussions with professors Chu-Wen Xiao
and Jian-Ming Shen. This project is supported by National Natural
Science Foundation of China (NSFC) with Grant Nos.11805059 and 11675051,
Joint Large Scale Scientific Facility Funds of the NSFC and Chinese
Academy of Sciences (CAS) under Contract No.U1932110, and Fundamental
Research Funds for the Central Universities.
This work has been supported in part by Grants No. FPA2017-84445-P
and SEV-2014-0398 (AEI/ERDF, EU) and by PROMETEO/2017/053 (GV).

\appendix
\global\long\def\theequation{\Alph{section}.\arabic{equation}}%

\global\long\def\thetable{\Alph{section}.\arabic{table}}%
 \setcounter{equation}{0} \setcounter{table}{0}

\section{Three body final state form factors and partial decay widths }

\label{app:A}

\subsection{Three body final state form factors}

\label{sub:1}
The cross section of the $e^+ e^- \rightarrow \pi^+(p_1) \pi^-(p_2) P(p_3)$ process (P a pseudoscalar meson) is driven by
the vector form factor in Eq.~(\ref{eq:fvv3}) through
\begin{equation} \label{eq:xs3body}
 \sigma_P(Q^2) = \frac{\alpha^2}{192 \, \pi \,Q^6} \, \int_{s_-}^{s_+} ds \int_{t_-}^{t_+} dt \, \phi(Q^2,s,t) \, |F_V^P(Q^2,s,t)|^2 ,
\end{equation}
where $Q=p_1+p_2+p_3$, $s=(Q-p_3)^2$, $t=(Q-p_1)^2$ and
\begin{equation} \label{eq:phiq2st}
 \phi(Q^2,s,t) = s t (Q^2-s-t)  +  s m_P^2(t-Q^2) -m_{\pi}^2 [m_P^4-m_P^2(2 Q^2+s)+Q^4-Q^2 s-2st] -sm_{\pi}^4\, ,
\end{equation}
being $m_P = m_{\pi},m_{\eta}$, depending on the final state. In Eq.~(\ref{eq:xs3body}) the integration limits are:
\begin{eqnarray}\label{eq:intlimits}
 s_- &=& 4 m_{\pi}^2 \, , \nonumber \\
s_+ &=& (\sqrt{Q^2}-m_P)^2 \, , \nonumber \\
t_{\pm} &=& \frac{1}{4\, s} \left\{ \left( Q^2-m_P^2 \right)^2 - \left[ \lambda^{1/2}(Q^2,s,m_P^2) \mp \lambda^{1/2}(s,m_{\pi}^2,m_{\pi}^2)\right]^2 \right\} \, ,
\end{eqnarray}
with $\lambda(a,b,c)$ the K\"all\'en's triangle function.
\par
The vector form factors relevant for the $e^{+}e^{-}\rightarrow\pi^{+}\pi^{-}\pi^{0}, \, \pi^{+}\pi^{-} \eta$
cross-sections are given by:
\begin{align}
F_{V}^{P}(Q^{2},s,t) & =F_{a}^{P}+F_{b}^{P}+F_{c}^{P}+F_{d}^{P}\,,\label{eq:apion}
\end{align}
with $P = \pi, \eta$. We give now the expressions for the form factors. When notation is not fully specified we refer
to Appendix~A.3 of Ref.~\cite{Dai:2013joa}.
\par
Hence the vector form factors are
\begin{align*}
{F}_{a}^{\pi} & =-\frac{N_{C}}{12\pi^{2}F^{3}},
\end{align*}

\begin{align*}
{F}_{b}^{\pi} & =\frac{8\sqrt{2}F_{V}(1+8\sqrt{2}\alpha_{V}\frac{m_{\pi}^{2}}{M_{V}^{2}})}{3M_{V}F^{3}}(\sqrt{2}\cos\theta_{V}+\sin\theta_{V})\,G_{R_{\pi}}(Q^{2})\times\left\{ (\sin\theta_{V}\cos\delta-\sqrt{3}\sin\delta^{\omega}(Q^{2}))\cos\delta\;\right.\\
 & \left.\times BW_{R}[\pi,\omega,Q^{2}]+(\sin\theta_{V}\sin\delta^{\rho}(Q^{2})+\sqrt{3}\cos\delta)\sin\delta^{\rho}(Q^{2})\;BW_{R}[\pi,\rho,Q^{2}]\right\} \\
 & +\frac{8\sqrt{2}F_{V}(1+8\sqrt{2}\alpha_{V}\frac{2m_{K}^{2}-m_{\pi}^{2}}{M_{V}^{2}})}{3M_{V}F^{3}}\cos\theta_{V}(\cos\theta_{V}-\sqrt{2}\sin\theta_{V})\,BW_{R}[\pi,\phi,Q^{2}]~G_{R_{\pi}}(Q^{2}),
\end{align*}

\begin{align*}
{F}_{c}^{\pi} & =-\frac{4\sqrt{2}G_{V}}{3M_{V}F^{3}}\left\{ (\cos\delta+\sqrt{6}\cos\theta_{V}\sin\delta^{\rho}(s)+\sqrt{3}\sin\delta^{\rho}(s)\sin\theta_{V})\cos\delta~BW_{R}[\pi,\rho,s]~C_{R\pi}(Q^{2},s)\right.\\
 & +BW_{R}[\pi,\rho,t]~C_{R\pi}(Q^{2},t)+BW_{R}[\pi,\rho,u]~C_{R\pi}(Q^{2},u)\\
 & \left.-\left[\sqrt{3}\cos\delta\left(\sqrt{2}\cos\theta_{V}+\sin\theta_{V}\right)-\sin\delta^{\omega}(s)\right]\sin\delta^{\omega}(s)~BW_{R}[\pi,\omega,s]~C_{R\pi}(Q^{2},s)\right\} ,
\end{align*}

\begin{align*}
{F}_{d}^{\pi} & =\frac{8G_{V}F_{V}(1+8\sqrt{2}\alpha_{V}\frac{m_{\pi}{}^{2}}{M_{V}^{2}})}{3F^{3}}(\sqrt{2}\cos\theta_{V}+\sin\theta_{V})\times\\
 & \left\{ (\sin\theta_{V}\cos\delta-\sqrt{3}\sin\delta^{\omega}(Q^{2}))\cos\delta(\cos^{2}\delta-\sin\delta^{\rho}(s)\sin\delta^{\omega}(Q^{2}))\right.\\
 & \times BW_{RR}[\pi,\omega,\rho,Q^{2},s]D_{R\pi}(Q^{2},s)\\
 & +(\sin\theta_{V}\cos\delta-\sqrt{3}\sin\delta^{\omega}(Q^{2}))\cos\delta\;BW_{RR}[\pi,\omega,\rho,Q^{2},t]~D_{R\pi}(Q^{2},t)\\
 & +(\sin\theta_{V}\cos\delta-\sqrt{3}\sin\delta^{\omega}(Q^{2}))\cos\delta\;BW_{RR}[\pi,\omega,\rho,Q^{2},u]~D_{R\pi}(Q^{2},u)\\
 & +(\sin\theta_{V}\cos\delta-\sqrt{3}\sin\delta^{\omega}(Q^{2}))[\sin\delta^{\omega}(Q^{2})+\sin\delta^{\omega}(s)]\cos\delta\sin\delta^{\omega}(s)\\
 & \times BW_{RR}[\pi,\omega,\omega,Q^{2},s]D_{R\pi}(Q^{2},s)\\
 & +(\sin\theta_{V}\sin\delta^{\rho}(Q^{2})+\sqrt{3}\cos\delta)[\sin\delta^{\rho}(Q^{2})+\sin\delta^{\rho}(s)]\cos^{2}\delta\\
 & \times\;BW_{RR}[\pi,\rho,\rho,Q^{2},s]~D_{R\pi}(Q^{2},s)\\
 & +(\sin\theta_{V}\sin\delta^{\rho}(Q^{2})+\sqrt{3}\cos\delta)\sin\delta^{\rho}(Q^{2})\;BW_{RR}[\pi,\rho,\rho,Q^{2},t]~D_{R\pi}(Q^{2},t)\\
 & +(\sin\theta_{V}\sin\delta^{\rho}(Q^{2})+\sqrt{3}\cos\delta)\sin\delta^{\rho}(Q^{2})\;BW_{RR}[\pi,\rho,\rho,Q^{2},u]~D_{R\pi}(Q^{2},u)\\
 & -(\sin\theta_{V}\sin\delta^{\rho}(Q^{2})+\sqrt{3}\cos\delta)(\cos^{2}\delta-\sin\delta^{\rho}(Q^{2})\sin\delta^{\omega}(s))\sin\delta^{\omega}(s)\\
 & \left.\times BW_{RR}[\pi,\rho,\omega,Q^{2},s]D_{R\pi}(Q^{2},s)\right\} +\frac{8G_{V}F_{V}(1+8\sqrt{2}\alpha_{V}\frac{2m_{K}{}^{2}-m_{\pi}{}^{2}}{M_{V}^{2}})}{3F^{3}}\\
 & \times(\cos\theta_{V}-\sqrt{2}\sin\theta_{V})\,\cos\theta_{V}\\
 & \times\left\{ \cos^{2}\delta BW_{RR}[\pi,\phi,\rho,Q^{2},s]D_{R\pi}(Q^{2},s)+\sin^{2}\delta^{\omega}(s)BW_{RR}[\pi,\phi,\omega,Q^{2},s]D_{R\pi}(Q^{2},s)\right.\\
 & \left.+BW_{RR}[\pi,\phi,\rho,Q^{2},t]~D_{R\pi}(Q^{2},t)+\;BW_{RR}[\pi,\phi,\rho,Q^{2},u]~D_{R\pi}(Q^{2},u)\right\} .
\end{align*}

\begin{align*}
{F}_{a}^{\eta} & =-\frac{N_{C}}{12\sqrt{3}\pi^{2}F^{3}}(-\sqrt{2}\sin\theta_{P}+\cos\theta_{P}),
\end{align*}

\begin{align*}
{F}_{b}^{\eta} & =\frac{8\sqrt{6}F_{V}(1+8\sqrt{2}\alpha_{V}\frac{m_{\pi}^{2}}{M_{V}^{2}})}{3M_{V}F^{3}}\left(\cos\delta+\frac{1}{\sqrt{3}}\sin\delta^{\rho}(Q^{2})\sin\theta_{V}\right)\\
 & \times\cos\delta(-\sqrt{2}\sin\theta_{P}+\cos\theta_{P})BW_{R}[\eta,\rho,Q^{2}]~G_{R\eta}(Q^{2},s)\ \\
 & -\frac{8\sqrt{6}F_{V}(1+8\sqrt{2}\alpha_{V}\frac{m_{\pi}^{2}}{M_{V}^{2}})}{3M_{V}F^{3}}\left(-\sin\delta^{\omega}(Q^{2})+\frac{1}{\sqrt{3}}\cos\delta\sin\theta_{V}\right)\sin\delta^{\omega}(Q^{2})\\
 & \times(-\sqrt{2}\sin\theta_{P}+\cos\theta_{P})\times BW_{R}[\eta,\omega,Q^{2}]~G_{R\eta}(Q^{2},s),
\end{align*}

\begin{align*}
{F}_{c}^{\eta} & =-\frac{4\sqrt{2}G_{V}}{3M_{V}F^{3}}\cos\delta\{\sqrt{3}\cos\delta(\cos\theta_{P}-\sqrt{2}\sin\theta_{P})+\sin\delta^{\rho}(s)\ [\;\sqrt{2}\cos\theta_{V}\cos\theta_{P}\\
 & -\sin\theta_{V}(\cos\theta_{P}+\sqrt{2}\sin\theta_{P})\;]\;\}BW_{R}[\eta,\rho,s]~C_{R\eta1}(Q^{2},s,m_{\eta}^{2})\\
 & -\frac{4\sqrt{2}G_{V}}{9M_{V}F^{3}}\cos\delta\{4\sin\delta^{\rho}(s)[\sqrt{2}\cos(\theta_{V}+\theta_{P})-2\cos\theta_{P}\sin\theta_{V}+\cos\theta_{V}\sin\theta_{P}]m_{K}^{2}+\\
 & [3\sqrt{3}\cos\delta(\cos\theta_{P}-\sqrt{2}\sin\theta_{P})-\sin\delta^{\rho}(s)(~\sqrt{2}\cos(\theta_{V}+\theta_{P})-5\cos\theta_{P}\sin\theta_{V}\\
 & +4\cos\theta_{V}\sin\theta_{P}~)]m_{\pi}^{2}\;\}~BW_{R}[\eta,\rho,s]~C_{R\eta2}\\
 & +\frac{4\sqrt{2}G_{V}}{3M_{V}F^{3}}\sin\delta^{\omega}(s)\{\sqrt{3}\sin\delta^{\omega}(s)(-\cos\theta_{P}+\sqrt{2}\sin\theta_{P})+\cos\delta[\sqrt{2}\cos\theta_{V}\cos\theta_{P}\\
 & -\sin\theta_{V}(\cos\theta_{P}+\sqrt{2}\sin\theta_{P})\;]\;\}~BW_{R}[\eta,\omega,s]~C_{R\eta1}(Q^{2},s,m_{\eta}^{2})\\
 & +\frac{4\sqrt{2}G_{V}}{9M_{V}F^{3}}\sin\delta^{\omega}(s)\{4\cos\delta[\sqrt{2}\cos(\theta_{V}+\theta_{P})-2\cos\theta_{P}\sin\theta_{V}+\cos\theta_{V}\sin\theta_{P}]m_{K}^{2}-\\
 & [3\sqrt{3}\sin\delta^{\omega}(s)(\cos\theta_{P}-\sqrt{2}\sin\theta_{P})+\cos\delta(\sqrt{2}\cos(\theta_{V}+\theta_{P})-5\cos\theta_{P}\sin\theta_{V}\\
 & +4\cos\theta_{V}\sin\theta_{P})]m_{\pi}^{2}\}~BW_{R}[\eta,\omega,s]~C_{R\eta2},
\end{align*}

\begin{align*}
{F}_{d}^{\eta} & =\frac{8F_{V}(1+8\sqrt{2}\alpha_{V}\frac{m_{\pi}^{2}}{M_{V}^{2}})G_{V}}{\sqrt{6}F^{3}}\cos\delta\left(\cos\delta+\frac{1}{\sqrt{3}}\sin\delta^{\rho}(Q^{2})\sin\theta_{V}\right)\\
 & \{\cos^{2}\delta(\sqrt{2}\cos\theta_{P}-2\sin\theta_{P})+\sin\delta^{\rho}(Q^{2})\sin\delta^{\rho}(s)[\cos\theta_{P}\sin\theta_{V}(4\cos\theta_{V}\\
 & -\sqrt{2}\sin\theta_{V})-2\sin\theta_{P}]\}BW_{RR}[\eta,\rho,\rho,Q^{2},s]~D_{R\eta1}(Q^{2},s,m_{\eta}^{2})\\
 & +\frac{2F_{V}(1+8\sqrt{2}\alpha_{V}\frac{m_{\pi}^{2}}{M_{V}^{2}})G_{V}}{3\sqrt{6}F^{3}}\cos\delta\left(\cos\delta+\frac{1}{\sqrt{3}}\sin\delta^{\rho}(Q^{2})\sin\theta_{V}\right)\\
 & \times\bigg\{8\sin\delta^{\rho}(Q^{2})\sin\delta^{\rho}(s)[\cos\theta_{P}(-3\sqrt{2}+\sqrt{2}\cos2\theta_{V}+4\sin2\theta_{V})\\
 & +(-3+\cos2\theta_{V}+2\sqrt{2}\sin2\theta_{V})\times\sin\theta_{P}]m_{K}^{2}+[12\cos^{2}\delta(\sqrt{2}\cos\theta_{P}-2\sin\theta_{P})\\
 & +\sin\delta^{\rho}(Q^{2})\sin\delta^{\rho}(s)(-9\sqrt{2}\cos(2\theta_{V}-\theta_{P})+18\sqrt{2}\cos\theta_{P}\\
 & +7\sqrt{2}\cos(2\theta_{V}+\theta_{P})-8\sin(2\theta_{V}+\theta_{P}))]m_{\pi}^{2}\bigg\} BW_{RR}[\eta,\rho,\rho,Q^{2},s]D_{R\eta2}\\
 & -\frac{8F_{V}(1+8\sqrt{2}\alpha_{V}\frac{m_{\pi}^{2}}{M_{V}^{2}})G_{V}}{\sqrt{6}F^{3}}\sin\delta^{\omega}(s)\left(-\sin\delta^{\omega}(Q^{2})+\frac{1}{\sqrt{3}}\cos\delta\sin\theta_{V}\right)\\
 & \times\bigg\{\cos\theta_{P}[\sqrt{2}\sin\delta^{\omega}(Q^{2})\times\sin\delta^{\omega}(s)+\cos^{2}\delta\sin\theta_{V}(4\cos\theta_{V}-\sqrt{2}\sin\theta_{V})]\\
 & -2\sin\theta_{P}(\cos^{2}\delta+\sin\delta^{\omega}(Q^{2})\sin\delta^{\omega}(s))\bigg\} BW_{RR}[\eta,\omega,\omega,Q^{2},s]~D_{R\eta1}(Q^{2},s,m_{\eta}^{2})\\
 & -\frac{2F_{V}(1+8\sqrt{2}\alpha_{V}\frac{m_{\pi}^{2}}{M_{V}^{2}})G_{V}}{3\sqrt{6}F^{3}}\sin\delta^{\omega}(s)\times\left(-\sin\delta^{\omega}(Q^{2})+\frac{1}{\sqrt{3}}\cos\delta\sin\theta_{V}\right)\\
 & \times\bigg\{8\cos^{2}\delta[\cos\theta_{P}(-3\sqrt{2}+\sqrt{2}\cos2\theta_{V}+4\sin2\theta_{V})+\\
 & (-3+\cos2\theta_{V}+2\sqrt{2}\sin2\theta_{V})\sin\theta_{P}]m_{K}^{2}\\
 & +[12\sin\delta^{\omega}(Q^{2})\sin\delta^{\omega}(s)(\sqrt{2}\cos\theta_{P}-2\sin\theta_{P})\\
 & +\cos^{2}\delta(-9\sqrt{2}\cos(2\theta_{V}-\theta_{P})+18\sqrt{2}\cos\theta_{P}+7\sqrt{2}\cos(2\theta_{V}+\theta_{P})\\
 & -8\sin(2\theta_{V}+\theta_{P}))\big]m_{\pi}^{2}\bigg\}\times BW_{RR}[\eta,\omega,\omega,Q^{2},s]~D_{R\eta2}\\
 & +\frac{2F_{V}(1+8\sqrt{2}\alpha_{V}\frac{m_{\pi}^{2}}{M_{V}^{2}})G_{V}}{\sqrt{6}F^{3}}\cos\delta\left(-\sin\delta^{\omega}(Q^{2})+\frac{1}{\sqrt{3}}\cos\delta\sin\theta_{V}\right)\\
 & \times\{(-\frac{1}{2}\cos^{2}\theta_{V}\cos\theta_{p}\sin\delta^{\rho}(s)+\frac{1}{2}\sin^{2}\theta_{V}\cos\theta_{p}\sin\delta^{\rho}(s)\\
 & -2\sqrt{2}\sin\theta_{V}\cos\theta_{V}\cos\theta_{P}\sin\delta^{\rho}(s)+\sqrt{2}\sin\theta_{P}\sin\delta^{\rho}(s)\\
 & +\frac{1}{2}\cos\theta_{P}\sin\delta^{\rho}(s)-\sqrt{2}\sin\theta_{P}\sin\delta^{\omega}(Q^{2})+\cos\theta_{P}\sin\delta^{\omega}(Q^{2}))(-4\sqrt{2}\cos\delta)\}\\
 & \times BW_{RR}[\eta,\omega,\rho,Q^{2},s]~D_{R\eta1}(Q^{2},s,m_{\eta}^{2})+\frac{2F_{V}(1+8\sqrt{2}\alpha_{V}\frac{m_{\pi}^{2}}{M_{V}^{2}})G_{V}}{3\sqrt{6}F^{3}}\cos\delta\\
 & \times\left(-\sin\delta^{\omega}(Q^{2})+\frac{1}{\sqrt{3}}\cos\delta\sin\theta_{V}\right)BW_{RR}[\eta,\omega,\rho,Q^{2},s]~D_{R\eta2}\\
 & \times\{-\sqrt{2}\cos\delta\ [m_{\pi}^{2}(\sin\delta^{\rho}(s)(4\sqrt{2}\sin(2\theta_{V}+\theta_{P})+9\cos(2\theta_{V}-\theta_{P})-7\cos(2\theta_{V}+\theta_{P})\\
 & -18\cos\theta_{P})+12\sin\delta^{\omega}(Q^{2})(\cos\theta_{P}-\sqrt{2}\sin\theta_{P}))-4m_{K}^{2}\sin\delta^{\rho}(s)\\
 & \times(2\cos\theta_{P}(2\sqrt{2}\sin2\theta_{V}+\cos2\theta_{V}-3)+\sin\theta_{P}(4\sin2\theta_{V}+\sqrt{2}\cos2\theta_{V}-3\sqrt{2}))]\}\\
 & -\frac{2F_{V}(1+8\sqrt{2}\alpha_{V}\frac{m_{\pi}^{2}}{M_{V}^{2}})G_{V}}{\sqrt{6}F^{3}}\left(\cos\delta+\frac{1}{\sqrt{3}}\sin\delta^{\rho}(Q^{2})\sin\theta_{V}\right)\\
 & \times BW_{RR}[\eta,\rho,\omega,Q^{2},s]D_{R\eta1}(Q^{2},s,m_{\eta}^{2})\\
 & \times\{-4\sqrt{2}\cos\delta\sin\delta^{\omega}(s)(\cos\theta_{P}(\sin\theta_{V}\sin\delta^{\rho}(Q^{2})(\sin\theta_{V}-2\sqrt{2}\cos\theta_{V})\\
 & +\sin\delta^{\omega}(s))+\sqrt{2}\sin\theta_{P}(\sin\delta^{\rho}(Q^{2})-\sin\delta^{\omega}(s)))\}\\
 & -\frac{2F_{V}(1+8\sqrt{2}\alpha_{V}\frac{m_{\pi}^{2}}{M_{V}^{2}})G_{V}}{3\sqrt{6}F^{3}}\left(\cos\delta+\frac{1}{\sqrt{3}}\sin\delta^{\rho}(Q^{2})\sin\theta_{V}\right)\\
 & \times\{-\sqrt{2}\cos\delta\sin\delta^{\omega}(s)[m_{\pi}^{2}(\sin\delta^{\rho}(Q^{2})(4\sqrt{2}\sin(2\theta_{V}+\theta_{P})+9\cos(2\theta_{V}-\theta_{P})\\
 & -7\cos(2\theta_{V}+\theta_{P})-18\cos\theta_{P})+12\sin\delta^{\omega}(s)(\cos\theta_{P}-\sqrt{2}\sin\theta_{P}))-4m_{K}^{2}\sin\delta^{\rho}(Q^{2})\\
 & \times(2\cos\theta_{P}(2\sqrt{2}\sin2\theta_{V}+\cos2\theta_{V}-3)+\sin\theta_{P}(4\sin2\theta_{V}+\sqrt{2}\cos2\theta_{V}-3\sqrt{2}))]\}\\
 & \times BW_{RR}[\eta,\rho,\omega,Q^{2},s]~D_{R\eta2}-\frac{4F_{V}(1+8\sqrt{2}\alpha_{V}\frac{2m_{K}{}^{2}-m_{\pi}^{2}}{M_{V}^{2}})G_{V}}{3\sqrt{2}F^{3}}\cos\delta\cos\theta_{V}\\
 & \times\cos\theta_{P}\sin\delta^{\rho}(s)\left(-4\cos2\theta_{V}+\sqrt{2}\sin2\theta_{V}\right)\times~BW_{RR}[\eta,\phi,\rho,Q^{2},s]~D_{R\eta1}(Q^{2},s,m_{\eta}^{2})\\
 & +\frac{4F_{V}(1+8\sqrt{2}\alpha_{V}\frac{2m_{K}^{2}-m_{\pi}^{2}}{M_{V}^{2}})G_{V}}{9\sqrt{2}F^{3}}\\
 & \times\cos\delta\cos\theta_{V}\sin\delta^{\rho}(s)\bigg\{4(2\sqrt{2}\cos2\theta_{V}-\sin2\theta_{V})\sin\theta_{P}(m_{K}^{2}-m_{\pi}^{2})\\
 & +\cos\theta_{P}(4\cos2\theta_{V}-\sqrt{2}\sin2\theta_{V})(4m_{K}^{2}-m_{\pi}^{2})~\bigg\}~BW_{RR}[\eta,\phi,\rho,Q^{2},s]~D_{R\eta2}\\
 & +\frac{4F_{V}(1+8\sqrt{2}\alpha_{V}\frac{2m_{K}^{2}-m_{\pi}^{2}}{M_{V}^{2}})G_{V}}{3\sqrt{2}F^{3}}\sin\delta^{\omega}(s)\cos\theta_{V}\cos\delta\cos\theta_{P}\\
 & \times(-4\cos2\theta_{V}+\sqrt{2}\sin2\theta_{V})\times BW_{RR}[\eta,\phi,\omega,Q^{2},s]~D_{R\eta1}(Q^{2},s,m_{\eta}^{2})\\
 & -\frac{4F_{V}(1+8\sqrt{2}\alpha_{V}\frac{2m_{K}{}^{2}-m_{\pi}^{2}}{M_{V}^{2}})G_{V}}{9\sqrt{2}F^{3}}\sin\delta^{\omega}(s)\cos\theta_{V}\cos\delta\\
 & \times\bigg\{4(2\sqrt{2}\cos2\theta_{V}-\sin2\theta_{V})\sin\theta_{P}(m_{K}^{2}-m_{\pi}^{2})+\cos\theta_{P}\\
 & \times(4\cos2\theta_{V}-\sqrt{2}\sin2\theta_{V})(4m_{K}^{2}-m_{\pi}^{2})\bigg\}~BW_{RR}[\eta,\phi,\omega,Q^{2},s]~D_{R\eta2}.
\end{align*}

\subsection{Decay widths involving vector resonances}

\subsubsection{Two-body decays}

\begin{align*}
\Gamma_{\omega\to\pi\pi} & =\frac{G_{V}^{2}\,M_{\omega}^{3}}{48\pi F^{4}}\sin^{2}\delta^{\omega}(M_{\omega}^{2})\left(1-\frac{4m_{\pi}^{2}}{M_{\omega}^{2}}\right)^{3/2},
\end{align*}

\begin{align*}
\Gamma_{\rho\to\ell^{+}\ell^{-}} & =\frac{4\,\alpha^{2}\,\pi\,F_{V}^{2}}{3\,M_{\rho}}\left(1+8\sqrt{2}\alpha_{V}\frac{m_{\pi}^{2}}{M_{V}^{2}}\right)^{2}\left(\cos\delta+\frac{1}{\sqrt{3}}\sin\theta_{V}\sin\delta^{\rho}(M_{\rho}^{2})\right)^{2}\left(1+\frac{2m_{\ell}^{2}}{M_{\rho}^{2}}\right)\left(1-\frac{4m_{\ell}^{2}}{M_{\rho}^{2}}\right)^{1/2}\ ,
\end{align*}

\begin{align*}
\Gamma_{\omega\to\ell^{+}\ell^{-}} & =\frac{4\,\alpha^{2}\,\pi\,F_{V}^{2}}{27\,M_{\omega}}\left(1+8\sqrt{2}\alpha_{V}\frac{m_{\pi}^{2}}{M_{V}^{2}}\right)^{2}(\sqrt{3}\sin\theta_{V}\cos\delta-3\sin\delta^{\omega}(M_{\omega}^{2}))^{2}\left(1+\frac{2m_{\ell}^{2}}{M_{\omega}^{2}}\right)\left(1-\frac{4m_{\ell}^{2}}{M_{\omega}^{2}}\right)^{1/2}\ ,
\end{align*}

\begin{align*}
F_{\rho^{0}\to\pi^{0}\gamma} & =\frac{2\sqrt{2}}{3M_{V}F}C_{R\pi}(0,M_{\rho}^{2})\left(\cos\delta+\sqrt{6}\cos\theta_{V}\sin\delta^{\rho}(M_{\rho}^{2})+\sqrt{3}\sin\delta^{\rho}(M_{\rho}^{2})\sin\theta_{V}\right)\\
 & -\frac{4F_{V}\left(1+8\sqrt{2}\alpha_{V}\frac{m_{\pi}^{2}}{M_{V}^{2}}\right)}{3M_{\rho}^{2}F}D_{R\pi}(0,M_{\rho}^{2})(\sin\theta_{V}\sin\delta^{\rho}(0)+\sqrt{3}\cos\delta)\\
 & \times[\sin\delta^{\rho}(M_{\rho}^{2})+\sin\delta^{\rho}(0)]\cos\delta(\sqrt{2}\cos\theta_{V}+\sin\theta_{V})-\frac{4F_{V}\left(1+8\sqrt{2}\alpha_{V}\frac{m_{\pi}^{2}}{M_{V}^{2}}\right)}{3M_{\omega}^{2}F}\\
 & \times D_{R\pi}(0,M_{\rho}^{2})(\sin\theta_{V}\cos\delta-\sqrt{3}\sin\delta^{\omega}(0))[\cos^{2}\delta-\sin\delta^{\rho}(M_{\rho}^{2})\sin\delta^{\omega}(0)]\\
 & \times(\sqrt{2}\cos\theta_{V}+\sin\theta_{V})\\
 & -\frac{4F_{V}\left(1+8\sqrt{2}\alpha_{V}\frac{2m_{K}^{2}-m_{\pi}^{2}}{M_{V}^{2}}\right)}{3M_{\phi}^{2}F}D_{R\pi}(0,M_{\rho}^{2})\cos\theta_{V}\cos\delta(\cos\theta_{V}-\sqrt{2}\sin\theta_{V})\,,
\end{align*}

\begin{align*}
F_{\rho^{\pm}\to\pi^{\pm}\gamma} & =\frac{2\sqrt{2}}{3M_{V}F}C_{R\pi}(0,M_{\rho}^{2})-\frac{4F_{V}\left(1+8\sqrt{2}\alpha_{V}\frac{m_{\pi}^{2}}{M_{V}^{2}}\right)}{3M_{\rho}^{2}F}D_{R\pi}(0,M_{\rho}^{2})\sin\delta^{\rho}(0)\\
 & \times(\sqrt{2}\cos\theta_{V}+\sin\theta_{V})(\sin\theta_{V}\sin\delta^{\rho}(0)+\sqrt{3}\cos\delta)-\frac{4F_{V}\left(1+8\sqrt{2}\alpha_{V}\frac{m_{\pi}^{2}}{M_{V}^{2}}\right)}{3M_{\omega}^{2}F}\\
 & \times D_{R\pi}(0,M_{\rho}^{2})\cos\delta(\sqrt{2}\cos\theta_{V}+\sin\theta_{V})(\sin\theta_{V}\cos\delta-\sqrt{3}\sin\delta^{\omega}(0))\\
 & -\frac{4F_{V}\left(1+8\sqrt{2}\alpha_{V}\frac{2m_{K}^{2}-m_{\pi}^{2}}{M_{V}^{2}}\right)}{3M_{\phi}^{2}F}D_{R\pi}(0,M_{\rho}^{2})\cos\theta_{V}(\cos\theta_{V}-\sqrt{2}\sin\theta_{V})\,,
\end{align*}

\begin{align*}
F_{\phi\to\pi^{0}\gamma} & =\frac{2\sqrt{6}}{3M_{V}F}C_{R\pi}(0,M_{\phi}^{2})(\cos\theta_{V}-\sqrt{2}\sin\theta_{V})-\frac{4F_{V}\left(1+8\sqrt{2}\alpha_{V}\frac{m_{\pi}^{2}}{M_{V}^{2}}\right)}{3M_{\rho}^{2}F}D_{R\pi}(0,M_{\phi}^{2})\\
 & \times(\sin\theta_{V}\sin\delta^{\rho}(0)+\sqrt{3}\cos\delta)\cos\delta(\cos\theta_{V}-\sqrt{2}\sin\theta_{V})-\frac{4F_{V}\left(1+8\sqrt{2}\alpha_{V}\frac{m_{\pi}^{2}}{M_{V}^{2}}\right)}{3M_{\omega}^{2}F}\\
 & \times D_{R\pi}(0,M_{\phi}^{2})(\sin\theta_{V}\cos\delta-\sqrt{3}\sin\delta^{\omega}(0))\sin\delta^{\omega}(0)(-\cos\theta_{V}+\sqrt{2}\sin\theta_{V})\,,
\end{align*}

\begin{align*}
F_{\omega\to\pi^{0}\gamma} & =\frac{2\sqrt{2}}{3M_{V}F}C_{R\pi}(0,M_{\omega}^{2})\left(\sqrt{3}\cos\delta(\sqrt{2}\cos\theta_{V}+\sin\theta_{V})-\sin\delta^{\omega}(M_{\omega}^{2})\right)-\frac{4F_{V}\left(1+8\sqrt{2}\alpha_{V}\frac{m_{\pi}^{2}}{M_{V}^{2}}\right)}{3M_{\rho}^{2}F}\\
 & \times D_{R\pi}(0,M_{\omega}^{2})(\sin\theta_{V}\sin\delta^{\rho}(0)+\sqrt{3}\cos\delta)(\cos^{2}\delta-\sin\delta^{\rho}(0)\sin\delta^{\omega}(M_{\omega}^{2}))(\sqrt{2}\cos\theta_{V}+\sin\theta_{V})\\
 & +\frac{4F_{V}\left(1+8\sqrt{2}\alpha_{V}\frac{m_{\pi}^{2}}{M_{V}^{2}}\right)}{3M_{\omega}^{2}F}D_{R\pi}(0,M_{\omega}^{2})(\sin\theta_{V}\cos\delta-\sqrt{3}\sin\delta^{\rho}(0))(\sin\delta^{\omega}(M_{\omega}^{2})+\sin\delta^{\omega}(0))\\
 & \times\cos\delta(\sqrt{2}\cos\theta_{V}+\sin\theta_{V})\\
 & -\frac{4F_{V}\left(1+8\sqrt{2}\alpha_{V}\frac{2m_{K}^{2}-m_{\pi}^{2}}{M_{V}^{2}}\right)}{3M_{\phi}^{2}F}D_{R\pi}(0,M_{\omega}^{2})\cos\theta_{V}\sin\delta^{\omega}(M_{\omega}^{2})(-\cos\theta_{V}+\sqrt{2}\sin\theta_{V})\,,
\end{align*}

\begin{align*}
F_{\omega\to\eta\gamma} & =\frac{2\sqrt{2}}{3M_{V}F}C_{R\eta1}(0,M_{\omega}^{2},m_{\eta}^{2})\\
 & \left\{ \sqrt{3}\sin\delta^{\omega}(M_{\omega}^{2})(-\cos\theta_{P}+\sqrt{2}\sin\theta_{P})+\cos\delta[\sqrt{2}\cos\theta_{V}\cos\theta_{P}-\sin\theta_{V}(\cos\theta_{P}+\sqrt{2}\sin\theta_{P})]\right\} \\
 & +\frac{2\sqrt{2}}{9M_{V}F}C_{R\eta2}\left\{ 4\cos\delta\left(\sqrt{2}\cos(\theta_{V}+\theta_{P})-2\cos\theta_{P}\sin\theta_{V}+\cos\theta_{V}\sin\theta_{P}\right)m_{K}^{2}\right.\\
 & -\left(3\sqrt{3}\sin\delta^{\omega}(M_{\omega}^{2})(\cos\theta_{P}-\sqrt{2}\sin\theta_{P})\right.\\
 & \left.\left.+\cos\delta[\sqrt{2}\cos(\theta_{V}+\theta_{P})-5\cos\theta_{P}\sin\theta_{V}+4\cos\theta_{V}\sin\theta_{P}]\right)m_{\pi}^{2}\right\} \\
 & -\frac{F_{V}\left(1+8\sqrt{2}\alpha_{V}\frac{m_{\pi}^{2}}{M_{V}^{2}}\right)}{3\sqrt{2}M_{\rho}^{2}F}D_{R\eta1}(0,M_{\omega}^{2},m_{\eta}^{2})(\sin\theta_{V}\sin\delta^{\rho}(0)+\sqrt{3}\cos\delta)\\
 & \times\{(-4\sqrt{2}\cos\delta)(-\frac{1}{2}\cos^{2}\theta_{V}\cos\theta_{p}\sin\delta^{\rho}(0)+\frac{1}{2}\sin^{2}\theta_{V}\cos\theta_{p}\sin\delta^{\rho}(0)\\
 & -2\sqrt{2}\sin\theta_{V}\cos\theta_{V}\cos\theta_{P}\sin\delta^{\rho}(0)+\sqrt{2}\sin\theta_{P}\sin\delta^{\rho}(0)+\frac{1}{2}\cos\theta_{P}\sin\delta^{\rho}(0)\\
 & -\sqrt{2}\sin\theta_{P}\sin\delta^{\omega}(M_{\omega}^{2})+\cos\theta_{P}\sin\delta^{\omega}(M_{\omega}^{2}))\}\\
 & -\frac{F_{V}\left(1+8\sqrt{2}\alpha_{V}\frac{m_{\pi}^{2}}{M_{V}^{2}}\right)}{9\sqrt{2}M_{\rho}^{2}F}D_{R\eta2}(\sin\theta_{V}\sin\delta^{\rho}(0)+\sqrt{3}\cos\delta)\\
 & \times\{-\sqrt{2}\cos\delta\ [m_{\pi}^{2}(\sin\delta^{\rho}(0)(4\sqrt{2}\sin(2\theta_{V}+\theta_{P})+9\cos(2\theta_{V}-\theta_{P})-7\cos(2\theta_{V}+\theta_{P})-18\cos\theta_{P})\\
 & +12\sin\delta^{\omega}(M_{\omega}^{2})(\cos\theta_{P}-\sqrt{2}\sin\theta_{P}))\\
 & -4m_{K}^{2}\sin\delta^{\rho}(0)(2\cos\theta_{P}(2\sqrt{2}\sin2\theta_{V}+\cos2\theta_{V}-3)+\sin\theta_{P}(4\sin2\theta_{V}+\sqrt{2}\cos2\theta_{V}-3\sqrt{2}))]\}\\
 & -\frac{2\sqrt{2}F_{V}\left(1+8\sqrt{2}\alpha_{V}\frac{m_{\pi}^{2}}{M_{V}^{2}}\right)}{3M_{\omega}^{2}F}D_{R\eta1}(0,M_{\omega}^{2},m_{\eta}^{2})(\sin\theta_{V}\cos\delta-\sqrt{3}\sin\delta^{\omega}(0))\\
 & \{\cos\theta_{P}[\sqrt{2}\sin\delta^{\omega}(0)\sin\delta^{\omega}(M_{\omega}^{2})+\cos^{2}\delta\sin\theta_{V}(4\cos\theta_{V}-\sqrt{2}\sin\theta_{V})]-2[\sin\delta^{\omega}(0)\sin\delta^{\omega}(M_{\omega}^{2})\\
 & +\cos^{2}\delta]\sin\theta_{P}\}-\frac{F_{V}\left(1+8\sqrt{2}\alpha_{V}\frac{m_{\pi}^{2}}{M_{V}^{2}}\right)}{9\sqrt{2}M_{\omega}^{2}F}D_{R\eta2}(\sin\theta_{V}\cos\delta-\sqrt{3}\sin\delta^{\omega}(0))\\
 & \left\{ 8\cos^{2}\delta\left(\cos\theta_{P}(-3\sqrt{2}+\sqrt{2}\cos2\theta_{V}+4\sin2\theta_{V})\right.\right.\\
 & \left.\left.+(-3+\cos2\theta_{V}+2\sqrt{2}\sin2\theta_{V})\sin\theta_{P}\right)m_{K}^{2}\right.+\left(12\sin\delta^{\omega}(0)\sin\delta^{\omega}(M_{\omega}^{2})(\sqrt{2}\cos\theta_{P}-2\sin\theta_{P})\right.\\
 & \left.+\cos^{2}\delta[-9\sqrt{2}\cos(2\theta_{V}-\theta_{P})+18\sqrt{2}\cos\theta_{P}\right.\\
 & \left.\left.+7\sqrt{2}\cos(2\theta_{V}+\theta_{P})-8\sin(2\theta_{V}+\theta_{P})]\right)m_{\pi}^{2}\right\} \\
 & +\frac{\sqrt{2}F_{V}\left(1+8\sqrt{2}\alpha_{V}\frac{2m_{K}^{2}-m_{\pi}^{2}}{M_{V}^{2}}\right)}{3M_{\phi}^{2}F}D_{R\eta1}(0,M_{\omega}^{2},m_{\eta}^{2})\\
 & \left\{ \cos\theta_{V}\cos\delta\cos\theta_{P}(-4\cos2\theta_{V}+\sqrt{2}\sin2\theta_{V})\right\} \\
 & -\frac{\sqrt{2}F_{V}\left(1+8\sqrt{2}\alpha_{V}\frac{2m_{K}^{2}-m_{\pi}^{2}}{M_{V}^{2}}\right)}{9M_{\phi}^{2}F}D_{R\eta2}\cos\theta_{V}\cos\delta\\
 & \left\{ 4(2\sqrt{2}\cos2\theta_{V}-\sin2\theta_{V})\sin\theta_{P}(m_{K}^{2}-m_{\pi}^{2})\right.\\
 & \left.+\cos\theta_{P}(4\cos2\theta_{V}-\sqrt{2}\sin2\theta_{V})(4m_{K}^{2}-m_{\pi}^{2})\right\} \,,
\end{align*}

\begin{align*}
F_{\rho^{0}\to\eta\gamma} & =\frac{2\sqrt{2}}{3M_{V}F}C_{R\eta1}(0,M_{\rho}^{2},m_{\eta}^{2})\left\{ \sqrt{3}\cos\delta(\cos\theta_{P}-\sqrt{2}\sin\theta_{P})\right.\\
 & \left.+\sin\delta^{\rho}(M_{\rho}^{2})[\sqrt{2}\cos\theta_{V}\cos\theta_{P}-\sin\theta_{V}(\cos\theta_{P}+\sqrt{2}\sin\theta_{P})]\right\} \\
 & +\frac{2\sqrt{2}}{9M_{V}F}C_{R\eta2}\left\{ 4\sin\delta^{\rho}(M_{\rho}^{2})\left(\sqrt{2}\cos(\theta_{V}+\theta_{P})-2\cos\theta_{P}\sin\theta_{V}+\cos\theta_{V}\sin\theta_{P}\right)m_{K}^{2}\right.\\
 & +\left(3\sqrt{3}\cos\delta(\cos\theta_{P}-\sqrt{2}\sin\theta_{P})\right.\\
 & \left.\left.-\sin\delta^{\rho}(M_{\rho}^{2})[\sqrt{2}\cos(\theta_{V}+\theta_{P})-5\cos\theta_{P}\sin\theta_{V}+4\cos\theta_{V}\sin\theta_{P}]\right)m_{\pi}^{2}\right\} \\
 & -\frac{2\sqrt{2}F_{V}\left(1+8\sqrt{2}\alpha_{V}\frac{m_{\pi}^{2}}{M_{V}^{2}}\right)}{3M_{\rho}^{2}F}D_{R\eta1}(0,M_{\rho}^{2},m_{\eta}^{2})(\sin\theta_{V}\sin\delta^{\rho}(0)+\sqrt{3}\cos\delta)\\
 & \left\{ \cos^{2}\delta(\sqrt{2}\cos\theta_{P}-2\sin\theta_{P})+\sin\delta^{\rho}(M_{\rho}^{2})\sin\delta^{\rho}(0)[\cos\theta_{P}\sin\theta_{V}(4\cos\theta_{V}-\sqrt{2}\sin\theta_{V})-2\sin\theta_{P}]\right\} \\
 & -\frac{F_{V}\left(1+8\sqrt{2}\alpha_{V}\frac{m_{\pi}^{2}}{M_{V}^{2}}\right)}{9\sqrt{2}M_{\rho}^{2}F}D_{R\eta2}(\sin\theta_{V}\sin\delta^{\rho}(0)+\sqrt{3}\cos\delta)\\
 & \left\{ 8\sin\delta^{\rho}(M_{\rho}^{2})\sin\delta^{\rho}(0)\left(\cos\theta_{P}(-3\sqrt{2}+\sqrt{2}\cos2\theta_{V}+4\sin2\theta_{V})\right.\right.\\
 & \left.+(-3+\cos2\theta_{V}+2\sqrt{2}\sin2\theta_{V})\sin\theta_{P}\right)m_{K}^{2}+\left(12\cos^{2}\delta(\sqrt{2}\cos\theta_{P}-2\sin\theta_{P})\right.\\
 & \left.+\sin\delta^{\rho}(M_{\rho}^{2})\sin\delta^{\rho}(0)[-9\sqrt{2}\cos(2\theta_{V}-\theta_{P})+18\sqrt{2}\cos\theta_{P}\right.\\
 & \left.\left.+7\sqrt{2}\cos(2\theta_{V}+\theta_{P})-8\sin(2\theta_{V}+\theta_{P})]\right)m_{\pi}^{2}\right\} \\
 & -\frac{F_{V}\left(1+8\sqrt{2}\alpha_{V}\frac{m_{\pi}^{2}}{M_{V}^{2}}\right)}{3\sqrt{2}M_{\omega}^{2}F}D_{R\eta1}(0,M_{\rho}^{2},m_{\eta}^{2})(\sin\theta_{V}\cos\delta-\sqrt{3}\sin\delta^{\omega}(0))\\
 & \times\{(-4\sqrt{2}\cos\delta)(-\frac{1}{2}\cos^{2}\theta_{V}\cos\theta_{p}\sin\delta^{\rho}(M_{\rho}^{2})+\frac{1}{2}\sin^{2}\theta_{V}\cos\theta_{p}\sin\delta^{\rho}(M_{\rho}^{2})\\
 & -2\sqrt{2}\sin\theta_{V}\cos\theta_{V}\cos\theta_{P}\sin\delta^{\rho}(M_{\rho}^{2})+\sqrt{2}\sin\theta_{P}\sin\delta^{\rho}(M_{\rho}^{2})+\frac{1}{2}\cos\theta_{P}\sin\delta^{\rho}(M_{\rho}^{2})\\
 & -\sqrt{2}\sin\theta_{P}\sin\delta^{\omega}(0)+\cos\theta_{P}\sin\delta^{\omega}(0))\}\\
 & -\frac{F_{V}\left(1+8\sqrt{2}\alpha_{V}\frac{m_{\pi}^{2}}{M_{V}^{2}}\right)}{9\sqrt{2}M_{\omega}^{2}F}D_{R\eta2}(\sin\theta_{V}\cos\delta-\sqrt{3}\sin\delta^{\omega}(0))\\
 & \times\left\{ -\sqrt{2}\cos\delta\ [m_{\pi}^{2}(\sin\delta^{\rho}(M_{\rho}^{2})(4\sqrt{2}\sin(2\theta_{V}+\theta_{P})+9\cos(2\theta_{V}-\theta_{P})-7\cos(2\theta_{V}+\theta_{P})\right.\\
 & -18\cos\theta_{P})+12\sin\delta^{\omega}(0)(\cos\theta_{P}-\sqrt{2}\sin\theta_{P}))-4m_{K}^{2}\sin\delta^{\rho}(M_{\rho}^{2})(2\cos\theta_{P}(2\sqrt{2}\sin2\theta_{V}\\
 & \left.+\cos2\theta_{V}-3)+\sin\theta_{P}(4\sin2\theta_{V}+\sqrt{2}\cos2\theta_{V}-3\sqrt{2}))]\right\} \\
 & +\frac{\sqrt{2}F_{V}\left(1+8\sqrt{2}\alpha_{V}\frac{2m_{K}^{2}-m_{\pi}^{2}}{M_{V}^{2}}\right)}{3M_{\phi}^{2}F}D_{R\eta1}(0,M_{\rho}^{2},m_{\eta}^{2})\cos\theta_{V}\cos\theta_{P}\sin\delta^{\rho}(M_{\rho}^{2})\\
 & \times(-4\cos2\theta_{V}+\sqrt{2}\sin2\theta_{V})-\frac{\sqrt{2}F_{V}\left(1+8\sqrt{2}\alpha_{V}\frac{2m_{K}^{2}-m_{\pi}^{2}}{M_{V}^{2}}\right)}{9M_{\phi}^{2}F}D_{R\eta2}\cos\theta_{V}\sin\delta^{\rho}(M_{\rho}^{2})\\
 & \left\{ 4(2\sqrt{2}\cos2\theta_{V}-\sin2\theta_{V})\sin\theta_{P}(m_{K}^{2}-m_{\pi}^{2})\right.\\
 & \left.+\cos\theta_{P}(4\cos2\theta_{V}-\sqrt{2}\sin2\theta_{V})(4m_{K}^{2}-m_{\pi}^{2})\right\} \,,
\end{align*}
\begin{align*}
F_{\phi\to\eta\gamma} & =\frac{2\sqrt{2}}{3M_{V}F}C_{R\eta1}(0,M_{\phi}^{2},m_{\eta}^{2})\left\{ -\sqrt{2}\cos\theta_{P}\sin\theta_{V}-\cos\theta_{V}(\cos\theta_{P}+\sqrt{2}\sin\theta_{P})\right\} \\
 & +\frac{\sqrt{2}}{9M_{V}F}C_{R\eta2}\left\{ -4\left(3\cos(\theta_{V}-\theta_{P})+\cos(\theta_{V}+\theta_{P})+2\sqrt{2}\sin(\theta_{V}+\theta_{P})\right)m_{K}^{2}\right.\\
 & \left.+\left(9\cos(\theta_{V}-\theta_{P})+\cos(\theta_{V}+\theta_{P})+2\sqrt{2}\sin(\theta_{V}+\theta_{P})\right)m_{\pi}^{2}\right\} \\
 & +\frac{\sqrt{2}F_{V}\left(1+8\sqrt{2}\alpha_{V}\frac{m_{\pi}^{2}}{M_{V}^{2}}\right)}{3M_{\rho}^{2}F}D_{R\eta1}(0,M_{\phi}^{2},m_{\eta}^{2})(\sin\theta_{V}\sin\delta^{\rho}(0)+\sqrt{3}\cos\delta)\\
 & \cos\theta_{P}\sin\delta^{\rho}(0)(-4\cos2\theta_{V}+\sqrt{2}\sin2\theta_{V})\\
 & -\frac{\sqrt{2}F_{V}\left(1+8\sqrt{2}\alpha_{V}\frac{m_{\pi}^{2}}{M_{V}^{2}}\right)}{9M_{\rho}^{2}F}D_{R\eta2}(\sin\theta_{V}\sin\delta^{\rho}(0)+\sqrt{3}\cos\delta)\sin\delta^{\rho}(0)\\
 & \left\{ 4(2\sqrt{2}\cos2\theta_{V}-\sin2\theta_{V})\sin\theta_{P}(m_{K}^{2}-m_{\pi}^{2})\right.\\
 & \left.+\cos\theta_{P}(4\cos2\theta_{V}-\sqrt{2}\sin2\theta_{V})(4m_{K}^{2}-m_{\pi}^{2})\right\} \\
 & +\frac{\sqrt{2}F_{V}\left(1+8\sqrt{2}\alpha_{V}\frac{m_{\pi}^{2}}{M_{V}^{2}}\right)}{3M_{\omega}^{2}F}D_{R\eta1}(0,M_{\phi}^{2},m_{\eta}^{2})(\sin\theta_{V}\cos\delta-\sqrt{3}\sin\delta^{\omega}(0))\\
 & \cos\delta\cos\theta_{P}(-4\cos2\theta_{V}+\sqrt{2}\sin2\theta_{V})\\
 & -\frac{\sqrt{2}F_{V}\left(1+8\sqrt{2}\alpha_{V}\frac{m_{\pi}^{2}}{M_{V}^{2}}\right)}{9M_{\omega}^{2}F}D_{R\eta2}(\sin\theta_{V}\cos\delta-\sqrt{3}\sin\delta^{\omega}(0))\cos\delta\\
 & \left\{ 4(2\sqrt{2}\cos2\theta_{V}-\sin2\theta_{V})\sin\theta_{P}(m_{K}^{2}-m_{\pi}^{2})\right.\\
 & \left.+\cos\theta_{P}(4\cos2\theta_{V}-\sqrt{2}\sin2\theta_{V})(4m_{K}^{2}-m_{\pi}^{2})\right\} \\
 & -\frac{2\sqrt{2}F_{V}\left(1+8\sqrt{2}\alpha_{V}\frac{2m_{K}^{2}-m_{\pi}^{2}}{M_{V}^{2}}\right)}{3M_{\phi}^{2}F}D_{R\eta1}(0,M_{\phi}^{2},m_{\eta}^{2})\cos\theta_{V}\\
 & \left\{ -\cos\theta_{V}\cos\theta_{P}(\sqrt{2}\cos\theta_{V}+4\sin\theta_{V})-2\sin\theta_{P}\right\} \\
 & -\frac{\sqrt{2}F_{V}\left(1+8\sqrt{2}\alpha_{V}\frac{2m_{K}^{2}-m_{\pi}^{2}}{M_{V}^{2}}\right)}{9M_{\phi}^{2}F}D_{R\eta2}\cos\theta_{V}\\
 & \left\{ (\sqrt{2}\cos\theta_{V}-2\sin\theta_{V})^{2}(\sqrt{2}\cos\theta_{P}-2\sin\theta_{P})m_{\pi}^{2}\right.\\
 & \left.-4(\sqrt{2}\cos\theta_{V}+\sin\theta_{V})^{2}(\sqrt{2}\cos\theta_{P}+\sin\theta_{P})(2m_{K}^{2}-m_{\pi}^{2})\right\} \,,
\end{align*}
\begin{align*}
F_{\eta'\to\omega\gamma} & =\frac{2\sqrt{2}}{3M_{V}F}C_{R\eta1}(0,M_{\omega}^{2},m_{\eta'}^{2})\left\{ \cos\delta\sin\theta_{V}(\sqrt{2}\cos\theta_{P}-\sin\theta_{P})\right.\\
 & \left.+\sqrt{2}\cos\delta\cos\theta_{V}\sin\theta_{P}-\sqrt{3}\sin\delta^{\omega}(M_{\omega}^{2})(\sqrt{2}\cos\theta_{P}+\sin\theta_{P})\right\} \\
 & +\frac{\sqrt{2}}{9M_{V}F}C_{R\eta2}\left\{ 4\cos\delta\left(-3\cos(\theta_{V}-\theta_{P})+\cos(\theta_{V}+\theta_{P})+2\sqrt{2}\sin(\theta_{V}+\theta_{P})\right)m_{K}^{2}\right.\\
 & +\left(-6\sqrt{3}\sin\delta^{\omega}(M_{\omega}^{2})(\sqrt{2}\cos\theta_{P}+\sin\theta_{P})\right.\\
 & \left.\left.-\cos\delta[-9\cos(\theta_{V}-\theta_{P})+\cos(\theta_{V}+\theta_{P})+2\sqrt{2}\sin(\theta_{V}+\theta_{P})]\right)m_{\pi}^{2}\right\} \\
 & -\frac{F_{V}\left(1+8\sqrt{2}\alpha_{V}\frac{m_{\pi}^{2}}{M_{V}^{2}}\right)}{3\sqrt{2}M_{\rho}^{2}F}D_{R\eta1}(0,M_{\omega}^{2},m_{\eta'}^{2})(\sin\theta_{V}\sin\delta^{\rho}(0)+\sqrt{3}\cos\delta)\\
 & \times\{(-4\sqrt{2}\cos\delta)\{\sin\theta_{P}[\sin\theta_{V}\sin\delta^{\rho}(0)(\sin\theta_{V}-2\sqrt{2}\cos\theta_{V})\\
 & +\sin\delta^{\omega}(M_{\omega}^{2})]+\sqrt{2}\cos\theta_{P}(\sin\delta^{\omega}(M_{\omega}^{2})-\sin\delta^{\rho}(0))\}\}\\
 & -\frac{F_{V}\left(1+8\sqrt{2}\alpha_{V}\frac{m_{\pi}^{2}}{M_{V}^{2}}\right)}{9\sqrt{2}M_{\rho}^{2}F}D_{R\eta2}(\sin\theta_{V}\sin\delta^{\rho}(0)+\sqrt{3}\cos\delta)\\
 & \times\{(-2\sqrt{2}\cos\delta)\{\sin\delta^{\rho}(0)[2m_{K}^{2}\ (\cos\theta_{P}\ (4\sin2\theta_{V}+\sqrt{2}\cos2\theta_{V}-3\sqrt{2})\\
 & -2\sin\theta_{P}(2\sqrt{2}\sin2\theta_{V}+\cos2\theta_{V}-3))+m_{\pi}^{2}\ (-2\sqrt{2}\cos(2\theta_{V}+\theta_{P})-8\sin2\theta_{V}\cos\theta_{P}\\
 & \left.+(\cos2\theta_{V}-9)\sin\theta_{P})]+6m_{\pi}^{2}\sin\delta^{\omega}(M_{\omega}^{2})(\sin\theta_{P}+\sqrt{2}\cos\theta_{P})\right\} \}\\
 & -\frac{2\sqrt{2}F_{V}\left(1+8\sqrt{2}\alpha_{V}\frac{m_{\pi}^{2}}{M_{V}^{2}}\right)}{3M_{\omega}^{2}F}D_{R\eta1}(0,M_{\omega}^{2},m_{\eta'}^{2})(\sin\theta_{V}\cos\delta-\sqrt{3}\sin\delta^{\omega}(0))\{\sin\delta^{\omega}(M_{\omega}^{2})\\
 & \times\sin\delta^{\omega}(0)(2\cos\theta_{P}+\sqrt{2}\sin\theta_{P})+\cos^{2}\delta[2\cos\theta_{P}+\sin\theta_{V}(4\cos\theta_{V}-\sqrt{2}\sin\theta_{V})\sin\theta_{P}]\}\\
 & -\frac{\sqrt{2}F_{V}\left(1+8\sqrt{2}\alpha_{V}\frac{m_{\pi}^{2}}{M_{V}^{2}}\right)}{9M_{\omega}^{2}F}D_{R\eta2}(\sin\theta_{V}\cos\delta-\sqrt{3}\sin\delta^{\omega}(0))\\
 & \left\{ -4\cos^{2}\delta\left(\cos\theta_{P}(-3+\cos2\theta_{V}+2\sqrt{2}\sin2\theta_{V})-(-3\sqrt{2}+\sqrt{2}\cos2\theta_{V}\right.\right.\\
 & \left.+4\sin2\theta_{V})\sin\theta_{P}\right)m_{K}^{2}+\left(6\sin\delta^{\omega}(M_{\omega}^{2})\sin\delta^{\omega}(0)(2\cos\theta_{P}+\sqrt{2}\sin\theta_{P})\right.\\
 & \left.\left.+\cos^{2}\delta\left(4\cos(2\theta_{V}+\theta_{P})+\sqrt{2}[8\cos\theta_{P}\sin2\theta_{V}-(-9+\cos2\theta_{V})\sin\theta_{P}]\right)\right)m_{\pi}^{2}\right\} \\
 & +\frac{\sqrt{2}F_{V}\left(1+8\sqrt{2}\alpha_{V}\frac{2m_{K}^{2}-m_{\pi}^{2}}{M_{V}^{2}}\right)}{3M_{\phi}^{2}F}D_{R\eta1}(0,M_{\omega}^{2},m_{\eta'}^{2})\\
 & \left\{ \cos\theta_{V}\cos\delta(-4\cos2\theta_{V}+\sqrt{2}\sin2\theta_{V})\sin\theta_{P}\right\} \\
 & -\frac{\sqrt{2}F_{V}\left(1+8\sqrt{2}\alpha_{V}\frac{2m_{K}^{2}-m_{\pi}^{2}}{M_{V}^{2}}\right)}{9M_{\phi}^{2}F}D_{R\eta2}\cos\theta_{V}\cos\delta\\
 & \left\{ -4\cos\theta_{P}(2\sqrt{2}\cos2\theta_{V}-\sin2\theta_{V})(m_{K}^{2}-m_{\pi}^{2})\right.\\
 & \left.+(4\cos2\theta_{V}-\sqrt{2}\sin2\theta_{V})\sin\theta_{P}(4m_{K}^{2}-m_{\pi}^{2})\right\} \,,
\end{align*}
\begin{align*}
F_{\eta'\to\rho\gamma} & =\frac{2\sqrt{2}}{3M_{V}F}C_{R\eta1}(0,M_{\rho}^{2},m_{\eta'}^{2})\left\{ \sqrt{3}\cos\delta(\sqrt{2}\cos\theta_{P}+\sin\theta_{P})\right.\\
 & \left.+\sin\delta^{\rho}(M_{\rho}^{2})[\sqrt{2}\cos\theta_{P}\sin\theta_{V}+(\sqrt{2}\cos\theta_{V}-\sin\theta_{V})\sin\theta_{P}]\right\} \\
 & +\frac{\sqrt{2}}{9M_{V}F}C_{R\eta2}\left\{ 4\sin\delta^{\rho}(M_{\rho}^{2})\left(-3\cos(\theta_{V}-\theta_{P})+\cos(\theta_{V}+\theta_{P})+2\sqrt{2}\sin(\theta_{V}+\theta_{P})\right)m_{K}^{2}\right.\\
 & +\left(6\sqrt{3}\cos\delta(\sqrt{2}\cos\theta_{P}+\sin\theta_{P})\right.\\
 & \left.\left.-\sin\delta^{\rho}(M_{\rho}^{2})[-9\cos(\theta_{V}-\theta_{P})+\cos(\theta_{V}+\theta_{P})+2\sqrt{2}\sin(\theta_{V}+\theta_{P})]\right)m_{\pi}^{2}\right\} \\
 & -\frac{2\sqrt{2}F_{V}\left(1+8\sqrt{2}\alpha_{V}\frac{m_{\pi}^{2}}{M_{V}^{2}}\right)}{3M_{\rho}^{2}F}D_{R\eta1}(0,M_{\rho}^{2},m_{\eta'}^{2})(\sin\theta_{V}\sin\delta^{\rho}(0)+\sqrt{3}\cos\delta)\\
 & \left\{ \cos^{2}\delta(2\cos\theta_{P}+\sqrt{2}\sin\theta_{P})+\sin\delta^{\rho}(0)\sin\delta^{\rho}(M_{\rho}^{2})[2\cos\theta_{P}+\sin\theta_{V}(4\cos\theta_{V}-\sqrt{2}\sin\theta_{V})\sin\theta_{P}]\right\} \\
 & -\frac{\sqrt{2}F_{V}\left(1+8\sqrt{2}\alpha_{V}\frac{m_{\pi}^{2}}{M_{V}^{2}}\right)}{9M_{\rho}^{2}F}D_{R\eta2}(\sin\theta_{V}\sin\delta^{\rho}(0)+\sqrt{3}\cos\delta)\\
 & \left\{ -4\sin\delta^{\rho}(0)\sin\delta^{\rho}(M_{\rho}^{2})\left(\cos\theta_{P}(-3+\cos2\theta_{V}+2\sqrt{2}\sin2\theta_{V})\right.\right.\\
 & \left.-(-3\sqrt{2}+\sqrt{2}\cos2\theta_{V}+4\sin2\theta_{V})\sin\theta_{P}\right)m_{K}^{2}+\left(6\cos^{2}\delta(2\cos\theta_{P}+\sqrt{2}\sin\theta_{P})\right.\\
 & \left.\left.+\sin\delta^{\rho}(0)\sin\delta^{\rho}(M_{\rho}^{2})\left(4\cos(2\theta_{V}+\theta_{P})+\sqrt{2}[8\cos\theta_{P}\sin2\theta_{V}-(-9+\cos2\theta_{V})\sin\theta_{P}]\right)\right)m_{\pi}^{2}\right\} \\
 & -\frac{\sqrt{2}F_{V}\left(1+8\sqrt{2}\alpha_{V}\frac{m_{\pi}^{2}}{M_{V}^{2}}\right)}{6M_{\omega}^{2}F}D_{R\eta1}(0,M_{\rho}^{2},m_{\eta'}^{2})(\sin\theta_{V}\cos\delta-\sqrt{3}\sin\delta^{\omega}(0))\\
 & \times\{(-4\sqrt{2}\cos\delta)\{\sin\theta_{P}[\sin\theta_{V}\sin\delta^{\rho}(M_{\rho}^{2})(\sin\theta_{V}-2\sqrt{2}\cos\theta_{V})\\
 & +\sin\delta^{\omega}(0)]+\sqrt{2}\cos\theta_{P}(\sin\delta^{\omega}(0)-\sin\delta^{\rho}(M_{\rho}^{2}))\}\}\\
 & -\frac{\sqrt{2}F_{V}\left(1+8\sqrt{2}\alpha_{V}\frac{m_{\pi}^{2}}{M_{V}^{2}}\right)}{18M_{\omega}^{2}F}D_{R\eta2}(\sin\theta_{V}\cos\delta-\sqrt{3}\sin\delta^{\omega}(0))\\
 & \times\{(-2\sqrt{2}\cos\delta)\{\sin\delta^{\rho}(M_{\rho}^{2})[2m_{K}^{2}\ (\cos\theta_{P}\ (4\sin2\theta_{V}+\sqrt{2}\cos2\theta_{V}-3\sqrt{2})\\
 & -2\sin\theta_{P}(2\sqrt{2}\sin2\theta_{V}+\cos2\theta_{V}-3))+m_{\pi}^{2}\ (-2\sqrt{2}\cos(2\theta_{V}+\theta_{P})-8\sin2\theta_{V}\cos\theta_{P}\\
 & \left.+(\cos2\theta_{V}-9)\sin\theta_{P})]+6m_{\pi}^{2}\sin\delta^{\omega}(0)(\sin\theta_{P}+\sqrt{2}\cos\theta_{P})\right\} \}\\
 & +\frac{\sqrt{2}F_{V}\left(1+8\sqrt{2}\alpha_{V}\frac{2m_{K}^{2}-m_{\pi}^{2}}{M_{V}^{2}}\right)}{3M_{\phi}^{2}F}D_{R\eta1}(0,M_{\rho}^{2},m_{\eta'}^{2})\\
 & \left\{ \cos\theta_{V}\sin\delta^{\rho}(M_{\rho}^{2})(-4\cos2\theta_{V}+\sqrt{2}\sin2\theta_{V})\sin\theta_{P}\right\} \\
 & -\frac{\sqrt{2}F_{V}\left(1+8\sqrt{2}\alpha_{V}\frac{2m_{K}^{2}-m_{\pi}^{2}}{M_{V}^{2}}\right)}{9M_{\phi}^{2}F}D_{R\eta2}\cos\theta_{V}\sin\delta^{\rho}(M_{\rho}^{2})\\
 & \left\{ -4\cos\theta_{P}(2\sqrt{2}\cos2\theta_{V}-\sin2\theta_{V})(m_{K}^{2}-m_{\pi}^{2})\right.\\
 & \left.+(4\cos2\theta_{V}-\sqrt{2}\sin2\theta_{V})\sin\theta_{P}(4m_{K}^{2}-m_{\pi}^{2})\right\} \,,
\end{align*}
\begin{align*}
F_{\phi\to\eta'\gamma} & =\frac{2\sqrt{2}}{3M_{V}F}C_{R\eta1}(0,M_{\phi}^{2},m_{\eta'}^{2})\left\{ \sqrt{2}\cos(\theta_{V}+\theta_{P})-\cos\theta_{V}\sin\theta_{P}\right\} \\
 & +\frac{\sqrt{2}}{9M_{V}F}C_{R\eta2}\left\{ 8\left(\sqrt{2}\cos(\theta_{V}+\theta_{P})+\cos\theta_{P}\sin\theta_{V}-2\cos\theta_{V}\sin\theta_{P}\right)m_{K}^{2}\right.\\
 & \left.+\left(-2\sqrt{2}\cos(\theta_{V}+\theta_{P})-9\sin(\theta_{V}-\theta_{P})+\sin(\theta_{V}+\theta_{P})\right)m_{\pi}^{2}\right\} \\
 & +\frac{\sqrt{2}F_{V}\left(1+8\sqrt{2}\alpha_{V}\frac{m_{\pi}^{2}}{M_{V}^{2}}\right)}{3M_{\rho}^{2}F}D_{R\eta1}(0,M_{\phi}^{2},m_{\eta'}^{2})(\sin\theta_{V}\sin\delta^{\rho}(0)+\sqrt{3}\cos\delta)\\
 & \left\{ \sin\delta^{\rho}(0)(-4\cos2\theta_{V}+\sqrt{2}\sin2\theta_{V})\sin\theta_{P}\right\} \\
 & -\frac{\sqrt{2}F_{V}\left(1+8\sqrt{2}\alpha_{V}\frac{m_{\pi}^{2}}{M_{V}^{2}}\right)}{9M_{\rho}^{2}F}D_{R\eta2}(\sin\theta_{V}\sin\delta^{\rho}(0)+\sqrt{3}\cos\delta)\sin\delta^{\rho}(0)\\
 & \left\{ -4\cos\theta_{P}(2\sqrt{2}\cos2\theta_{V}-\sin2\theta_{V})(m_{K}^{2}-m_{\pi}^{2})\right.\\
 & \left.+(4\cos2\theta_{V}-\sqrt{2}\sin2\theta_{V})\sin\theta_{P}(4m_{K}^{2}-m_{\pi}^{2})\right\} \\
 & +\frac{\sqrt{2}F_{V}\left(1+8\sqrt{2}\alpha_{V}\frac{m_{\pi}^{2}}{M_{V}^{2}}\right)}{3M_{\omega}^{2}F}D_{R\eta1}(0,M_{\phi}^{2},m_{\eta'}^{2})(\sin\theta_{V}\cos\delta-\sqrt{3}\sin\delta^{\omega}(0))\\
 & \left\{ \cos\delta(-4\cos2\theta_{V}+\sqrt{2}\sin2\theta_{V})\sin\theta_{P}\right\} \\
 & -\frac{\sqrt{2}F_{V}\left(1+8\sqrt{2}\alpha_{V}\frac{m_{\pi}^{2}}{M_{V}^{2}}\right)}{9M_{\omega}^{2}F}D_{R\eta2}(\sin\theta_{V}\cos\delta-\sqrt{3}\sin\delta^{\omega}(0))\cos\delta\\
 & \left\{ -4\cos\theta_{P}(2\sqrt{2}\cos2\theta_{V}-\sin2\theta_{V})(m_{K}^{2}-m_{\pi}^{2})\right.\\
 & \left.+(4\cos2\theta_{V}-\sqrt{2}\sin2\theta_{V})\sin\theta_{P}(4m_{K}^{2}-m_{\pi}^{2})\right\} \\
 & -\frac{2\sqrt{2}F_{V}\left(1+8\sqrt{2}\alpha_{V}\frac{2m_{K}^{2}-m_{\pi}^{2}}{M_{V}^{2}}\right)}{3M_{\phi}^{2}F}D_{R\eta1}(0,M_{\phi}^{2},m_{\eta'}^{2})\cos\theta_{V}\\
 & \left\{ 2\cos\theta_{P}-\cos\theta_{V}(\sqrt{2}\cos\theta_{V}+4\sin\theta_{V})\sin\theta_{P}\right\} \\
 & -\frac{\sqrt{2}F_{V}\left(1+8\sqrt{2}\alpha_{V}\frac{2m_{K}^{2}-m_{\pi}^{2}}{M_{V}^{2}}\right)}{9M_{\phi}^{2}F}D_{R\eta2}\cos\theta_{V}\\
 & \left\{ (\sqrt{2}\cos\theta_{V}-2\sin\theta_{V})^{2}(2\cos\theta_{P}+\sqrt{2}\sin\theta_{P})m_{\pi}^{2}\right.\\
 & \left.-4(\sqrt{2}\cos\theta_{V}+\sin\theta_{V})^{2}(-\cos\theta_{P}+\sqrt{2}\sin\theta_{P})(2m_{K}^{2}-m_{\pi}^{2})\right\} \,.
\end{align*}

\subsubsection{Three-body decays}
The three pion decays of the vector resonances are given by:
\begin{equation} \label{eq:wp3pio}
\Gamma ( V \rightarrow \pi^+(p_1) \pi^-(p_2) \pi^0(p_3)) = \frac{1}{256 \, \pi^3 \, M_V^3} \int_{s_-}^{s_+} ds  \int_{t_-}^{t_+} dt \, {\cal P}(s,t) | \Omega_V |^2 \, ,
\end{equation}
for $V = \rho, \omega, \phi$, where $s=(p_1+p_2)^2$, $t=(p_1+p_3)^2$ and
\begin{equation} \label{eq:pw3pio}
 {\cal P}(s,t) = \frac{1}{12} \left[ (3 m_{\pi}^2+M_V^2-s) s t - st^2-m_{\pi}^2 ( m_{\pi}^2-M_V^2)^2 \right] \,.
\end{equation}
The integration limits are:
\begin{eqnarray} \label{eq:intlimo}
 s_+ &=& (M_V-m_{\pi})^2 \, , \nonumber \\
s_- &=& 4 m_{\pi}^2 \, , \nonumber \\
t_{\mp} &=& \frac{1}{4s} \left[ \left( M_V^2-m_{\pi}^2 \right)^2- \left( \lambda^{1/2}(s,m_{\pi}^2,m_{\pi}^2) \pm \lambda^{1/2}(M_V^2,s,m_{\pi}^2) \right)^2 \right]  .
\end{eqnarray}
Finally $\Omega_V$ is defined by
\begin{equation} \label{eq:defmvo}
 {\cal M}_{V \rightarrow \pi^+ \pi^- \pi^0} = i \varepsilon_{\mu \nu \alpha \beta} \,  p_1^\mu \, p_2^{\nu} \, p_3^\alpha \, \varepsilon_V^\beta \, \Omega_V \, ,
\end{equation}
being $\varepsilon_V^{\mu}$ the polarization of the vector meson.
Within resonance chiral theory the corresponding reduced amplitudes, $\Omega_V$, are:

\begin{align*}
\Omega_{\omega} & =\left(\sqrt{\frac{2}{3}}\cos\theta_{V}+\sqrt{\frac{1}{3}}\sin\theta_{V}\right)\frac{8\cos\delta}{M_{\omega}F^{3}}\left\{ \frac{\sqrt{2}}{M_{V}}G_{R\pi}(M_{\omega}^{2})+G_{V}(\cos^{2}\delta+\sin\delta^{\rho}(s)\sin\delta^{\omega}(M_{\omega}^{2}))\right.\\
 & \times\,BW[\rho,s]~D_{R\pi}(M_{\omega}^{2},s)+G_{V}BW[\rho,t]~D_{R\pi}(M_{\omega}^{2},t)+G_{V}BW[\rho,u]~D_{R\pi}(M_{\omega}^{2},u)\\
 & +G_{V}\sin\delta^{\omega}(M_{\omega}^{2})\,(\sin\delta^{\omega}(M_{\omega}^{2})\,+\sin\delta^{\omega}(s)\,)BW[\omega,s]~D_{R\pi}(M_{\omega}^{2},s)\bigg\},
\end{align*}

\begin{align*}
\Omega_{\phi} & =\left(\sqrt{\frac{1}{3}}\cos\theta_{V}-\sqrt{\frac{2}{3}}\sin\theta_{V}\right)\frac{8}{M_{\phi}F^{3}}\left\{ \frac{\sqrt{2}}{M_{V}}G_{R\pi}(M_{\phi}^{2})\right.\\
 & +G_{V}\cos^{2}\delta\,BW[\rho,s]~D_{R\pi}(M_{\phi}^{2},s)+G_{V}BW[\rho,t]~D_{R\pi}(M_{\phi}^{2},t)\\
 & +G_{V}BW[\rho,u]~D_{R\pi}(M_{\phi}^{2},u)+G_{V}\sin^{2}\delta^{\omega}(s)\,BW[\omega,s]~D_{R\pi}(M_{\phi}^{2},s)\bigg\},
\end{align*}

\begin{align*}
\Omega_{\rho} & =\left(\sqrt{\frac{2}{3}}\cos\theta_{V}+\sqrt{\frac{1}{3}}\sin\theta_{V}\right)\frac{8\sin\delta^{\rho}(M_{\rho}^{2})}{M_{\rho}F^{3}}\Bigg\lbrace\frac{\sqrt{2}}{M_{V}}G_{R\pi}(M_{\rho}^{2})+2G_{V}\cos^{2}\delta\,BW_{R}[\rho,s]D_{R\pi}(M_{\rho}^{2},s)\\
 & +G_{V}BW_{R}[\rho,t]D_{R\pi}(M_{\rho}^{2},t)+G_{V}BW_{R}[\rho,u]D_{R\pi}(M_{\rho}^{2},u)\Bigg\rbrace-\left(\sqrt{\frac{2}{3}}\cos\theta_{V}+\sqrt{\frac{1}{3}}\sin\theta_{V}\right)\\
 & \times\frac{8\sin\delta^{\omega}(s)}{M_{\rho}F^{3}}G_{V}(\cos^{2}\delta-\sin\delta^{\rho}(M_{\rho}^{2})\sin\delta^{\omega}(s))\,BW_{R}[\omega,s]D_{R\pi}(M_{\rho}^{2},s),
\end{align*}
being $u = M_V^2 + 3 m_{\pi}^2 -s -t$.

\bibliographystyle{apsrev4-1}
\addcontentsline{toc}{section}{\refname}\bibliography{eepipipi}

\begin{thebibliography}{121}%
\makeatletter
\providecommand \@ifxundefined [1]{%
 \@ifx{#1\undefined}
}%
\providecommand \@ifnum [1]{%
 \ifnum #1\expandafter \@firstoftwo
 \else \expandafter \@secondoftwo
 \fi
}%
\providecommand \@ifx [1]{%
 \ifx #1\expandafter \@firstoftwo
 \else \expandafter \@secondoftwo
 \fi
}%
\providecommand \natexlab [1]{#1}%
\providecommand \enquote  [1]{``#1''}%
\providecommand \bibnamefont  [1]{#1}%
\providecommand \bibfnamefont [1]{#1}%
\providecommand \citenamefont [1]{#1}%
\providecommand \href@noop [0]{\@secondoftwo}%
\providecommand \href [0]{\begingroup \@sanitize@url \@href}%
\providecommand \@href[1]{\@@startlink{#1}\@@href}%
\providecommand \@@href[1]{\endgroup#1\@@endlink}%
\providecommand \@sanitize@url [0]{\catcode `\\12\catcode `\$12\catcode
  `\&12\catcode `\#12\catcode `\^12\catcode `\_12\catcode `\%12\relax}%
\providecommand \@@startlink[1]{}%
\providecommand \@@endlink[0]{}%
\providecommand \url  [0]{\begingroup\@sanitize@url \@url }%
\providecommand \@url [1]{\endgroup\@href {#1}{\urlprefix }}%
\providecommand \urlprefix  [0]{URL }%
\providecommand \Eprint [0]{\href }%
\providecommand \doibase [0]{http://dx.doi.org/}%
\providecommand \selectlanguage [0]{\@gobble}%
\providecommand \bibinfo  [0]{\@secondoftwo}%
\providecommand \bibfield  [0]{\@secondoftwo}%
\providecommand \translation [1]{[#1]}%
\providecommand \BibitemOpen [0]{}%
\providecommand \bibitemStop [0]{}%
\providecommand \bibitemNoStop [0]{.\EOS\space}%
\providecommand \EOS [0]{\spacefactor3000\relax}%
\providecommand \BibitemShut  [1]{\csname bibitem#1\endcsname}%
\let\auto@bib@innerbib\@empty
\bibitem [{\citenamefont {Weinberg}(1979)}]{Weinberg:1978kz}%
  \BibitemOpen
  \bibfield  {author} {\bibinfo {author} {\bibfnamefont {S.}~\bibnamefont
  {Weinberg}},\ }\href {\doibase 10.1016/0378-4371(79)90223-1} {\bibfield
  {journal} {\bibinfo  {journal} {Physica A}\ }\textbf {\bibinfo {volume}
  {96}},\ \bibinfo {pages} {327} (\bibinfo {year} {1979})}\BibitemShut
  {NoStop}%
\bibitem [{\citenamefont {Gasser}\ and\ \citenamefont
  {Leutwyler}(1984)}]{Gasser:1983yg}%
  \BibitemOpen
  \bibfield  {author} {\bibinfo {author} {\bibfnamefont {J.}~\bibnamefont
  {Gasser}}\ and\ \bibinfo {author} {\bibfnamefont {H.}~\bibnamefont
  {Leutwyler}},\ }\href {\doibase 10.1016/0003-4916(84)90242-2} {\bibfield
  {journal} {\bibinfo  {journal} {Annals Phys.}\ }\textbf {\bibinfo {volume}
  {158}},\ \bibinfo {pages} {142} (\bibinfo {year} {1984})}\BibitemShut
  {NoStop}%
\bibitem [{\citenamefont {Ecker}\ \emph
  {et~al.}(1989{\natexlab{a}})\citenamefont {Ecker}, \citenamefont {Gasser},
  \citenamefont {Pich},\ and\ \citenamefont {de~Rafael}}]{Ecker:1988te}%
  \BibitemOpen
  \bibfield  {author} {\bibinfo {author} {\bibfnamefont {G.}~\bibnamefont
  {Ecker}}, \bibinfo {author} {\bibfnamefont {J.}~\bibnamefont {Gasser}},
  \bibinfo {author} {\bibfnamefont {A.}~\bibnamefont {Pich}}, \ and\ \bibinfo
  {author} {\bibfnamefont {E.}~\bibnamefont {de~Rafael}},\ }\href {\doibase
  10.1016/0550-3213(89)90346-5} {\bibfield  {journal} {\bibinfo  {journal}
  {Nucl. Phys. B}\ }\textbf {\bibinfo {volume} {321}},\ \bibinfo {pages} {311}
  (\bibinfo {year} {1989}{\natexlab{a}})}\BibitemShut {NoStop}%
\bibitem [{\citenamefont {Ecker}\ \emph
  {et~al.}(1989{\natexlab{b}})\citenamefont {Ecker}, \citenamefont {Gasser},
  \citenamefont {Leutwyler}, \citenamefont {Pich},\ and\ \citenamefont
  {de~Rafael}}]{Ecker:1989yg}%
  \BibitemOpen
  \bibfield  {author} {\bibinfo {author} {\bibfnamefont {G.}~\bibnamefont
  {Ecker}}, \bibinfo {author} {\bibfnamefont {J.}~\bibnamefont {Gasser}},
  \bibinfo {author} {\bibfnamefont {H.}~\bibnamefont {Leutwyler}}, \bibinfo
  {author} {\bibfnamefont {A.}~\bibnamefont {Pich}}, \ and\ \bibinfo {author}
  {\bibfnamefont {E.}~\bibnamefont {de~Rafael}},\ }\href {\doibase
  10.1016/0370-2693(89)91627-4} {\bibfield  {journal} {\bibinfo  {journal}
  {Phys. Lett. B}\ }\textbf {\bibinfo {volume} {223}},\ \bibinfo {pages} {425}
  (\bibinfo {year} {1989}{\natexlab{b}})}\BibitemShut {NoStop}%
\bibitem [{\citenamefont {Cirigliano}\ \emph {et~al.}(2006)\citenamefont
  {Cirigliano}, \citenamefont {Ecker}, \citenamefont {Eidemuller},
  \citenamefont {Kaiser}, \citenamefont {Pich},\ and\ \citenamefont
  {Portol\'es}}]{Cirigliano:2006hb}%
  \BibitemOpen
  \bibfield  {author} {\bibinfo {author} {\bibfnamefont {V.}~\bibnamefont
  {Cirigliano}}, \bibinfo {author} {\bibfnamefont {G.}~\bibnamefont {Ecker}},
  \bibinfo {author} {\bibfnamefont {M.}~\bibnamefont {Eidemuller}}, \bibinfo
  {author} {\bibfnamefont {R.}~\bibnamefont {Kaiser}}, \bibinfo {author}
  {\bibfnamefont {A.}~\bibnamefont {Pich}}, \ and\ \bibinfo {author}
  {\bibfnamefont {J.}~\bibnamefont {Portol\'es}},\ }\href {\doibase
  10.1016/j.nuclphysb.2006.07.010} {\bibfield  {journal} {\bibinfo  {journal}
  {Nucl. Phys. B}\ }\textbf {\bibinfo {volume} {753}},\ \bibinfo {pages} {139}
  (\bibinfo {year} {2006})},\ \Eprint {http://arxiv.org/abs/hep-ph/0603205}
  {arXiv:hep-ph/0603205} \BibitemShut {NoStop}%
\bibitem [{\citenamefont {Portol\'es}(2010)}]{Portoles:2010yt}%
  \BibitemOpen
  \bibfield  {author} {\bibinfo {author} {\bibfnamefont {J.}~\bibnamefont
  {Portol\'es}},\ }\href {\doibase 10.1063/1.3541978} {\bibfield  {journal}
  {\bibinfo  {journal} {AIP Conf. Proc.}\ }\textbf {\bibinfo {volume} {1322}},\
  \bibinfo {pages} {178} (\bibinfo {year} {2010})},\ \Eprint
  {http://arxiv.org/abs/1010.3360} {arXiv:1010.3360 [hep-ph]} \BibitemShut
  {NoStop}%
\bibitem [{\citenamefont {'t~Hooft}(1974{\natexlab{a}})}]{tHooft:1973alw}%
  \BibitemOpen
  \bibfield  {author} {\bibinfo {author} {\bibfnamefont {G.}~\bibnamefont
  {'t~Hooft}},\ }\href {\doibase 10.1016/0550-3213(74)90154-0} {\bibfield
  {journal} {\bibinfo  {journal} {Nucl. Phys.}\ }\textbf {\bibinfo {volume}
  {B72}},\ \bibinfo {pages} {461} (\bibinfo {year}
  {1974}{\natexlab{a}})}\BibitemShut {NoStop}%
\bibitem [{\citenamefont {'t~Hooft}(1974{\natexlab{b}})}]{tHooft:1974pnl}%
  \BibitemOpen
  \bibfield  {author} {\bibinfo {author} {\bibfnamefont {G.}~\bibnamefont
  {'t~Hooft}},\ }\href {\doibase 10.1016/0550-3213(74)90088-1} {\bibfield
  {journal} {\bibinfo  {journal} {Nucl. Phys.}\ }\textbf {\bibinfo {volume}
  {B75}},\ \bibinfo {pages} {461} (\bibinfo {year}
  {1974}{\natexlab{b}})}\BibitemShut {NoStop}%
\bibitem [{\citenamefont {Witten}(1979)}]{Witten:1979kh}%
  \BibitemOpen
  \bibfield  {author} {\bibinfo {author} {\bibfnamefont {E.}~\bibnamefont
  {Witten}},\ }\href {\doibase 10.1016/0550-3213(79)90232-3} {\bibfield
  {journal} {\bibinfo  {journal} {Nucl. Phys.}\ }\textbf {\bibinfo {volume}
  {B160}},\ \bibinfo {pages} {57} (\bibinfo {year} {1979})}\BibitemShut
  {NoStop}%
\bibitem [{\citenamefont {Knecht}\ and\ \citenamefont
  {Nyffeler}(2001)}]{Knecht:2001xc}%
  \BibitemOpen
  \bibfield  {author} {\bibinfo {author} {\bibfnamefont {M.}~\bibnamefont
  {Knecht}}\ and\ \bibinfo {author} {\bibfnamefont {A.}~\bibnamefont
  {Nyffeler}},\ }\href {\doibase 10.1007/s100520100755} {\bibfield  {journal}
  {\bibinfo  {journal} {Eur. Phys. J. C}\ }\textbf {\bibinfo {volume} {21}},\
  \bibinfo {pages} {659} (\bibinfo {year} {2001})},\ \Eprint
  {http://arxiv.org/abs/hep-ph/0106034} {arXiv:hep-ph/0106034} \BibitemShut
  {NoStop}%
\bibitem [{\citenamefont {Ruiz-Femenia}\ \emph {et~al.}(2003)\citenamefont
  {Ruiz-Femenia}, \citenamefont {Pich},\ and\ \citenamefont
  {Portol\'es}}]{RuizFemenia:2003hm}%
  \BibitemOpen
  \bibfield  {author} {\bibinfo {author} {\bibfnamefont {P.}~\bibnamefont
  {Ruiz-Femenia}}, \bibinfo {author} {\bibfnamefont {A.}~\bibnamefont {Pich}},
  \ and\ \bibinfo {author} {\bibfnamefont {J.}~\bibnamefont {Portol\'es}},\
  }\href {\doibase 10.1088/1126-6708/2003/07/003} {\bibfield  {journal}
  {\bibinfo  {journal} {JHEP}\ }\textbf {\bibinfo {volume} {07}},\ \bibinfo
  {pages} {003} (\bibinfo {year} {2003})},\ \Eprint
  {http://arxiv.org/abs/hep-ph/0306157} {arXiv:hep-ph/0306157} \BibitemShut
  {NoStop}%
\bibitem [{\citenamefont {Cirigliano}\ \emph {et~al.}(2004)\citenamefont
  {Cirigliano}, \citenamefont {Ecker}, \citenamefont {Eidemuller},
  \citenamefont {Pich},\ and\ \citenamefont {Portol\'es}}]{Cirigliano:2004ue}%
  \BibitemOpen
  \bibfield  {author} {\bibinfo {author} {\bibfnamefont {V.}~\bibnamefont
  {Cirigliano}}, \bibinfo {author} {\bibfnamefont {G.}~\bibnamefont {Ecker}},
  \bibinfo {author} {\bibfnamefont {M.}~\bibnamefont {Eidemuller}}, \bibinfo
  {author} {\bibfnamefont {A.}~\bibnamefont {Pich}}, \ and\ \bibinfo {author}
  {\bibfnamefont {J.}~\bibnamefont {Portol\'es}},\ }\href {\doibase
  10.1016/j.physletb.2004.06.082} {\bibfield  {journal} {\bibinfo  {journal}
  {Phys. Lett. B}\ }\textbf {\bibinfo {volume} {596}},\ \bibinfo {pages} {96}
  (\bibinfo {year} {2004})},\ \Eprint {http://arxiv.org/abs/hep-ph/0404004}
  {arXiv:hep-ph/0404004} \BibitemShut {NoStop}%
\bibitem [{\citenamefont {Cirigliano}\ \emph {et~al.}(2005)\citenamefont
  {Cirigliano}, \citenamefont {Ecker}, \citenamefont {Eidemuller},
  \citenamefont {Kaiser}, \citenamefont {Pich},\ and\ \citenamefont
  {Portol\'es}}]{Cirigliano:2005xn}%
  \BibitemOpen
  \bibfield  {author} {\bibinfo {author} {\bibfnamefont {V.}~\bibnamefont
  {Cirigliano}}, \bibinfo {author} {\bibfnamefont {G.}~\bibnamefont {Ecker}},
  \bibinfo {author} {\bibfnamefont {M.}~\bibnamefont {Eidemuller}}, \bibinfo
  {author} {\bibfnamefont {R.}~\bibnamefont {Kaiser}}, \bibinfo {author}
  {\bibfnamefont {A.}~\bibnamefont {Pich}}, \ and\ \bibinfo {author}
  {\bibfnamefont {J.}~\bibnamefont {Portol\'es}},\ }\href {\doibase
  10.1088/1126-6708/2005/04/006} {\bibfield  {journal} {\bibinfo  {journal}
  {JHEP}\ }\textbf {\bibinfo {volume} {04}},\ \bibinfo {pages} {006} (\bibinfo
  {year} {2005})},\ \Eprint {http://arxiv.org/abs/hep-ph/0503108}
  {arXiv:hep-ph/0503108} \BibitemShut {NoStop}%
\bibitem [{\citenamefont {Husek}\ and\ \citenamefont
  {Leupold}(2015)}]{Husek:2015wta}%
  \BibitemOpen
  \bibfield  {author} {\bibinfo {author} {\bibfnamefont {T.}~\bibnamefont
  {Husek}}\ and\ \bibinfo {author} {\bibfnamefont {S.}~\bibnamefont
  {Leupold}},\ }\href {\doibase 10.1140/epjc/s10052-015-3778-x} {\bibfield
  {journal} {\bibinfo  {journal} {Eur. Phys. J. C}\ }\textbf {\bibinfo {volume}
  {75}},\ \bibinfo {pages} {586} (\bibinfo {year} {2015})},\ \Eprint
  {http://arxiv.org/abs/1507.00478} {arXiv:1507.00478 [hep-ph]} \BibitemShut
  {NoStop}%
\bibitem [{\citenamefont {Dai}\ \emph {et~al.}(2019)\citenamefont {Dai},
  \citenamefont {Fuentes-Mart\'\i{}n},\ and\ \citenamefont
  {Portol\'es}}]{Dai:2019lmj}%
  \BibitemOpen
  \bibfield  {author} {\bibinfo {author} {\bibfnamefont {L.-Y.}\ \bibnamefont
  {Dai}}, \bibinfo {author} {\bibfnamefont {J.}~\bibnamefont
  {Fuentes-Mart\'\i{}n}}, \ and\ \bibinfo {author} {\bibfnamefont
  {J.}~\bibnamefont {Portol\'es}},\ }\href {\doibase
  10.1103/PhysRevD.99.114015} {\bibfield  {journal} {\bibinfo  {journal} {Phys.
  Rev. D}\ }\textbf {\bibinfo {volume} {99}},\ \bibinfo {pages} {114015}
  (\bibinfo {year} {2019})},\ \Eprint {http://arxiv.org/abs/1902.10411}
  {arXiv:1902.10411 [hep-ph]} \BibitemShut {NoStop}%
\bibitem [{\citenamefont {Kadavy}\ \emph {et~al.}(2020)\citenamefont {Kadavy},
  \citenamefont {Kampf},\ and\ \citenamefont {Novotny}}]{Kadavy:2020hox}%
  \BibitemOpen
  \bibfield  {author} {\bibinfo {author} {\bibfnamefont {T.}~\bibnamefont
  {Kadavy}}, \bibinfo {author} {\bibfnamefont {K.}~\bibnamefont {Kampf}}, \
  and\ \bibinfo {author} {\bibfnamefont {J.}~\bibnamefont {Novotny}},\
  }\href@noop {} {\  (\bibinfo {year} {2020})},\ \Eprint
  {http://arxiv.org/abs/2006.13006} {arXiv:2006.13006 [hep-ph]} \BibitemShut
  {NoStop}%
\bibitem [{\citenamefont {Guo}\ \emph {et~al.}(2007)\citenamefont {Guo},
  \citenamefont {Sanz~Cillero},\ and\ \citenamefont {Zheng}}]{Guo:2007ff}%
  \BibitemOpen
  \bibfield  {author} {\bibinfo {author} {\bibfnamefont {Z.}~\bibnamefont
  {Guo}}, \bibinfo {author} {\bibfnamefont {J.}~\bibnamefont {Sanz~Cillero}}, \
  and\ \bibinfo {author} {\bibfnamefont {H.}~\bibnamefont {Zheng}},\ }\href
  {\doibase 10.1088/1126-6708/2007/06/030} {\bibfield  {journal} {\bibinfo
  {journal} {JHEP}\ }\textbf {\bibinfo {volume} {06}},\ \bibinfo {pages} {030}
  (\bibinfo {year} {2007})},\ \Eprint {http://arxiv.org/abs/hep-ph/0701232}
  {arXiv:hep-ph/0701232} \BibitemShut {NoStop}%
\bibitem [{\citenamefont {Jamin}\ \emph {et~al.}(2008)\citenamefont {Jamin},
  \citenamefont {Pich},\ and\ \citenamefont {Portol\'es}}]{Jamin:2008qg}%
  \BibitemOpen
  \bibfield  {author} {\bibinfo {author} {\bibfnamefont {M.}~\bibnamefont
  {Jamin}}, \bibinfo {author} {\bibfnamefont {A.}~\bibnamefont {Pich}}, \ and\
  \bibinfo {author} {\bibfnamefont {J.}~\bibnamefont {Portol\'es}},\ }\href
  {\doibase 10.1016/j.physletb.2008.04.049} {\bibfield  {journal} {\bibinfo
  {journal} {Phys. Lett. B}\ }\textbf {\bibinfo {volume} {664}},\ \bibinfo
  {pages} {78} (\bibinfo {year} {2008})},\ \Eprint
  {http://arxiv.org/abs/0803.1786} {arXiv:0803.1786 [hep-ph]} \BibitemShut
  {NoStop}%
\bibitem [{\citenamefont {Dumm}\ \emph
  {et~al.}(2010{\natexlab{a}})\citenamefont {Dumm}, \citenamefont {Roig},
  \citenamefont {Pich},\ and\ \citenamefont {Portol\'es}}]{Dumm:2009kj}%
  \BibitemOpen
  \bibfield  {author} {\bibinfo {author} {\bibfnamefont {D.}~\bibnamefont
  {Dumm}}, \bibinfo {author} {\bibfnamefont {P.}~\bibnamefont {Roig}}, \bibinfo
  {author} {\bibfnamefont {A.}~\bibnamefont {Pich}}, \ and\ \bibinfo {author}
  {\bibfnamefont {J.}~\bibnamefont {Portol\'es}},\ }\href {\doibase
  10.1103/PhysRevD.81.034031} {\bibfield  {journal} {\bibinfo  {journal} {Phys.
  Rev. D}\ }\textbf {\bibinfo {volume} {81}},\ \bibinfo {pages} {034031}
  (\bibinfo {year} {2010}{\natexlab{a}})},\ \Eprint
  {http://arxiv.org/abs/0911.2640} {arXiv:0911.2640 [hep-ph]} \BibitemShut
  {NoStop}%
\bibitem [{\citenamefont {Dumm}\ \emph
  {et~al.}(2010{\natexlab{b}})\citenamefont {Dumm}, \citenamefont {Roig},
  \citenamefont {Pich},\ and\ \citenamefont {Portol\'es}}]{Dumm:2009va}%
  \BibitemOpen
  \bibfield  {author} {\bibinfo {author} {\bibfnamefont {D.}~\bibnamefont
  {Dumm}}, \bibinfo {author} {\bibfnamefont {P.}~\bibnamefont {Roig}}, \bibinfo
  {author} {\bibfnamefont {A.}~\bibnamefont {Pich}}, \ and\ \bibinfo {author}
  {\bibfnamefont {J.}~\bibnamefont {Portol\'es}},\ }\href {\doibase
  10.1016/j.physletb.2010.01.059} {\bibfield  {journal} {\bibinfo  {journal}
  {Phys. Lett. B}\ }\textbf {\bibinfo {volume} {685}},\ \bibinfo {pages} {158}
  (\bibinfo {year} {2010}{\natexlab{b}})},\ \Eprint
  {http://arxiv.org/abs/0911.4436} {arXiv:0911.4436 [hep-ph]} \BibitemShut
  {NoStop}%
\bibitem [{\citenamefont {Guo}\ and\ \citenamefont {Roig}(2010)}]{Guo:2010dv}%
  \BibitemOpen
  \bibfield  {author} {\bibinfo {author} {\bibfnamefont {Z.-H.}\ \bibnamefont
  {Guo}}\ and\ \bibinfo {author} {\bibfnamefont {P.}~\bibnamefont {Roig}},\
  }\href {\doibase 10.1103/PhysRevD.82.113016} {\bibfield  {journal} {\bibinfo
  {journal} {Phys. Rev. D}\ }\textbf {\bibinfo {volume} {82}},\ \bibinfo
  {pages} {113016} (\bibinfo {year} {2010})},\ \Eprint
  {http://arxiv.org/abs/1009.2542} {arXiv:1009.2542 [hep-ph]} \BibitemShut
  {NoStop}%
\bibitem [{\citenamefont {Escribano}\ \emph {et~al.}(2013)\citenamefont
  {Escribano}, \citenamefont {Gonz\'alez-Solis},\ and\ \citenamefont
  {Roig}}]{Escribano:2013bca}%
  \BibitemOpen
  \bibfield  {author} {\bibinfo {author} {\bibfnamefont {R.}~\bibnamefont
  {Escribano}}, \bibinfo {author} {\bibfnamefont {S.}~\bibnamefont
  {Gonz\'alez-Solis}}, \ and\ \bibinfo {author} {\bibfnamefont
  {P.}~\bibnamefont {Roig}},\ }\href {\doibase 10.1007/JHEP10(2013)039}
  {\bibfield  {journal} {\bibinfo  {journal} {JHEP}\ }\textbf {\bibinfo
  {volume} {10}},\ \bibinfo {pages} {039} (\bibinfo {year} {2013})},\ \Eprint
  {http://arxiv.org/abs/1307.7908} {arXiv:1307.7908 [hep-ph]} \BibitemShut
  {NoStop}%
\bibitem [{\citenamefont {Nugent}\ \emph {et~al.}(2013)\citenamefont {Nugent},
  \citenamefont {Przedzinski}, \citenamefont {Roig}, \citenamefont
  {Shekhovtsova},\ and\ \citenamefont {Was}}]{Nugent:2013hxa}%
  \BibitemOpen
  \bibfield  {author} {\bibinfo {author} {\bibfnamefont {I.}~\bibnamefont
  {Nugent}}, \bibinfo {author} {\bibfnamefont {T.}~\bibnamefont {Przedzinski}},
  \bibinfo {author} {\bibfnamefont {P.}~\bibnamefont {Roig}}, \bibinfo {author}
  {\bibfnamefont {O.}~\bibnamefont {Shekhovtsova}}, \ and\ \bibinfo {author}
  {\bibfnamefont {Z.}~\bibnamefont {Was}},\ }\href {\doibase
  10.1103/PhysRevD.88.093012} {\bibfield  {journal} {\bibinfo  {journal} {Phys.
  Rev. D}\ }\textbf {\bibinfo {volume} {88}},\ \bibinfo {pages} {093012}
  (\bibinfo {year} {2013})},\ \Eprint {http://arxiv.org/abs/1310.1053}
  {arXiv:1310.1053 [hep-ph]} \BibitemShut {NoStop}%
\bibitem [{\citenamefont {Miranda}\ and\ \citenamefont
  {Roig}(2020)}]{Miranda:2020wdg}%
  \BibitemOpen
  \bibfield  {author} {\bibinfo {author} {\bibfnamefont {J.}~\bibnamefont
  {Miranda}}\ and\ \bibinfo {author} {\bibfnamefont {P.}~\bibnamefont {Roig}},\
  }\href@noop {} {\  (\bibinfo {year} {2020})},\ \Eprint
  {http://arxiv.org/abs/2007.11019} {arXiv:2007.11019 [hep-ph]} \BibitemShut
  {NoStop}%
\bibitem [{\citenamefont {Chen}\ \emph {et~al.}(2012)\citenamefont {Chen},
  \citenamefont {Guo},\ and\ \citenamefont {Zheng}}]{Chen:2012vw}%
  \BibitemOpen
  \bibfield  {author} {\bibinfo {author} {\bibfnamefont {Y.-H.}\ \bibnamefont
  {Chen}}, \bibinfo {author} {\bibfnamefont {Z.-H.}\ \bibnamefont {Guo}}, \
  and\ \bibinfo {author} {\bibfnamefont {H.-Q.}\ \bibnamefont {Zheng}},\ }\href
  {\doibase 10.1103/PhysRevD.85.054018} {\bibfield  {journal} {\bibinfo
  {journal} {Phys. Rev. D}\ }\textbf {\bibinfo {volume} {85}},\ \bibinfo
  {pages} {054018} (\bibinfo {year} {2012})},\ \Eprint
  {http://arxiv.org/abs/1201.2135} {arXiv:1201.2135 [hep-ph]} \BibitemShut
  {NoStop}%
\bibitem [{\citenamefont {Xiao}\ \emph {et~al.}(2015)\citenamefont {Xiao},
  \citenamefont {Dato}, \citenamefont {Hanhart}, \citenamefont {Kubis},
  \citenamefont {Meissner},\ and\ \citenamefont {Wirzba}}]{Xiao:2015uva}%
  \BibitemOpen
  \bibfield  {author} {\bibinfo {author} {\bibfnamefont {C.}~\bibnamefont
  {Xiao}}, \bibinfo {author} {\bibfnamefont {T.}~\bibnamefont {Dato}}, \bibinfo
  {author} {\bibfnamefont {C.}~\bibnamefont {Hanhart}}, \bibinfo {author}
  {\bibfnamefont {B.}~\bibnamefont {Kubis}}, \bibinfo {author} {\bibfnamefont
  {U.~G.}\ \bibnamefont {Meissner}}, \ and\ \bibinfo {author} {\bibfnamefont
  {A.}~\bibnamefont {Wirzba}},\ }\href@noop {} {\  (\bibinfo {year} {2015})},\
  \Eprint {http://arxiv.org/abs/1509.02194} {arXiv:1509.02194 [hep-ph]}
  \BibitemShut {NoStop}%
\bibitem [{\citenamefont {Dai}\ \emph {et~al.}(2018)\citenamefont {Dai},
  \citenamefont {Kang}, \citenamefont {Meissner}, \citenamefont {Song},\ and\
  \citenamefont {Yao}}]{Dai:2017tew}%
  \BibitemOpen
  \bibfield  {author} {\bibinfo {author} {\bibfnamefont {L.-Y.}\ \bibnamefont
  {Dai}}, \bibinfo {author} {\bibfnamefont {X.-W.}\ \bibnamefont {Kang}},
  \bibinfo {author} {\bibfnamefont {U.-G.}\ \bibnamefont {Meissner}}, \bibinfo
  {author} {\bibfnamefont {X.-Y.}\ \bibnamefont {Song}}, \ and\ \bibinfo
  {author} {\bibfnamefont {D.-L.}\ \bibnamefont {Yao}},\ }\href {\doibase
  10.1103/PhysRevD.97.036012} {\bibfield  {journal} {\bibinfo  {journal} {Phys.
  Rev. D}\ }\textbf {\bibinfo {volume} {97}},\ \bibinfo {pages} {036012}
  (\bibinfo {year} {2018})},\ \Eprint {http://arxiv.org/abs/1712.02119}
  {arXiv:1712.02119 [hep-ph]} \BibitemShut {NoStop}%
\bibitem [{\citenamefont {Dubinsky}\ \emph {et~al.}(2005)\citenamefont
  {Dubinsky}, \citenamefont {Korchin}, \citenamefont {Merenkov}, \citenamefont
  {Pancheri},\ and\ \citenamefont {Shekhovtsova}}]{Dubinsky:2004xv}%
  \BibitemOpen
  \bibfield  {author} {\bibinfo {author} {\bibfnamefont {S.}~\bibnamefont
  {Dubinsky}}, \bibinfo {author} {\bibfnamefont {A.}~\bibnamefont {Korchin}},
  \bibinfo {author} {\bibfnamefont {N.}~\bibnamefont {Merenkov}}, \bibinfo
  {author} {\bibfnamefont {G.}~\bibnamefont {Pancheri}}, \ and\ \bibinfo
  {author} {\bibfnamefont {O.}~\bibnamefont {Shekhovtsova}},\ }\href {\doibase
  10.1140/epjc/s2004-02109-7} {\bibfield  {journal} {\bibinfo  {journal} {Eur.
  Phys. J. C}\ }\textbf {\bibinfo {volume} {40}},\ \bibinfo {pages} {41}
  (\bibinfo {year} {2005})},\ \Eprint {http://arxiv.org/abs/hep-ph/0411113}
  {arXiv:hep-ph/0411113} \BibitemShut {NoStop}%
\bibitem [{\citenamefont {Dai}\ \emph {et~al.}(2013)\citenamefont {Dai},
  \citenamefont {Portol\'es},\ and\ \citenamefont
  {Shekhovtsova}}]{Dai:2013joa}%
  \BibitemOpen
  \bibfield  {author} {\bibinfo {author} {\bibfnamefont {L.}~\bibnamefont
  {Dai}}, \bibinfo {author} {\bibfnamefont {J.}~\bibnamefont {Portol\'es}}, \
  and\ \bibinfo {author} {\bibfnamefont {O.}~\bibnamefont {Shekhovtsova}},\
  }\href {\doibase 10.1103/PhysRevD.88.056001} {\bibfield  {journal} {\bibinfo
  {journal} {Phys. Rev. D}\ }\textbf {\bibinfo {volume} {88}},\ \bibinfo
  {pages} {056001} (\bibinfo {year} {2013})},\ \Eprint
  {http://arxiv.org/abs/1305.5751} {arXiv:1305.5751 [hep-ph]} \BibitemShut
  {NoStop}%
\bibitem [{\citenamefont {Niecknig}\ \emph {et~al.}(2012)\citenamefont
  {Niecknig}, \citenamefont {Kubis},\ and\ \citenamefont
  {Schneider}}]{Niecknig:2012sj}%
  \BibitemOpen
  \bibfield  {author} {\bibinfo {author} {\bibfnamefont {F.}~\bibnamefont
  {Niecknig}}, \bibinfo {author} {\bibfnamefont {B.}~\bibnamefont {Kubis}}, \
  and\ \bibinfo {author} {\bibfnamefont {S.~P.}\ \bibnamefont {Schneider}},\
  }\href {\doibase 10.1140/epjc/s10052-012-2014-1} {\bibfield  {journal}
  {\bibinfo  {journal} {Eur. Phys. J. C}\ }\textbf {\bibinfo {volume} {72}},\
  \bibinfo {pages} {2014} (\bibinfo {year} {2012})},\ \Eprint
  {http://arxiv.org/abs/1203.2501} {arXiv:1203.2501 [hep-ph]} \BibitemShut
  {NoStop}%
\bibitem [{\citenamefont {Schneider}\ \emph {et~al.}(2012)\citenamefont
  {Schneider}, \citenamefont {Kubis},\ and\ \citenamefont
  {Niecknig}}]{Schneider:2012ez}%
  \BibitemOpen
  \bibfield  {author} {\bibinfo {author} {\bibfnamefont {S.~P.}\ \bibnamefont
  {Schneider}}, \bibinfo {author} {\bibfnamefont {B.}~\bibnamefont {Kubis}}, \
  and\ \bibinfo {author} {\bibfnamefont {F.}~\bibnamefont {Niecknig}},\ }\href
  {\doibase 10.1103/PhysRevD.86.054013} {\bibfield  {journal} {\bibinfo
  {journal} {Phys. Rev. D}\ }\textbf {\bibinfo {volume} {86}},\ \bibinfo
  {pages} {054013} (\bibinfo {year} {2012})},\ \Eprint
  {http://arxiv.org/abs/1206.3098} {arXiv:1206.3098 [hep-ph]} \BibitemShut
  {NoStop}%
\bibitem [{\citenamefont {Danilkin}\ \emph {et~al.}(2015)\citenamefont
  {Danilkin}, \citenamefont {Fern\'andez-Ram\'\i{}rez}, \citenamefont {Guo},
  \citenamefont {Mathieu}, \citenamefont {Schott}, \citenamefont {Shi},\ and\
  \citenamefont {Szczepaniak}}]{Danilkin:2014cra}%
  \BibitemOpen
  \bibfield  {author} {\bibinfo {author} {\bibfnamefont {I.}~\bibnamefont
  {Danilkin}}, \bibinfo {author} {\bibfnamefont {C.}~\bibnamefont
  {Fern\'andez-Ram\'\i{}rez}}, \bibinfo {author} {\bibfnamefont
  {P.}~\bibnamefont {Guo}}, \bibinfo {author} {\bibfnamefont {V.}~\bibnamefont
  {Mathieu}}, \bibinfo {author} {\bibfnamefont {D.}~\bibnamefont {Schott}},
  \bibinfo {author} {\bibfnamefont {M.}~\bibnamefont {Shi}}, \ and\ \bibinfo
  {author} {\bibfnamefont {A.}~\bibnamefont {Szczepaniak}},\ }\href {\doibase
  10.1103/PhysRevD.91.094029} {\bibfield  {journal} {\bibinfo  {journal} {Phys.
  Rev. D}\ }\textbf {\bibinfo {volume} {91}},\ \bibinfo {pages} {094029}
  (\bibinfo {year} {2015})},\ \Eprint {http://arxiv.org/abs/1409.7708}
  {arXiv:1409.7708 [hep-ph]} \BibitemShut {NoStop}%
\bibitem [{\citenamefont {Albaladejo}\ and\ \citenamefont
  {Moussallam}(2017)}]{Albaladejo:2017hhj}%
  \BibitemOpen
  \bibfield  {author} {\bibinfo {author} {\bibfnamefont {M.}~\bibnamefont
  {Albaladejo}}\ and\ \bibinfo {author} {\bibfnamefont {B.}~\bibnamefont
  {Moussallam}},\ }\href {\doibase 10.1140/epjc/s10052-017-5052-x} {\bibfield
  {journal} {\bibinfo  {journal} {Eur. Phys. J. C}\ }\textbf {\bibinfo {volume}
  {77}},\ \bibinfo {pages} {508} (\bibinfo {year} {2017})},\ \Eprint
  {http://arxiv.org/abs/1702.04931} {arXiv:1702.04931 [hep-ph]} \BibitemShut
  {NoStop}%
\bibitem [{\citenamefont {Isken}\ \emph {et~al.}(2017)\citenamefont {Isken},
  \citenamefont {Kubis}, \citenamefont {Schneider},\ and\ \citenamefont
  {Stoffer}}]{Isken:2017dkw}%
  \BibitemOpen
  \bibfield  {author} {\bibinfo {author} {\bibfnamefont {T.}~\bibnamefont
  {Isken}}, \bibinfo {author} {\bibfnamefont {B.}~\bibnamefont {Kubis}},
  \bibinfo {author} {\bibfnamefont {S.~P.}\ \bibnamefont {Schneider}}, \ and\
  \bibinfo {author} {\bibfnamefont {P.}~\bibnamefont {Stoffer}},\ }\href
  {\doibase 10.1140/epjc/s10052-017-5024-1} {\bibfield  {journal} {\bibinfo
  {journal} {Eur. Phys. J. C}\ }\textbf {\bibinfo {volume} {77}},\ \bibinfo
  {pages} {489} (\bibinfo {year} {2017})},\ \Eprint
  {http://arxiv.org/abs/1705.04339} {arXiv:1705.04339 [hep-ph]} \BibitemShut
  {NoStop}%
\bibitem [{\citenamefont {Colangelo}\ \emph {et~al.}(2018)\citenamefont
  {Colangelo}, \citenamefont {Lanz}, \citenamefont {Leutwyler},\ and\
  \citenamefont {Passemar}}]{Colangelo:2018jxw}%
  \BibitemOpen
  \bibfield  {author} {\bibinfo {author} {\bibfnamefont {G.}~\bibnamefont
  {Colangelo}}, \bibinfo {author} {\bibfnamefont {S.}~\bibnamefont {Lanz}},
  \bibinfo {author} {\bibfnamefont {H.}~\bibnamefont {Leutwyler}}, \ and\
  \bibinfo {author} {\bibfnamefont {E.}~\bibnamefont {Passemar}},\ }\href
  {\doibase 10.1140/epjc/s10052-018-6377-9} {\bibfield  {journal} {\bibinfo
  {journal} {Eur. Phys. J. C}\ }\textbf {\bibinfo {volume} {78}},\ \bibinfo
  {pages} {947} (\bibinfo {year} {2018})},\ \Eprint
  {http://arxiv.org/abs/1807.11937} {arXiv:1807.11937 [hep-ph]} \BibitemShut
  {NoStop}%
\bibitem [{\citenamefont {Yao}\ \emph {et~al.}(2020)\citenamefont {Yao},
  \citenamefont {Dai}, \citenamefont {Zheng},\ and\ \citenamefont
  {Zhou}}]{Yao:2020bxx}%
  \BibitemOpen
  \bibfield  {author} {\bibinfo {author} {\bibfnamefont {D.-L.}\ \bibnamefont
  {Yao}}, \bibinfo {author} {\bibfnamefont {L.-Y.}\ \bibnamefont {Dai}},
  \bibinfo {author} {\bibfnamefont {H.-Q.}\ \bibnamefont {Zheng}}, \ and\
  \bibinfo {author} {\bibfnamefont {Z.-Y.}\ \bibnamefont {Zhou}},\ }\href@noop
  {} {\  (\bibinfo {year} {2020})},\ \Eprint {http://arxiv.org/abs/2009.13495}
  {arXiv:2009.13495 [hep-ph]} \BibitemShut {NoStop}%
\bibitem [{\citenamefont {Bennett}\ \emph {et~al.}(2006)\citenamefont {Bennett}
  \emph {et~al.}}]{Bennett:2006fi}%
  \BibitemOpen
  \bibfield  {author} {\bibinfo {author} {\bibfnamefont {G.}~\bibnamefont
  {Bennett}} \emph {et~al.} (\bibinfo {collaboration} {Muon g-2}),\ }\href
  {\doibase 10.1103/PhysRevD.73.072003} {\bibfield  {journal} {\bibinfo
  {journal} {Phys. Rev. D}\ }\textbf {\bibinfo {volume} {73}},\ \bibinfo
  {pages} {072003} (\bibinfo {year} {2006})},\ \Eprint
  {http://arxiv.org/abs/hep-ex/0602035} {arXiv:hep-ex/0602035} \BibitemShut
  {NoStop}%
\bibitem [{\citenamefont {Zyla}\ \emph {et~al.}(2020)\citenamefont {Zyla} \emph
  {et~al.}}]{Zyla:2020zbs}%
  \BibitemOpen
  \bibfield  {author} {\bibinfo {author} {\bibfnamefont {P.}~\bibnamefont
  {Zyla}} \emph {et~al.} (\bibinfo {collaboration} {Particle Data Group}),\
  }\href {\doibase 10.1093/ptep/ptaa104} {\bibfield  {journal} {\bibinfo
  {journal} {PTEP}\ }\textbf {\bibinfo {volume} {2020}},\ \bibinfo {pages}
  {083C01} (\bibinfo {year} {2020})}\BibitemShut {NoStop}%
\bibitem [{\citenamefont {Aoyama}\ \emph {et~al.}(2020)\citenamefont {Aoyama}
  \emph {et~al.}}]{Aoyama:2020ynm}%
  \BibitemOpen
  \bibfield  {author} {\bibinfo {author} {\bibfnamefont {T.}~\bibnamefont
  {Aoyama}} \emph {et~al.},\ }\href {\doibase 10.1016/j.physrep.2020.07.006}
  {\bibfield  {journal} {\bibinfo  {journal} {Phys. Rept.}\ }\textbf {\bibinfo
  {volume} {887}},\ \bibinfo {pages} {1} (\bibinfo {year} {2020})},\ \Eprint
  {http://arxiv.org/abs/2006.04822} {arXiv:2006.04822 [hep-ph]} \BibitemShut
  {NoStop}%
\bibitem [{\citenamefont {Jegerlehner}(2017)}]{Jegerlehner:2017gek}%
  \BibitemOpen
  \bibfield  {author} {\bibinfo {author} {\bibfnamefont {F.}~\bibnamefont
  {Jegerlehner}},\ }\href {\doibase 10.1007/978-3-319-63577-4} {\emph {\bibinfo
  {title} {{The Anomalous Magnetic Moment of the Muon}}}},\ Vol.\ \bibinfo
  {volume} {274}\ (\bibinfo  {publisher} {Springer},\ \bibinfo {address}
  {Cham},\ \bibinfo {year} {2017})\BibitemShut {NoStop}%
\bibitem [{\citenamefont {Aoyama}\ \emph {et~al.}(2012)\citenamefont {Aoyama},
  \citenamefont {Hayakawa}, \citenamefont {Kinoshita},\ and\ \citenamefont
  {Nio}}]{Aoyama:2012wk}%
  \BibitemOpen
  \bibfield  {author} {\bibinfo {author} {\bibfnamefont {T.}~\bibnamefont
  {Aoyama}}, \bibinfo {author} {\bibfnamefont {M.}~\bibnamefont {Hayakawa}},
  \bibinfo {author} {\bibfnamefont {T.}~\bibnamefont {Kinoshita}}, \ and\
  \bibinfo {author} {\bibfnamefont {M.}~\bibnamefont {Nio}},\ }\href {\doibase
  10.1103/PhysRevLett.109.111808} {\bibfield  {journal} {\bibinfo  {journal}
  {Phys. Rev. Lett.}\ }\textbf {\bibinfo {volume} {109}},\ \bibinfo {pages}
  {111808} (\bibinfo {year} {2012})},\ \Eprint {http://arxiv.org/abs/1205.5370}
  {arXiv:1205.5370 [hep-ph]} \BibitemShut {NoStop}%
\bibitem [{\citenamefont {Aoyama}\ \emph {et~al.}(2019)\citenamefont {Aoyama},
  \citenamefont {Kinoshita},\ and\ \citenamefont {Nio}}]{Aoyama:2019ryr}%
  \BibitemOpen
  \bibfield  {author} {\bibinfo {author} {\bibfnamefont {T.}~\bibnamefont
  {Aoyama}}, \bibinfo {author} {\bibfnamefont {T.}~\bibnamefont {Kinoshita}}, \
  and\ \bibinfo {author} {\bibfnamefont {M.}~\bibnamefont {Nio}},\ }\href
  {\doibase 10.3390/atoms7010028} {\bibfield  {journal} {\bibinfo  {journal}
  {Atoms}\ }\textbf {\bibinfo {volume} {7}},\ \bibinfo {pages} {28} (\bibinfo
  {year} {2019})}\BibitemShut {NoStop}%
\bibitem [{\citenamefont {Jackiw}\ and\ \citenamefont
  {Weinberg}(1972)}]{Jackiw:1972jz}%
  \BibitemOpen
  \bibfield  {author} {\bibinfo {author} {\bibfnamefont {R.}~\bibnamefont
  {Jackiw}}\ and\ \bibinfo {author} {\bibfnamefont {S.}~\bibnamefont
  {Weinberg}},\ }\href {\doibase 10.1103/PhysRevD.5.2396} {\bibfield  {journal}
  {\bibinfo  {journal} {Phys. Rev. D}\ }\textbf {\bibinfo {volume} {5}},\
  \bibinfo {pages} {2396} (\bibinfo {year} {1972})}\BibitemShut {NoStop}%
\bibitem [{\citenamefont {Knecht}\ \emph {et~al.}(2002)\citenamefont {Knecht},
  \citenamefont {Peris}, \citenamefont {Perrottet},\ and\ \citenamefont
  {De~Rafael}}]{Knecht:2002hr}%
  \BibitemOpen
  \bibfield  {author} {\bibinfo {author} {\bibfnamefont {M.}~\bibnamefont
  {Knecht}}, \bibinfo {author} {\bibfnamefont {S.}~\bibnamefont {Peris}},
  \bibinfo {author} {\bibfnamefont {M.}~\bibnamefont {Perrottet}}, \ and\
  \bibinfo {author} {\bibfnamefont {E.}~\bibnamefont {De~Rafael}},\ }\href
  {\doibase 10.1088/1126-6708/2002/11/003} {\bibfield  {journal} {\bibinfo
  {journal} {JHEP}\ }\textbf {\bibinfo {volume} {11}},\ \bibinfo {pages} {003}
  (\bibinfo {year} {2002})},\ \Eprint {http://arxiv.org/abs/hep-ph/0205102}
  {arXiv:hep-ph/0205102} \BibitemShut {NoStop}%
\bibitem [{\citenamefont {Czarnecki}\ \emph {et~al.}(2003)\citenamefont
  {Czarnecki}, \citenamefont {Marciano},\ and\ \citenamefont
  {Vainshtein}}]{Czarnecki:2002nt}%
  \BibitemOpen
  \bibfield  {author} {\bibinfo {author} {\bibfnamefont {A.}~\bibnamefont
  {Czarnecki}}, \bibinfo {author} {\bibfnamefont {W.~J.}\ \bibnamefont
  {Marciano}}, \ and\ \bibinfo {author} {\bibfnamefont {A.}~\bibnamefont
  {Vainshtein}},\ }\href {\doibase 10.1103/PhysRevD.67.073006} {\bibfield
  {journal} {\bibinfo  {journal} {Phys. Rev. D}\ }\textbf {\bibinfo {volume}
  {67}},\ \bibinfo {pages} {073006} (\bibinfo {year} {2003})},\ \bibinfo {note}
  {[Erratum: Phys.Rev.D 73, 119901 (2006)]},\ \Eprint
  {http://arxiv.org/abs/hep-ph/0212229} {arXiv:hep-ph/0212229} \BibitemShut
  {NoStop}%
\bibitem [{\citenamefont {Gnendiger}\ \emph {et~al.}(2013)\citenamefont
  {Gnendiger}, \citenamefont {St\"ockinger},\ and\ \citenamefont
  {St\"ockinger-Kim}}]{Gnendiger:2013pva}%
  \BibitemOpen
  \bibfield  {author} {\bibinfo {author} {\bibfnamefont {C.}~\bibnamefont
  {Gnendiger}}, \bibinfo {author} {\bibfnamefont {D.}~\bibnamefont
  {St\"ockinger}}, \ and\ \bibinfo {author} {\bibfnamefont {H.}~\bibnamefont
  {St\"ockinger-Kim}},\ }\href {\doibase 10.1103/PhysRevD.88.053005} {\bibfield
   {journal} {\bibinfo  {journal} {Phys. Rev. D}\ }\textbf {\bibinfo {volume}
  {88}},\ \bibinfo {pages} {053005} (\bibinfo {year} {2013})},\ \Eprint
  {http://arxiv.org/abs/1306.5546} {arXiv:1306.5546 [hep-ph]} \BibitemShut
  {NoStop}%
\bibitem [{\citenamefont {Prades}\ \emph {et~al.}(2009)\citenamefont {Prades},
  \citenamefont {de~Rafael},\ and\ \citenamefont {Vainshtein}}]{Prades:2009tw}%
  \BibitemOpen
  \bibfield  {author} {\bibinfo {author} {\bibfnamefont {J.}~\bibnamefont
  {Prades}}, \bibinfo {author} {\bibfnamefont {E.}~\bibnamefont {de~Rafael}}, \
  and\ \bibinfo {author} {\bibfnamefont {A.}~\bibnamefont {Vainshtein}},\
  }\enquote {\bibinfo {title} {{The Hadronic Light-by-Light Scattering
  Contribution to the Muon and Electron Anomalous Magnetic Moments}},}\ \
  (\bibinfo {year} {2009})\ pp.\ \bibinfo {pages} {303--317},\ \Eprint
  {http://arxiv.org/abs/0901.0306} {arXiv:0901.0306 [hep-ph]} \BibitemShut
  {NoStop}%
\bibitem [{\citenamefont {Colangelo}\ \emph {et~al.}(2020)\citenamefont
  {Colangelo}, \citenamefont {Hagelstein}, \citenamefont {Hoferichter},
  \citenamefont {Laub},\ and\ \citenamefont {Stoffer}}]{Colangelo:2019uex}%
  \BibitemOpen
  \bibfield  {author} {\bibinfo {author} {\bibfnamefont {G.}~\bibnamefont
  {Colangelo}}, \bibinfo {author} {\bibfnamefont {F.}~\bibnamefont
  {Hagelstein}}, \bibinfo {author} {\bibfnamefont {M.}~\bibnamefont
  {Hoferichter}}, \bibinfo {author} {\bibfnamefont {L.}~\bibnamefont {Laub}}, \
  and\ \bibinfo {author} {\bibfnamefont {P.}~\bibnamefont {Stoffer}},\ }\href
  {\doibase 10.1007/JHEP03(2020)101} {\bibfield  {journal} {\bibinfo  {journal}
  {JHEP}\ }\textbf {\bibinfo {volume} {03}},\ \bibinfo {pages} {101} (\bibinfo
  {year} {2020})},\ \Eprint {http://arxiv.org/abs/1910.13432} {arXiv:1910.13432
  [hep-ph]} \BibitemShut {NoStop}%
\bibitem [{\citenamefont {Danilkin}\ \emph {et~al.}(2020)\citenamefont
  {Danilkin}, \citenamefont {Deineka},\ and\ \citenamefont
  {Vanderhaeghen}}]{Danilkin:2019opj}%
  \BibitemOpen
  \bibfield  {author} {\bibinfo {author} {\bibfnamefont {I.}~\bibnamefont
  {Danilkin}}, \bibinfo {author} {\bibfnamefont {O.}~\bibnamefont {Deineka}}, \
  and\ \bibinfo {author} {\bibfnamefont {M.}~\bibnamefont {Vanderhaeghen}},\
  }\href {\doibase 10.1103/PhysRevD.101.054008} {\bibfield  {journal} {\bibinfo
   {journal} {Phys. Rev. D}\ }\textbf {\bibinfo {volume} {101}},\ \bibinfo
  {pages} {054008} (\bibinfo {year} {2020})},\ \Eprint
  {http://arxiv.org/abs/1909.04158} {arXiv:1909.04158 [hep-ph]} \BibitemShut
  {NoStop}%
\bibitem [{\citenamefont {Borsanyi}\ \emph {et~al.}(2020)\citenamefont
  {Borsanyi} \emph {et~al.}}]{Borsanyi:2020mff}%
  \BibitemOpen
  \bibfield  {author} {\bibinfo {author} {\bibfnamefont {S.}~\bibnamefont
  {Borsanyi}} \emph {et~al.},\ }\href@noop {} {\  (\bibinfo {year} {2020})},\
  \Eprint {http://arxiv.org/abs/2002.12347} {arXiv:2002.12347 [hep-lat]}
  \BibitemShut {NoStop}%
\bibitem [{\citenamefont {Blum}\ \emph
  {et~al.}(2017{\natexlab{a}})\citenamefont {Blum}, \citenamefont {Christ},
  \citenamefont {Hayakawa}, \citenamefont {Izubuchi}, \citenamefont {Jin},
  \citenamefont {Jung},\ and\ \citenamefont {Lehner}}]{Blum:2016lnc}%
  \BibitemOpen
  \bibfield  {author} {\bibinfo {author} {\bibfnamefont {T.}~\bibnamefont
  {Blum}}, \bibinfo {author} {\bibfnamefont {N.}~\bibnamefont {Christ}},
  \bibinfo {author} {\bibfnamefont {M.}~\bibnamefont {Hayakawa}}, \bibinfo
  {author} {\bibfnamefont {T.}~\bibnamefont {Izubuchi}}, \bibinfo {author}
  {\bibfnamefont {L.}~\bibnamefont {Jin}}, \bibinfo {author} {\bibfnamefont
  {C.}~\bibnamefont {Jung}}, \ and\ \bibinfo {author} {\bibfnamefont
  {C.}~\bibnamefont {Lehner}},\ }\href {\doibase
  10.1103/PhysRevLett.118.022005} {\bibfield  {journal} {\bibinfo  {journal}
  {Phys. Rev. Lett.}\ }\textbf {\bibinfo {volume} {118}},\ \bibinfo {pages}
  {022005} (\bibinfo {year} {2017}{\natexlab{a}})},\ \Eprint
  {http://arxiv.org/abs/1610.04603} {arXiv:1610.04603 [hep-lat]} \BibitemShut
  {NoStop}%
\bibitem [{\citenamefont {Blum}\ \emph
  {et~al.}(2017{\natexlab{b}})\citenamefont {Blum}, \citenamefont {Christ},
  \citenamefont {Hayakawa}, \citenamefont {Izubuchi}, \citenamefont {Jin},
  \citenamefont {Jung},\ and\ \citenamefont {Lehner}}]{Blum:2017cer}%
  \BibitemOpen
  \bibfield  {author} {\bibinfo {author} {\bibfnamefont {T.}~\bibnamefont
  {Blum}}, \bibinfo {author} {\bibfnamefont {N.}~\bibnamefont {Christ}},
  \bibinfo {author} {\bibfnamefont {M.}~\bibnamefont {Hayakawa}}, \bibinfo
  {author} {\bibfnamefont {T.}~\bibnamefont {Izubuchi}}, \bibinfo {author}
  {\bibfnamefont {L.}~\bibnamefont {Jin}}, \bibinfo {author} {\bibfnamefont
  {C.}~\bibnamefont {Jung}}, \ and\ \bibinfo {author} {\bibfnamefont
  {C.}~\bibnamefont {Lehner}},\ }\href {\doibase 10.1103/PhysRevD.96.034515}
  {\bibfield  {journal} {\bibinfo  {journal} {Phys. Rev. D}\ }\textbf {\bibinfo
  {volume} {96}},\ \bibinfo {pages} {034515} (\bibinfo {year}
  {2017}{\natexlab{b}})},\ \Eprint {http://arxiv.org/abs/1705.01067}
  {arXiv:1705.01067 [hep-lat]} \BibitemShut {NoStop}%
\bibitem [{\citenamefont {Asmussen}\ \emph {et~al.}(2019)\citenamefont
  {Asmussen}, \citenamefont {Chao}, \citenamefont {G\'erardin}, \citenamefont
  {Green}, \citenamefont {Hudspith}, \citenamefont {Meyer},\ and\ \citenamefont
  {Nyffeler}}]{Asmussen:2019act}%
  \BibitemOpen
  \bibfield  {author} {\bibinfo {author} {\bibfnamefont {N.}~\bibnamefont
  {Asmussen}}, \bibinfo {author} {\bibfnamefont {E.-H.}\ \bibnamefont {Chao}},
  \bibinfo {author} {\bibfnamefont {A.}~\bibnamefont {G\'erardin}}, \bibinfo
  {author} {\bibfnamefont {J.~R.}\ \bibnamefont {Green}}, \bibinfo {author}
  {\bibfnamefont {R.~J.}\ \bibnamefont {Hudspith}}, \bibinfo {author}
  {\bibfnamefont {H.~B.}\ \bibnamefont {Meyer}}, \ and\ \bibinfo {author}
  {\bibfnamefont {A.}~\bibnamefont {Nyffeler}},\ }\href {\doibase
  10.22323/1.363.0195} {\bibfield  {journal} {\bibinfo  {journal} {PoS}\
  }\textbf {\bibinfo {volume} {LATTICE2019}},\ \bibinfo {pages} {195} (\bibinfo
  {year} {2019})},\ \Eprint {http://arxiv.org/abs/1911.05573} {arXiv:1911.05573
  [hep-lat]} \BibitemShut {NoStop}%
\bibitem [{\citenamefont {Dai}\ and\ \citenamefont
  {Pennington}(2014{\natexlab{a}})}]{Dai:2014lza}%
  \BibitemOpen
  \bibfield  {author} {\bibinfo {author} {\bibfnamefont {L.-Y.}\ \bibnamefont
  {Dai}}\ and\ \bibinfo {author} {\bibfnamefont {M.}~\bibnamefont
  {Pennington}},\ }\href {\doibase 10.1016/j.physletb.2014.07.005} {\bibfield
  {journal} {\bibinfo  {journal} {Phys. Lett. B}\ }\textbf {\bibinfo {volume}
  {736}},\ \bibinfo {pages} {11} (\bibinfo {year} {2014}{\natexlab{a}})},\
  \Eprint {http://arxiv.org/abs/1403.7514} {arXiv:1403.7514 [hep-ph]}
  \BibitemShut {NoStop}%
\bibitem [{\citenamefont {Dai}\ and\ \citenamefont
  {Pennington}(2014{\natexlab{b}})}]{Dai:2014zta}%
  \BibitemOpen
  \bibfield  {author} {\bibinfo {author} {\bibfnamefont {L.-Y.}\ \bibnamefont
  {Dai}}\ and\ \bibinfo {author} {\bibfnamefont {M.~R.}\ \bibnamefont
  {Pennington}},\ }\href {\doibase 10.1103/PhysRevD.90.036004} {\bibfield
  {journal} {\bibinfo  {journal} {Phys. Rev. D}\ }\textbf {\bibinfo {volume}
  {90}},\ \bibinfo {pages} {036004} (\bibinfo {year} {2014}{\natexlab{b}})},\
  \Eprint {http://arxiv.org/abs/1404.7524} {arXiv:1404.7524 [hep-ph]}
  \BibitemShut {NoStop}%
\bibitem [{\citenamefont {Dai}\ and\ \citenamefont
  {Pennington}(2016)}]{Dai:2016ytz}%
  \BibitemOpen
  \bibfield  {author} {\bibinfo {author} {\bibfnamefont {L.-Y.}\ \bibnamefont
  {Dai}}\ and\ \bibinfo {author} {\bibfnamefont {M.~R.}\ \bibnamefont
  {Pennington}},\ }\href {\doibase 10.1103/PhysRevD.94.116021} {\bibfield
  {journal} {\bibinfo  {journal} {Phys. Rev. D}\ }\textbf {\bibinfo {volume}
  {94}},\ \bibinfo {pages} {116021} (\bibinfo {year} {2016})},\ \Eprint
  {http://arxiv.org/abs/1611.04441} {arXiv:1611.04441 [hep-ph]} \BibitemShut
  {NoStop}%
\bibitem [{\citenamefont {Dai}\ and\ \citenamefont
  {Pennington}(2017)}]{Dai:2017cvz}%
  \BibitemOpen
  \bibfield  {author} {\bibinfo {author} {\bibfnamefont {L.-Y.}\ \bibnamefont
  {Dai}}\ and\ \bibinfo {author} {\bibfnamefont {M.}~\bibnamefont
  {Pennington}},\ }\href {\doibase 10.1103/PhysRevD.95.056007} {\bibfield
  {journal} {\bibinfo  {journal} {Phys. Rev. D}\ }\textbf {\bibinfo {volume}
  {95}},\ \bibinfo {pages} {056007} (\bibinfo {year} {2017})},\ \Eprint
  {http://arxiv.org/abs/1701.04460} {arXiv:1701.04460 [hep-ph]} \BibitemShut
  {NoStop}%
\bibitem [{Note1()}]{Note1}%
  \BibitemOpen
  \bibinfo {note} {We note that in the early works \cite
  {Terazawa:1968jh,Terazawa:1969ih}, the upper limit of HVP contribution has
  been given.}\BibitemShut {Stop}%
\bibitem [{\citenamefont {Davier}\ \emph {et~al.}(2020)\citenamefont {Davier},
  \citenamefont {Hoecker}, \citenamefont {Malaescu},\ and\ \citenamefont
  {Zhang}}]{Davier:2019can}%
  \BibitemOpen
  \bibfield  {author} {\bibinfo {author} {\bibfnamefont {M.}~\bibnamefont
  {Davier}}, \bibinfo {author} {\bibfnamefont {A.}~\bibnamefont {Hoecker}},
  \bibinfo {author} {\bibfnamefont {B.}~\bibnamefont {Malaescu}}, \ and\
  \bibinfo {author} {\bibfnamefont {Z.}~\bibnamefont {Zhang}},\ }\href
  {\doibase 10.1140/epjc/s10052-020-7792-2} {\bibfield  {journal} {\bibinfo
  {journal} {Eur. Phys. J. C}\ }\textbf {\bibinfo {volume} {80}},\ \bibinfo
  {pages} {241} (\bibinfo {year} {2020})},\ \bibinfo {note} {[Erratum:
  Eur.Phys.J.C 80, 410 (2020)]},\ \Eprint {http://arxiv.org/abs/1908.00921}
  {arXiv:1908.00921 [hep-ph]} \BibitemShut {NoStop}%
\bibitem [{\citenamefont {Keshavarzi}\ \emph {et~al.}(2020)\citenamefont
  {Keshavarzi}, \citenamefont {Nomura},\ and\ \citenamefont
  {Teubner}}]{Keshavarzi:2019abf}%
  \BibitemOpen
  \bibfield  {author} {\bibinfo {author} {\bibfnamefont {A.}~\bibnamefont
  {Keshavarzi}}, \bibinfo {author} {\bibfnamefont {D.}~\bibnamefont {Nomura}},
  \ and\ \bibinfo {author} {\bibfnamefont {T.}~\bibnamefont {Teubner}},\ }\href
  {\doibase 10.1103/PhysRevD.101.014029} {\bibfield  {journal} {\bibinfo
  {journal} {Phys. Rev. D}\ }\textbf {\bibinfo {volume} {101}},\ \bibinfo
  {pages} {014029} (\bibinfo {year} {2020})},\ \Eprint
  {http://arxiv.org/abs/1911.00367} {arXiv:1911.00367 [hep-ph]} \BibitemShut
  {NoStop}%
\bibitem [{\citenamefont {Kurz}\ \emph {et~al.}(2014)\citenamefont {Kurz},
  \citenamefont {Liu}, \citenamefont {Marquard},\ and\ \citenamefont
  {Steinhauser}}]{Kurz:2014wya}%
  \BibitemOpen
  \bibfield  {author} {\bibinfo {author} {\bibfnamefont {A.}~\bibnamefont
  {Kurz}}, \bibinfo {author} {\bibfnamefont {T.}~\bibnamefont {Liu}}, \bibinfo
  {author} {\bibfnamefont {P.}~\bibnamefont {Marquard}}, \ and\ \bibinfo
  {author} {\bibfnamefont {M.}~\bibnamefont {Steinhauser}},\ }\href {\doibase
  10.1016/j.physletb.2014.05.043} {\bibfield  {journal} {\bibinfo  {journal}
  {Phys. Lett. B}\ }\textbf {\bibinfo {volume} {734}},\ \bibinfo {pages} {144}
  (\bibinfo {year} {2014})},\ \Eprint {http://arxiv.org/abs/1403.6400}
  {arXiv:1403.6400 [hep-ph]} \BibitemShut {NoStop}%
\bibitem [{\citenamefont {Aul'chenko}\ \emph {et~al.}(2015)\citenamefont
  {Aul'chenko} \emph {et~al.}}]{Aulchenko:2015mwt}%
  \BibitemOpen
  \bibfield  {author} {\bibinfo {author} {\bibfnamefont {V.~M.}\ \bibnamefont
  {Aul'chenko}} \emph {et~al.},\ }\href {\doibase 10.1134/S1063776115060023}
  {\bibfield  {journal} {\bibinfo  {journal} {J. Exp. Theor. Phys.}\ }\textbf
  {\bibinfo {volume} {121}},\ \bibinfo {pages} {27} (\bibinfo {year} {2015})},\
  \bibinfo {note} {[Zh. Eksp. Teor. Fiz.148,no.1,34(2015)]}\BibitemShut
  {NoStop}%
\bibitem [{\citenamefont {Ablikim}\ \emph
  {et~al.}(2019{\natexlab{a}})\citenamefont {Ablikim} \emph
  {et~al.}}]{Ablikim:2019sjw}%
  \BibitemOpen
  \bibfield  {author} {\bibinfo {author} {\bibfnamefont {M.}~\bibnamefont
  {Ablikim}} \emph {et~al.} (\bibinfo {collaboration} {BESIII}),\ }\href@noop
  {} {\  (\bibinfo {year} {2019}{\natexlab{a}})},\ \Eprint
  {http://arxiv.org/abs/1912.11208} {arXiv:1912.11208 [hep-ex]} \BibitemShut
  {NoStop}%
\bibitem [{\citenamefont {Aulchenko}\ \emph {et~al.}(2015)\citenamefont
  {Aulchenko} \emph {et~al.}}]{Aulchenko:2014vkn}%
  \BibitemOpen
  \bibfield  {author} {\bibinfo {author} {\bibfnamefont {V.~M.}\ \bibnamefont
  {Aulchenko}} \emph {et~al.} (\bibinfo {collaboration} {SND}),\ }\href
  {\doibase 10.1103/PhysRevD.91.052013} {\bibfield  {journal} {\bibinfo
  {journal} {Phys. Rev.}\ }\textbf {\bibinfo {volume} {D91}},\ \bibinfo {pages}
  {052013} (\bibinfo {year} {2015})},\ \Eprint {http://arxiv.org/abs/1412.1971}
  {arXiv:1412.1971 [hep-ex]} \BibitemShut {NoStop}%
\bibitem [{\citenamefont {Achasov}\ \emph {et~al.}(2018)\citenamefont {Achasov}
  \emph {et~al.}}]{Achasov:2017kqm}%
  \BibitemOpen
  \bibfield  {author} {\bibinfo {author} {\bibfnamefont {M.~N.}\ \bibnamefont
  {Achasov}} \emph {et~al.},\ }\href {\doibase 10.1103/PhysRevD.97.012008}
  {\bibfield  {journal} {\bibinfo  {journal} {Phys. Rev.}\ }\textbf {\bibinfo
  {volume} {D97}},\ \bibinfo {pages} {012008} (\bibinfo {year} {2018})},\
  \Eprint {http://arxiv.org/abs/1711.08862} {arXiv:1711.08862 [hep-ex]}
  \BibitemShut {NoStop}%
\bibitem [{\citenamefont {Gribanov}\ \emph {et~al.}(2020)\citenamefont
  {Gribanov} \emph {et~al.}}]{Gribanov:2019qgw}%
  \BibitemOpen
  \bibfield  {author} {\bibinfo {author} {\bibfnamefont {S.}~\bibnamefont
  {Gribanov}} \emph {et~al.},\ }\href {\doibase 10.1007/JHEP01(2020)112}
  {\bibfield  {journal} {\bibinfo  {journal} {JHEP}\ }\textbf {\bibinfo
  {volume} {01}},\ \bibinfo {pages} {112} (\bibinfo {year} {2020})},\ \Eprint
  {http://arxiv.org/abs/1907.08002} {arXiv:1907.08002 [hep-ex]} \BibitemShut
  {NoStop}%
\bibitem [{\citenamefont {Ablikim}\ \emph {et~al.}(2020)\citenamefont {Ablikim}
  \emph {et~al.}}]{Ablikim:2020wyk}%
  \BibitemOpen
  \bibfield  {author} {\bibinfo {author} {\bibfnamefont {M.}~\bibnamefont
  {Ablikim}} \emph {et~al.} (\bibinfo {collaboration} {BESIII}),\ }\href@noop
  {} {\  (\bibinfo {year} {2020})},\ \Eprint {http://arxiv.org/abs/2012.07360}
  {arXiv:2012.07360 [hep-ex]} \BibitemShut {NoStop}%
\bibitem [{\citenamefont {Lees}\ \emph {et~al.}(2012)\citenamefont {Lees} \emph
  {et~al.}}]{Lees:2012cj}%
  \BibitemOpen
  \bibfield  {author} {\bibinfo {author} {\bibfnamefont {J.~P.}\ \bibnamefont
  {Lees}} \emph {et~al.} (\bibinfo {collaboration} {BaBar}),\ }\href {\doibase
  10.1103/PhysRevD.86.032013} {\bibfield  {journal} {\bibinfo  {journal} {Phys.
  Rev.}\ }\textbf {\bibinfo {volume} {D86}},\ \bibinfo {pages} {032013}
  (\bibinfo {year} {2012})},\ \Eprint {http://arxiv.org/abs/1205.2228}
  {arXiv:1205.2228 [hep-ex]} \BibitemShut {NoStop}%
\bibitem [{\citenamefont {Ambrosino}\ \emph {et~al.}(2009)\citenamefont
  {Ambrosino} \emph {et~al.}}]{Ambrosino:2008aa}%
  \BibitemOpen
  \bibfield  {author} {\bibinfo {author} {\bibfnamefont {F.}~\bibnamefont
  {Ambrosino}} \emph {et~al.} (\bibinfo {collaboration} {KLOE}),\ }\href
  {\doibase 10.1016/j.physletb.2008.10.060} {\bibfield  {journal} {\bibinfo
  {journal} {Phys. Lett.}\ }\textbf {\bibinfo {volume} {B670}},\ \bibinfo
  {pages} {285} (\bibinfo {year} {2009})},\ \Eprint
  {http://arxiv.org/abs/0809.3950} {arXiv:0809.3950 [hep-ex]} \BibitemShut
  {NoStop}%
\bibitem [{\citenamefont {Ambrosino}\ \emph {et~al.}(2011)\citenamefont
  {Ambrosino} \emph {et~al.}}]{Ambrosino:2010bv}%
  \BibitemOpen
  \bibfield  {author} {\bibinfo {author} {\bibfnamefont {F.}~\bibnamefont
  {Ambrosino}} \emph {et~al.} (\bibinfo {collaboration} {KLOE}),\ }\href
  {\doibase 10.1016/j.physletb.2011.04.055} {\bibfield  {journal} {\bibinfo
  {journal} {Phys. Lett.}\ }\textbf {\bibinfo {volume} {B700}},\ \bibinfo
  {pages} {102} (\bibinfo {year} {2011})},\ \Eprint
  {http://arxiv.org/abs/1006.5313} {arXiv:1006.5313 [hep-ex]} \BibitemShut
  {NoStop}%
\bibitem [{\citenamefont {Babusci}\ \emph {et~al.}(2013)\citenamefont {Babusci}
  \emph {et~al.}}]{Babusci:2012rp}%
  \BibitemOpen
  \bibfield  {author} {\bibinfo {author} {\bibfnamefont {D.}~\bibnamefont
  {Babusci}} \emph {et~al.} (\bibinfo {collaboration} {KLOE}),\ }\href
  {\doibase 10.1016/j.physletb.2013.02.029} {\bibfield  {journal} {\bibinfo
  {journal} {Phys. Lett.}\ }\textbf {\bibinfo {volume} {B720}},\ \bibinfo
  {pages} {336} (\bibinfo {year} {2013})},\ \Eprint
  {http://arxiv.org/abs/1212.4524} {arXiv:1212.4524 [hep-ex]} \BibitemShut
  {NoStop}%
\bibitem [{\citenamefont {Anastasi}\ \emph {et~al.}(2018)\citenamefont
  {Anastasi} \emph {et~al.}}]{Anastasi:2017eio}%
  \BibitemOpen
  \bibfield  {author} {\bibinfo {author} {\bibfnamefont {A.}~\bibnamefont
  {Anastasi}} \emph {et~al.} (\bibinfo {collaboration} {KLOE-2}),\ }\href
  {\doibase 10.1007/JHEP03(2018)173} {\bibfield  {journal} {\bibinfo  {journal}
  {JHEP}\ }\textbf {\bibinfo {volume} {03}},\ \bibinfo {pages} {173} (\bibinfo
  {year} {2018})},\ \Eprint {http://arxiv.org/abs/1711.03085} {arXiv:1711.03085
  [hep-ex]} \BibitemShut {NoStop}%
\bibitem [{\citenamefont {Achasov}\ \emph {et~al.}(2020)\citenamefont {Achasov}
  \emph {et~al.}}]{Achasov:2020iys}%
  \BibitemOpen
  \bibfield  {author} {\bibinfo {author} {\bibfnamefont {M.}~\bibnamefont
  {Achasov}} \emph {et~al.} (\bibinfo {collaboration} {SND}),\ }\href@noop {}
  {\  (\bibinfo {year} {2020})},\ \Eprint {http://arxiv.org/abs/2004.00263}
  {arXiv:2004.00263 [hep-ex]} \BibitemShut {NoStop}%
\bibitem [{\citenamefont {Ablikim}\ \emph {et~al.}(2016)\citenamefont {Ablikim}
  \emph {et~al.}}]{Ablikim:2015orh}%
  \BibitemOpen
  \bibfield  {author} {\bibinfo {author} {\bibfnamefont {M.}~\bibnamefont
  {Ablikim}} \emph {et~al.} (\bibinfo {collaboration} {BESIII}),\ }\href
  {\doibase 10.1016/j.physletb.2015.11.043} {\bibfield  {journal} {\bibinfo
  {journal} {Phys. Lett. B}\ }\textbf {\bibinfo {volume} {753}},\ \bibinfo
  {pages} {629} (\bibinfo {year} {2016})},\ \Eprint
  {http://arxiv.org/abs/1507.08188} {arXiv:1507.08188 [hep-ex]} \BibitemShut
  {NoStop}%
\bibitem [{\citenamefont {Xiao}\ \emph {et~al.}(2018)\citenamefont {Xiao},
  \citenamefont {Dobbs}, \citenamefont {Tomaradze}, \citenamefont {Seth},\ and\
  \citenamefont {Bonvicini}}]{Xiao:2017dqv}%
  \BibitemOpen
  \bibfield  {author} {\bibinfo {author} {\bibfnamefont {T.}~\bibnamefont
  {Xiao}}, \bibinfo {author} {\bibfnamefont {S.}~\bibnamefont {Dobbs}},
  \bibinfo {author} {\bibfnamefont {A.}~\bibnamefont {Tomaradze}}, \bibinfo
  {author} {\bibfnamefont {K.~K.}\ \bibnamefont {Seth}}, \ and\ \bibinfo
  {author} {\bibfnamefont {G.}~\bibnamefont {Bonvicini}},\ }\href {\doibase
  10.1103/PhysRevD.97.032012} {\bibfield  {journal} {\bibinfo  {journal} {Phys.
  Rev. D}\ }\textbf {\bibinfo {volume} {97}},\ \bibinfo {pages} {032012}
  (\bibinfo {year} {2018})},\ \Eprint {http://arxiv.org/abs/1712.04530}
  {arXiv:1712.04530 [hep-ex]} \BibitemShut {NoStop}%
\bibitem [{\citenamefont {Aul'chenko}\ \emph {et~al.}(2005)\citenamefont
  {Aul'chenko} \emph {et~al.}}]{Aulchenko:2006na}%
  \BibitemOpen
  \bibfield  {author} {\bibinfo {author} {\bibfnamefont {V.}~\bibnamefont
  {Aul'chenko}} \emph {et~al.} (\bibinfo {collaboration} {CMD-2}),\ }\href
  {\doibase 10.1134/1.2175241} {\bibfield  {journal} {\bibinfo  {journal} {JETP
  Lett.}\ }\textbf {\bibinfo {volume} {82}},\ \bibinfo {pages} {743} (\bibinfo
  {year} {2005})},\ \Eprint {http://arxiv.org/abs/hep-ex/0603021}
  {arXiv:hep-ex/0603021} \BibitemShut {NoStop}%
\bibitem [{\citenamefont {Aul'chenko}\ \emph {et~al.}(2006)\citenamefont
  {Aul'chenko} \emph {et~al.}}]{Akhmetshin:2006wh}%
  \BibitemOpen
  \bibfield  {author} {\bibinfo {author} {\bibfnamefont {V.}~\bibnamefont
  {Aul'chenko}} \emph {et~al.},\ }\href {\doibase 10.1134/S0021364006200021}
  {\bibfield  {journal} {\bibinfo  {journal} {JETP Lett.}\ }\textbf {\bibinfo
  {volume} {84}},\ \bibinfo {pages} {413} (\bibinfo {year} {2006})},\ \Eprint
  {http://arxiv.org/abs/hep-ex/0610016} {arXiv:hep-ex/0610016} \BibitemShut
  {NoStop}%
\bibitem [{\citenamefont {Akhmetshin}\ \emph {et~al.}(2007)\citenamefont
  {Akhmetshin} \emph {et~al.}}]{Akhmetshin:2006bx}%
  \BibitemOpen
  \bibfield  {author} {\bibinfo {author} {\bibfnamefont {R.}~\bibnamefont
  {Akhmetshin}} \emph {et~al.} (\bibinfo {collaboration} {CMD-2}),\ }\href
  {\doibase 10.1016/j.physletb.2007.01.073} {\bibfield  {journal} {\bibinfo
  {journal} {Phys. Lett. B}\ }\textbf {\bibinfo {volume} {648}},\ \bibinfo
  {pages} {28} (\bibinfo {year} {2007})},\ \Eprint
  {http://arxiv.org/abs/hep-ex/0610021} {arXiv:hep-ex/0610021} \BibitemShut
  {NoStop}%
\bibitem [{\citenamefont {Bisello}\ \emph {et~al.}(1989)\citenamefont {Bisello}
  \emph {et~al.}}]{Bisello:1988hq}%
  \BibitemOpen
  \bibfield  {author} {\bibinfo {author} {\bibfnamefont {D.}~\bibnamefont
  {Bisello}} \emph {et~al.} (\bibinfo {collaboration} {DM2}),\ }\href {\doibase
  10.1016/0370-2693(89)90060-9} {\bibfield  {journal} {\bibinfo  {journal}
  {Phys. Lett. B}\ }\textbf {\bibinfo {volume} {220}},\ \bibinfo {pages} {321}
  (\bibinfo {year} {1989})}\BibitemShut {NoStop}%
\bibitem [{\citenamefont {Barkov}\ \emph {et~al.}(1985)\citenamefont {Barkov}
  \emph {et~al.}}]{Barkov:1985ac}%
  \BibitemOpen
  \bibfield  {author} {\bibinfo {author} {\bibfnamefont {L.}~\bibnamefont
  {Barkov}} \emph {et~al.},\ }\href {\doibase 10.1016/0550-3213(85)90399-2}
  {\bibfield  {journal} {\bibinfo  {journal} {Nucl. Phys. B}\ }\textbf
  {\bibinfo {volume} {256}},\ \bibinfo {pages} {365} (\bibinfo {year}
  {1985})}\BibitemShut {NoStop}%
\bibitem [{\citenamefont {Achasov}\ \emph {et~al.}(2001)\citenamefont {Achasov}
  \emph {et~al.}}]{Achasov:2000am}%
  \BibitemOpen
  \bibfield  {author} {\bibinfo {author} {\bibfnamefont {M.}~\bibnamefont
  {Achasov}} \emph {et~al.},\ }\href {\doibase 10.1103/PhysRevD.63.072002}
  {\bibfield  {journal} {\bibinfo  {journal} {Phys. Rev. D}\ }\textbf {\bibinfo
  {volume} {63}},\ \bibinfo {pages} {072002} (\bibinfo {year} {2001})},\
  \Eprint {http://arxiv.org/abs/hep-ex/0009036} {arXiv:hep-ex/0009036}
  \BibitemShut {NoStop}%
\bibitem [{\citenamefont {Achasov}\ \emph {et~al.}(2007)\citenamefont {Achasov}
  \emph {et~al.}}]{Achasov:2007kg}%
  \BibitemOpen
  \bibfield  {author} {\bibinfo {author} {\bibfnamefont {M.}~\bibnamefont
  {Achasov}} \emph {et~al.},\ }\href {\doibase 10.1103/PhysRevD.76.072012}
  {\bibfield  {journal} {\bibinfo  {journal} {Phys. Rev. D}\ }\textbf {\bibinfo
  {volume} {76}},\ \bibinfo {pages} {072012} (\bibinfo {year} {2007})},\
  \Eprint {http://arxiv.org/abs/0707.2279} {arXiv:0707.2279 [hep-ex]}
  \BibitemShut {NoStop}%
\bibitem [{\citenamefont {Achasov}\ \emph {et~al.}(2016)\citenamefont {Achasov}
  \emph {et~al.}}]{Achasov:2016lbc}%
  \BibitemOpen
  \bibfield  {author} {\bibinfo {author} {\bibfnamefont {M.}~\bibnamefont
  {Achasov}} \emph {et~al.},\ }\href {\doibase 10.1103/PhysRevD.94.112006}
  {\bibfield  {journal} {\bibinfo  {journal} {Phys. Rev. D}\ }\textbf {\bibinfo
  {volume} {94}},\ \bibinfo {pages} {112006} (\bibinfo {year} {2016})},\
  \Eprint {http://arxiv.org/abs/1608.08757} {arXiv:1608.08757 [hep-ex]}
  \BibitemShut {NoStop}%
\bibitem [{\citenamefont {Ablikim}\ \emph
  {et~al.}(2019{\natexlab{b}})\citenamefont {Ablikim} \emph
  {et~al.}}]{Ablikim:2018iyx}%
  \BibitemOpen
  \bibfield  {author} {\bibinfo {author} {\bibfnamefont {M.}~\bibnamefont
  {Ablikim}} \emph {et~al.} (\bibinfo {collaboration} {BESIII}),\ }\href
  {\doibase 10.1103/PhysRevD.99.032001} {\bibfield  {journal} {\bibinfo
  {journal} {Phys. Rev. D}\ }\textbf {\bibinfo {volume} {99}},\ \bibinfo
  {pages} {032001} (\bibinfo {year} {2019}{\natexlab{b}})},\ \Eprint
  {http://arxiv.org/abs/1811.08742} {arXiv:1811.08742 [hep-ex]} \BibitemShut
  {NoStop}%
\bibitem [{\citenamefont {Lees}\ \emph {et~al.}(2013)\citenamefont {Lees} \emph
  {et~al.}}]{Lees:2013gzt}%
  \BibitemOpen
  \bibfield  {author} {\bibinfo {author} {\bibfnamefont {J.}~\bibnamefont
  {Lees}} \emph {et~al.} (\bibinfo {collaboration} {BaBar}),\ }\href {\doibase
  10.1103/PhysRevD.88.032013} {\bibfield  {journal} {\bibinfo  {journal} {Phys.
  Rev. D}\ }\textbf {\bibinfo {volume} {88}},\ \bibinfo {pages} {032013}
  (\bibinfo {year} {2013})},\ \Eprint {http://arxiv.org/abs/1306.3600}
  {arXiv:1306.3600 [hep-ex]} \BibitemShut {NoStop}%
\bibitem [{\citenamefont {Akhmetshin}\ \emph {et~al.}(2008)\citenamefont
  {Akhmetshin} \emph {et~al.}}]{Akhmetshin:2008gz}%
  \BibitemOpen
  \bibfield  {author} {\bibinfo {author} {\bibfnamefont {R.}~\bibnamefont
  {Akhmetshin}} \emph {et~al.} (\bibinfo {collaboration} {CMD-2}),\ }\href
  {\doibase 10.1016/j.physletb.2008.09.053} {\bibfield  {journal} {\bibinfo
  {journal} {Phys. Lett. B}\ }\textbf {\bibinfo {volume} {669}},\ \bibinfo
  {pages} {217} (\bibinfo {year} {2008})},\ \Eprint
  {http://arxiv.org/abs/0804.0178} {arXiv:0804.0178 [hep-ex]} \BibitemShut
  {NoStop}%
\bibitem [{\citenamefont {Kozyrev}\ \emph {et~al.}(2018)\citenamefont {Kozyrev}
  \emph {et~al.}}]{Kozyrev:2017agm}%
  \BibitemOpen
  \bibfield  {author} {\bibinfo {author} {\bibfnamefont {E.}~\bibnamefont
  {Kozyrev}} \emph {et~al.},\ }\href {\doibase 10.1016/j.physletb.2018.01.079}
  {\bibfield  {journal} {\bibinfo  {journal} {Phys. Lett. B}\ }\textbf
  {\bibinfo {volume} {779}},\ \bibinfo {pages} {64} (\bibinfo {year} {2018})},\
  \Eprint {http://arxiv.org/abs/1710.02989} {arXiv:1710.02989 [hep-ex]}
  \BibitemShut {NoStop}%
\bibitem [{\citenamefont {Witten}(1983)}]{Witten:1983tw}%
  \BibitemOpen
  \bibfield  {author} {\bibinfo {author} {\bibfnamefont {E.}~\bibnamefont
  {Witten}},\ }\href {\doibase 10.1016/0550-3213(83)90063-9} {\bibfield
  {journal} {\bibinfo  {journal} {Nucl. Phys. B}\ }\textbf {\bibinfo {volume}
  {223}},\ \bibinfo {pages} {422} (\bibinfo {year} {1983})}\BibitemShut
  {NoStop}%
\bibitem [{\citenamefont {Wess}\ and\ \citenamefont
  {Zumino}(1971)}]{Wess:1971yu}%
  \BibitemOpen
  \bibfield  {author} {\bibinfo {author} {\bibfnamefont {J.}~\bibnamefont
  {Wess}}\ and\ \bibinfo {author} {\bibfnamefont {B.}~\bibnamefont {Zumino}},\
  }\href {\doibase 10.1016/0370-2693(71)90582-X} {\bibfield  {journal}
  {\bibinfo  {journal} {Phys. Lett. B}\ }\textbf {\bibinfo {volume} {37}},\
  \bibinfo {pages} {95} (\bibinfo {year} {1971})}\BibitemShut {NoStop}%
\bibitem [{Note2()}]{Note2}%
  \BibitemOpen
  \bibinfo {note} {Up to ${\protect \cal O}(p^4)$ at least, this setting
  depends on the realization of the spin-1 resonance fields. In Ref.~\cite
  {Ecker:1989yg}, it was proven that this assumption is correct if one uses the
  antisymmetric formulation for those fields, as we do.}\BibitemShut {Stop}%
\bibitem [{\citenamefont {Gasser}\ and\ \citenamefont
  {Leutwyler}(1982)}]{Gasser:1982ap}%
  \BibitemOpen
  \bibfield  {author} {\bibinfo {author} {\bibfnamefont {J.}~\bibnamefont
  {Gasser}}\ and\ \bibinfo {author} {\bibfnamefont {H.}~\bibnamefont
  {Leutwyler}},\ }\href {\doibase 10.1016/0370-1573(82)90035-7} {\bibfield
  {journal} {\bibinfo  {journal} {Phys. Rept.}\ }\textbf {\bibinfo {volume}
  {87}},\ \bibinfo {pages} {77} (\bibinfo {year} {1982})}\BibitemShut {NoStop}%
\bibitem [{\citenamefont {Arganda}\ \emph {et~al.}(2008)\citenamefont
  {Arganda}, \citenamefont {Herrero},\ and\ \citenamefont
  {Portol\'es}}]{Arganda:2008jj}%
  \BibitemOpen
  \bibfield  {author} {\bibinfo {author} {\bibfnamefont {E.}~\bibnamefont
  {Arganda}}, \bibinfo {author} {\bibfnamefont {M.}~\bibnamefont {Herrero}}, \
  and\ \bibinfo {author} {\bibfnamefont {J.}~\bibnamefont {Portol\'es}},\
  }\href {\doibase 10.1088/1126-6708/2008/06/079} {\bibfield  {journal}
  {\bibinfo  {journal} {JHEP}\ }\textbf {\bibinfo {volume} {06}},\ \bibinfo
  {pages} {079} (\bibinfo {year} {2008})},\ \Eprint
  {http://arxiv.org/abs/0803.2039} {arXiv:0803.2039 [hep-ph]} \BibitemShut
  {NoStop}%
\bibitem [{\citenamefont {Guerrero}\ and\ \citenamefont
  {Pich}(1997)}]{Guerrero:1997ku}%
  \BibitemOpen
  \bibfield  {author} {\bibinfo {author} {\bibfnamefont {F.}~\bibnamefont
  {Guerrero}}\ and\ \bibinfo {author} {\bibfnamefont {A.}~\bibnamefont
  {Pich}},\ }\href {\doibase 10.1016/S0370-2693(97)01070-8} {\bibfield
  {journal} {\bibinfo  {journal} {Phys. Lett. B}\ }\textbf {\bibinfo {volume}
  {412}},\ \bibinfo {pages} {382} (\bibinfo {year} {1997})},\ \Eprint
  {http://arxiv.org/abs/hep-ph/9707347} {arXiv:hep-ph/9707347} \BibitemShut
  {NoStop}%
\bibitem [{\citenamefont {Pich}\ and\ \citenamefont
  {Portol\'es}(2001)}]{Pich:2001pj}%
  \BibitemOpen
  \bibfield  {author} {\bibinfo {author} {\bibfnamefont {A.}~\bibnamefont
  {Pich}}\ and\ \bibinfo {author} {\bibfnamefont {J.}~\bibnamefont
  {Portol\'es}},\ }\href {\doibase 10.1103/PhysRevD.63.093005} {\bibfield
  {journal} {\bibinfo  {journal} {Phys. Rev. D}\ }\textbf {\bibinfo {volume}
  {63}},\ \bibinfo {pages} {093005} (\bibinfo {year} {2001})},\ \Eprint
  {http://arxiv.org/abs/hep-ph/0101194} {arXiv:hep-ph/0101194} \BibitemShut
  {NoStop}%
\bibitem [{\citenamefont {Miranda}\ and\ \citenamefont
  {Roig}(2018)}]{Miranda:2018cpf}%
  \BibitemOpen
  \bibfield  {author} {\bibinfo {author} {\bibfnamefont {J.}~\bibnamefont
  {Miranda}}\ and\ \bibinfo {author} {\bibfnamefont {P.}~\bibnamefont {Roig}},\
  }\href {\doibase 10.1007/JHEP11(2018)038} {\bibfield  {journal} {\bibinfo
  {journal} {JHEP}\ }\textbf {\bibinfo {volume} {11}},\ \bibinfo {pages} {038}
  (\bibinfo {year} {2018})},\ \Eprint {http://arxiv.org/abs/1806.09547}
  {arXiv:1806.09547 [hep-ph]} \BibitemShut {NoStop}%
\bibitem [{\citenamefont {G\'omez~Dumm}\ \emph {et~al.}(2000)\citenamefont
  {G\'omez~Dumm}, \citenamefont {Pich},\ and\ \citenamefont
  {Porto\'es}}]{GomezDumm:2000fz}%
  \BibitemOpen
  \bibfield  {author} {\bibinfo {author} {\bibfnamefont {D.}~\bibnamefont
  {G\'omez~Dumm}}, \bibinfo {author} {\bibfnamefont {A.}~\bibnamefont {Pich}},
  \ and\ \bibinfo {author} {\bibfnamefont {J.}~\bibnamefont {Porto\'es}},\
  }\href {\doibase 10.1103/PhysRevD.62.054014} {\bibfield  {journal} {\bibinfo
  {journal} {Phys. Rev. D}\ }\textbf {\bibinfo {volume} {62}},\ \bibinfo
  {pages} {054014} (\bibinfo {year} {2000})},\ \Eprint
  {http://arxiv.org/abs/hep-ph/0003320} {arXiv:hep-ph/0003320} \BibitemShut
  {NoStop}%
\bibitem [{Note3()}]{Note3}%
  \BibitemOpen
  \bibinfo {note} {Notice that $X=\pi ,\eta $ appear in the three pseudoscalar
  final state $\pi ^{+}\pi ^{-}\pi ^{0}$ and $\pi ^{+}\pi ^{-}\eta $,
  respectively and $X=\pi \pi ,KK$ denote the two pseudoscalar final state $\pi
  ^{+}\pi ^{-}$and $K^{+}K^{-}$, respectively.}\BibitemShut {Stop}%
\bibitem [{\citenamefont {Cordier}\ \emph {et~al.}(1980)\citenamefont {Cordier}
  \emph {et~al.}}]{Cordier:1979qg}%
  \BibitemOpen
  \bibfield  {author} {\bibinfo {author} {\bibfnamefont {A.}~\bibnamefont
  {Cordier}} \emph {et~al.},\ }\href {\doibase 10.1016/0550-3213(80)90157-1}
  {\bibfield  {journal} {\bibinfo  {journal} {Nucl. Phys. B}\ }\textbf
  {\bibinfo {volume} {172}},\ \bibinfo {pages} {13} (\bibinfo {year}
  {1980})}\BibitemShut {NoStop}%
\bibitem [{\citenamefont {Dolinsky}\ \emph {et~al.}(1991)\citenamefont
  {Dolinsky} \emph {et~al.}}]{Dolinsky:1991vq}%
  \BibitemOpen
  \bibfield  {author} {\bibinfo {author} {\bibfnamefont {S.}~\bibnamefont
  {Dolinsky}} \emph {et~al.},\ }\href {\doibase 10.1016/0370-1573(91)90127-8}
  {\bibfield  {journal} {\bibinfo  {journal} {Phys. Rept.}\ }\textbf {\bibinfo
  {volume} {202}},\ \bibinfo {pages} {99} (\bibinfo {year} {1991})}\BibitemShut
  {NoStop}%
\bibitem [{\citenamefont {Antonelli}\ \emph {et~al.}(1992)\citenamefont
  {Antonelli} \emph {et~al.}}]{Antonelli:1992jx}%
  \BibitemOpen
  \bibfield  {author} {\bibinfo {author} {\bibfnamefont {A.}~\bibnamefont
  {Antonelli}} \emph {et~al.} (\bibinfo {collaboration} {DM2}),\ }\href
  {\doibase 10.1007/BF01589702} {\bibfield  {journal} {\bibinfo  {journal} {Z.
  Phys. C}\ }\textbf {\bibinfo {volume} {56}},\ \bibinfo {pages} {15} (\bibinfo
  {year} {1992})}\BibitemShut {NoStop}%
\bibitem [{\citenamefont {Akhmetshin}\ \emph {et~al.}(2004)\citenamefont
  {Akhmetshin} \emph {et~al.}}]{Akhmetshin:2003zn}%
  \BibitemOpen
  \bibfield  {author} {\bibinfo {author} {\bibfnamefont {R.}~\bibnamefont
  {Akhmetshin}} \emph {et~al.} (\bibinfo {collaboration} {CMD-2}),\ }\href
  {\doibase 10.1016/j.physletb.2003.10.108} {\bibfield  {journal} {\bibinfo
  {journal} {Phys. Lett. B}\ }\textbf {\bibinfo {volume} {578}},\ \bibinfo
  {pages} {285} (\bibinfo {year} {2004})},\ \Eprint
  {http://arxiv.org/abs/hep-ex/0308008} {arXiv:hep-ex/0308008} \BibitemShut
  {NoStop}%
\bibitem [{\citenamefont {Akhmetshin}\ \emph {et~al.}(1998)\citenamefont
  {Akhmetshin} \emph {et~al.}}]{Akhmetshin:1998se}%
  \BibitemOpen
  \bibfield  {author} {\bibinfo {author} {\bibfnamefont {R.}~\bibnamefont
  {Akhmetshin}} \emph {et~al.},\ }\href {\doibase
  10.1016/S0370-2693(98)00826-0} {\bibfield  {journal} {\bibinfo  {journal}
  {Phys. Lett. B}\ }\textbf {\bibinfo {volume} {434}},\ \bibinfo {pages} {426}
  (\bibinfo {year} {1998})}\BibitemShut {NoStop}%
\bibitem [{\citenamefont {Akhmetshin}\ \emph
  {et~al.}(2000{\natexlab{a}})\citenamefont {Akhmetshin} \emph
  {et~al.}}]{Akhmetshin:2000ca}%
  \BibitemOpen
  \bibfield  {author} {\bibinfo {author} {\bibfnamefont {R.}~\bibnamefont
  {Akhmetshin}} \emph {et~al.} (\bibinfo {collaboration} {CMD-2}),\ }\href
  {\doibase 10.1016/S0370-2693(00)00123-4} {\bibfield  {journal} {\bibinfo
  {journal} {Phys. Lett. B}\ }\textbf {\bibinfo {volume} {476}},\ \bibinfo
  {pages} {33} (\bibinfo {year} {2000}{\natexlab{a}})},\ \Eprint
  {http://arxiv.org/abs/hep-ex/0002017} {arXiv:hep-ex/0002017} \BibitemShut
  {NoStop}%
\bibitem [{\citenamefont {Achasov}\ \emph {et~al.}(2003)\citenamefont {Achasov}
  \emph {et~al.}}]{Achasov:2003ir}%
  \BibitemOpen
  \bibfield  {author} {\bibinfo {author} {\bibfnamefont {M.}~\bibnamefont
  {Achasov}} \emph {et~al.},\ }\href {\doibase 10.1103/PhysRevD.68.052006}
  {\bibfield  {journal} {\bibinfo  {journal} {Phys. Rev. D}\ }\textbf {\bibinfo
  {volume} {68}},\ \bibinfo {pages} {052006} (\bibinfo {year} {2003})},\
  \Eprint {http://arxiv.org/abs/hep-ex/0305049} {arXiv:hep-ex/0305049}
  \BibitemShut {NoStop}%
\bibitem [{\citenamefont {Achasov}\ \emph {et~al.}(2002)\citenamefont {Achasov}
  \emph {et~al.}}]{Achasov:2002ud}%
  \BibitemOpen
  \bibfield  {author} {\bibinfo {author} {\bibfnamefont {M.}~\bibnamefont
  {Achasov}} \emph {et~al.},\ }\href {\doibase 10.1103/PhysRevD.66.032001}
  {\bibfield  {journal} {\bibinfo  {journal} {Phys. Rev. D}\ }\textbf {\bibinfo
  {volume} {66}},\ \bibinfo {pages} {032001} (\bibinfo {year} {2002})},\
  \Eprint {http://arxiv.org/abs/hep-ex/0201040} {arXiv:hep-ex/0201040}
  \BibitemShut {NoStop}%
\bibitem [{\citenamefont {Aubert}\ \emph {et~al.}(2004)\citenamefont {Aubert}
  \emph {et~al.}}]{Aubert:2004kj}%
  \BibitemOpen
  \bibfield  {author} {\bibinfo {author} {\bibfnamefont {B.}~\bibnamefont
  {Aubert}} \emph {et~al.} (\bibinfo {collaboration} {BaBar}),\ }\href
  {\doibase 10.1103/PhysRevD.70.072004} {\bibfield  {journal} {\bibinfo
  {journal} {Phys. Rev. D}\ }\textbf {\bibinfo {volume} {70}},\ \bibinfo
  {pages} {072004} (\bibinfo {year} {2004})},\ \Eprint
  {http://arxiv.org/abs/hep-ex/0408078} {arXiv:hep-ex/0408078} \BibitemShut
  {NoStop}%
\bibitem [{\citenamefont {Antonelli}\ \emph {et~al.}(1988)\citenamefont
  {Antonelli} \emph {et~al.}}]{Antonelli:1988fw}%
  \BibitemOpen
  \bibfield  {author} {\bibinfo {author} {\bibfnamefont {A.}~\bibnamefont
  {Antonelli}} \emph {et~al.} (\bibinfo {collaboration} {DM2}),\ }\href
  {\doibase 10.1016/0370-2693(88)91250-6} {\bibfield  {journal} {\bibinfo
  {journal} {Phys. Lett. B}\ }\textbf {\bibinfo {volume} {212}},\ \bibinfo
  {pages} {133} (\bibinfo {year} {1988})}\BibitemShut {NoStop}%
\bibitem [{\citenamefont {Akhmetshin}\ \emph
  {et~al.}(2000{\natexlab{b}})\citenamefont {Akhmetshin} \emph
  {et~al.}}]{Akhmetshin:2000wv}%
  \BibitemOpen
  \bibfield  {author} {\bibinfo {author} {\bibfnamefont {R.}~\bibnamefont
  {Akhmetshin}} \emph {et~al.} (\bibinfo {collaboration} {CMD-2}),\ }\href
  {\doibase 10.1016/S0370-2693(00)00937-0} {\bibfield  {journal} {\bibinfo
  {journal} {Phys. Lett. B}\ }\textbf {\bibinfo {volume} {489}},\ \bibinfo
  {pages} {125} (\bibinfo {year} {2000}{\natexlab{b}})},\ \Eprint
  {http://arxiv.org/abs/hep-ex/0009013} {arXiv:hep-ex/0009013} \BibitemShut
  {NoStop}%
\bibitem [{\citenamefont {Aubert}\ \emph {et~al.}(2007)\citenamefont {Aubert}
  \emph {et~al.}}]{Aubert:2007ef}%
  \BibitemOpen
  \bibfield  {author} {\bibinfo {author} {\bibfnamefont {B.}~\bibnamefont
  {Aubert}} \emph {et~al.} (\bibinfo {collaboration} {BaBar}),\ }\href
  {\doibase 10.1103/PhysRevD.76.092005} {\bibfield  {journal} {\bibinfo
  {journal} {Phys. Rev. D}\ }\textbf {\bibinfo {volume} {76}},\ \bibinfo
  {pages} {092005} (\bibinfo {year} {2007})},\ \bibinfo {note} {[Erratum:
  Phys.Rev.D 77, 119902 (2008)]},\ \Eprint {http://arxiv.org/abs/0708.2461}
  {arXiv:0708.2461 [hep-ex]} \BibitemShut {NoStop}%
\bibitem [{\citenamefont {Achasov}\ \emph {et~al.}(2014)\citenamefont {Achasov}
  \emph {et~al.}}]{ACHASOV:2014nra}%
  \BibitemOpen
  \bibfield  {author} {\bibinfo {author} {\bibfnamefont {M.~N.}\ \bibnamefont
  {Achasov}} \emph {et~al.},\ }\bibfield  {booktitle} {\emph {\bibinfo
  {booktitle} {{Proceedings, 9th International Workshop on e+ e- collisions
  from Phi to Psi (PHIPSI13): Rome, Italy, September 9-12, 2013}}},\ }\href
  {\doibase 10.1142/S2010194514603883} {\bibfield  {journal} {\bibinfo
  {journal} {Int. J. Mod. Phys. Conf. Ser.}\ }\textbf {\bibinfo {volume}
  {35}},\ \bibinfo {pages} {1460388} (\bibinfo {year} {2014})}\BibitemShut
  {NoStop}%
\bibitem [{\citenamefont {James}\ and\ \citenamefont
  {Roos}(1975)}]{James:1975dr}%
  \BibitemOpen
  \bibfield  {author} {\bibinfo {author} {\bibfnamefont {F.}~\bibnamefont
  {James}}\ and\ \bibinfo {author} {\bibfnamefont {M.}~\bibnamefont {Roos}},\
  }\href {\doibase 10.1016/0010-4655(75)90039-9} {\bibfield  {journal}
  {\bibinfo  {journal} {Comput. Phys. Commun.}\ }\textbf {\bibinfo {volume}
  {10}},\ \bibinfo {pages} {343} (\bibinfo {year} {1975})}\BibitemShut
  {NoStop}%
\bibitem [{\citenamefont {Moussallam}(1995)}]{Moussallam:1994xp}%
  \BibitemOpen
  \bibfield  {author} {\bibinfo {author} {\bibfnamefont {B.}~\bibnamefont
  {Moussallam}},\ }\href {\doibase 10.1103/PhysRevD.51.4939} {\bibfield
  {journal} {\bibinfo  {journal} {Phys. Rev. D}\ }\textbf {\bibinfo {volume}
  {51}},\ \bibinfo {pages} {4939} (\bibinfo {year} {1995})},\ \Eprint
  {http://arxiv.org/abs/hep-ph/9407402} {arXiv:hep-ph/9407402} \BibitemShut
  {NoStop}%
\bibitem [{\citenamefont {Moussallam}(1997)}]{Moussallam:1997xx}%
  \BibitemOpen
  \bibfield  {author} {\bibinfo {author} {\bibfnamefont {B.}~\bibnamefont
  {Moussallam}},\ }\href {\doibase 10.1016/S0550-3213(97)00464-1} {\bibfield
  {journal} {\bibinfo  {journal} {Nucl. Phys. B}\ }\textbf {\bibinfo {volume}
  {504}},\ \bibinfo {pages} {381} (\bibinfo {year} {1997})},\ \Eprint
  {http://arxiv.org/abs/hep-ph/9701400} {arXiv:hep-ph/9701400} \BibitemShut
  {NoStop}%
\bibitem [{\citenamefont {Knecht}\ \emph {et~al.}(1999)\citenamefont {Knecht},
  \citenamefont {Peris}, \citenamefont {Perrottet},\ and\ \citenamefont
  {de~Rafael}}]{Knecht:1999gb}%
  \BibitemOpen
  \bibfield  {author} {\bibinfo {author} {\bibfnamefont {M.}~\bibnamefont
  {Knecht}}, \bibinfo {author} {\bibfnamefont {S.}~\bibnamefont {Peris}},
  \bibinfo {author} {\bibfnamefont {M.}~\bibnamefont {Perrottet}}, \ and\
  \bibinfo {author} {\bibfnamefont {E.}~\bibnamefont {de~Rafael}},\ }\href
  {\doibase 10.1103/PhysRevLett.83.5230} {\bibfield  {journal} {\bibinfo
  {journal} {Phys. Rev. Lett.}\ }\textbf {\bibinfo {volume} {83}},\ \bibinfo
  {pages} {5230} (\bibinfo {year} {1999})},\ \Eprint
  {http://arxiv.org/abs/hep-ph/9908283} {arXiv:hep-ph/9908283} \BibitemShut
  {NoStop}%
\bibitem [{\citenamefont {Bijnens}\ \emph {et~al.}(2003)\citenamefont
  {Bijnens}, \citenamefont {Gamiz}, \citenamefont {Lipartia},\ and\
  \citenamefont {Prades}}]{Bijnens:2003rc}%
  \BibitemOpen
  \bibfield  {author} {\bibinfo {author} {\bibfnamefont {J.}~\bibnamefont
  {Bijnens}}, \bibinfo {author} {\bibfnamefont {E.}~\bibnamefont {Gamiz}},
  \bibinfo {author} {\bibfnamefont {E.}~\bibnamefont {Lipartia}}, \ and\
  \bibinfo {author} {\bibfnamefont {J.}~\bibnamefont {Prades}},\ }\href
  {\doibase 10.1088/1126-6708/2003/04/055} {\bibfield  {journal} {\bibinfo
  {journal} {JHEP}\ }\textbf {\bibinfo {volume} {04}},\ \bibinfo {pages} {055}
  (\bibinfo {year} {2003})},\ \Eprint {http://arxiv.org/abs/hep-ph/0304222}
  {arXiv:hep-ph/0304222} \BibitemShut {NoStop}%
\bibitem [{\citenamefont {Gourdin}\ and\ \citenamefont
  {De~Rafael}(1969)}]{Gourdin:1969dm}%
  \BibitemOpen
  \bibfield  {author} {\bibinfo {author} {\bibfnamefont {M.}~\bibnamefont
  {Gourdin}}\ and\ \bibinfo {author} {\bibfnamefont {E.}~\bibnamefont
  {De~Rafael}},\ }\href {\doibase 10.1016/0550-3213(69)90333-2} {\bibfield
  {journal} {\bibinfo  {journal} {Nucl. Phys. B}\ }\textbf {\bibinfo {volume}
  {10}},\ \bibinfo {pages} {667} (\bibinfo {year} {1969})}\BibitemShut
  {NoStop}%
\bibitem [{\citenamefont {Colangelo}\ \emph {et~al.}(2019)\citenamefont
  {Colangelo}, \citenamefont {Hoferichter},\ and\ \citenamefont
  {Stoffer}}]{Colangelo:2018mtw}%
  \BibitemOpen
  \bibfield  {author} {\bibinfo {author} {\bibfnamefont {G.}~\bibnamefont
  {Colangelo}}, \bibinfo {author} {\bibfnamefont {M.}~\bibnamefont
  {Hoferichter}}, \ and\ \bibinfo {author} {\bibfnamefont {P.}~\bibnamefont
  {Stoffer}},\ }\href {\doibase 10.1007/JHEP02(2019)006} {\bibfield  {journal}
  {\bibinfo  {journal} {JHEP}\ }\textbf {\bibinfo {volume} {02}},\ \bibinfo
  {pages} {006} (\bibinfo {year} {2019})},\ \Eprint
  {http://arxiv.org/abs/1810.00007} {arXiv:1810.00007 [hep-ph]} \BibitemShut
  {NoStop}%
\bibitem [{\citenamefont {Hoferichter}\ \emph {et~al.}(2019)\citenamefont
  {Hoferichter}, \citenamefont {Hoid},\ and\ \citenamefont
  {Kubis}}]{Hoferichter:2019mqg}%
  \BibitemOpen
  \bibfield  {author} {\bibinfo {author} {\bibfnamefont {M.}~\bibnamefont
  {Hoferichter}}, \bibinfo {author} {\bibfnamefont {B.-L.}\ \bibnamefont
  {Hoid}}, \ and\ \bibinfo {author} {\bibfnamefont {B.}~\bibnamefont {Kubis}},\
  }\href {\doibase 10.1007/JHEP08(2019)137} {\bibfield  {journal} {\bibinfo
  {journal} {JHEP}\ }\textbf {\bibinfo {volume} {08}},\ \bibinfo {pages} {137}
  (\bibinfo {year} {2019})},\ \Eprint {http://arxiv.org/abs/1907.01556}
  {arXiv:1907.01556 [hep-ph]} \BibitemShut {NoStop}%
\bibitem [{\citenamefont {Krause}(1997)}]{Krause:1996rf}%
  \BibitemOpen
  \bibfield  {author} {\bibinfo {author} {\bibfnamefont {B.}~\bibnamefont
  {Krause}},\ }\href {\doibase 10.1016/S0370-2693(96)01346-9} {\bibfield
  {journal} {\bibinfo  {journal} {Phys. Lett. B}\ }\textbf {\bibinfo {volume}
  {390}},\ \bibinfo {pages} {392} (\bibinfo {year} {1997})},\ \Eprint
  {http://arxiv.org/abs/hep-ph/9607259} {arXiv:hep-ph/9607259} \BibitemShut
  {NoStop}%
\bibitem [{\citenamefont {Terazawa}(1968)}]{Terazawa:1968jh}%
  \BibitemOpen
  \bibfield  {author} {\bibinfo {author} {\bibfnamefont {H.}~\bibnamefont
  {Terazawa}},\ }\href {\doibase 10.1143/PTP.39.1326} {\bibfield  {journal}
  {\bibinfo  {journal} {Prog. Theor. Phys.}\ }\textbf {\bibinfo {volume}
  {39}},\ \bibinfo {pages} {1326} (\bibinfo {year} {1968})}\BibitemShut
  {NoStop}%
\bibitem [{\citenamefont {Terazawa}(1969)}]{Terazawa:1969ih}%
  \BibitemOpen
  \bibfield  {author} {\bibinfo {author} {\bibfnamefont {H.}~\bibnamefont
  {Terazawa}},\ }\href {\doibase 10.1103/PhysRev.177.2159} {\bibfield
  {journal} {\bibinfo  {journal} {Phys. Rev.}\ }\textbf {\bibinfo {volume}
  {177}},\ \bibinfo {pages} {2159} (\bibinfo {year} {1969})}\BibitemShut
  {NoStop}%
\end{thebibliography}%

\end{document}